\documentclass[10pt,twocolumn]{article}

\usepackage{iccv}
\usepackage{times}
\usepackage{epsfig}
\usepackage{graphicx}
\usepackage{amsmath}
\usepackage{amssymb}
\usepackage{cases}
\usepackage{bm}
\usepackage{amsmath}
\usepackage{subfigure}
\usepackage{makecell}
\usepackage{setspace}
\usepackage{booktabs}
\usepackage{stackengine}
\usepackage{multirow}
\usepackage{float}
\usepackage{url}

\usepackage[pagebackref=true,breaklinks=true,letterpaper=true,colorlinks,bookmarks=false]{hyperref}

 \iccvfinalcopy 

\def\iccvPaperID{2818} 

\ificcvfinal\fi
\begin{document}

\title{Adaptive Unfolding Total Variation Network for Low-Light Image Enhancement}

\author{Chuanjun Zheng \quad Daming Shi\thanks{Corresponding author} \quad Wentian Shi\\
Shenzhen University\\
{\tt\small \{zhengchuanjun2019,shiwentian2018\}@email.szu.edu.cn,  dshi@szu.edu.cn}

}

\maketitle
\ificcvfinal\thispagestyle{empty}\fi

\begin{abstract}
Real-world low-light images suffer from two main degradations, namely, inevitable noise and poor visibility. Since the noise exhibits different levels, its estimation has been implemented in recent works when enhancing low-light images from raw Bayer space. When it comes to sRGB color space, the noise estimation becomes more complicated due to the effect of the image processing pipeline. Nevertheless, most existing enhancing algorithms in sRGB space only focus on the low visibility problem or suppress the noise under a hypothetical noise level, leading them impractical due to the lack of robustness.  To address this issue, we propose an adaptive unfolding total variation network (UTVNet), which approximates the noise level from the real sRGB low-light image by learning the balancing parameter in the model-based denoising method with total variation regularization. Meanwhile, we learn the noise level map by unrolling the corresponding minimization process for providing the inferences of smoothness and fidelity constraints. Guided by the noise level map,  our UTVNet can recover finer details and is more capable to suppress noise in real captured low-light scenes. Extensive experiments on real-world low-light images clearly demonstrate the superior performance of UTVNet over state-of-the-art methods. 

\end{abstract}

\begin{figure}
\begin{center}
	\subfigcapskip=-3pt 
	\subfigure[ Input]{
		\includegraphics[width=0.32\linewidth]{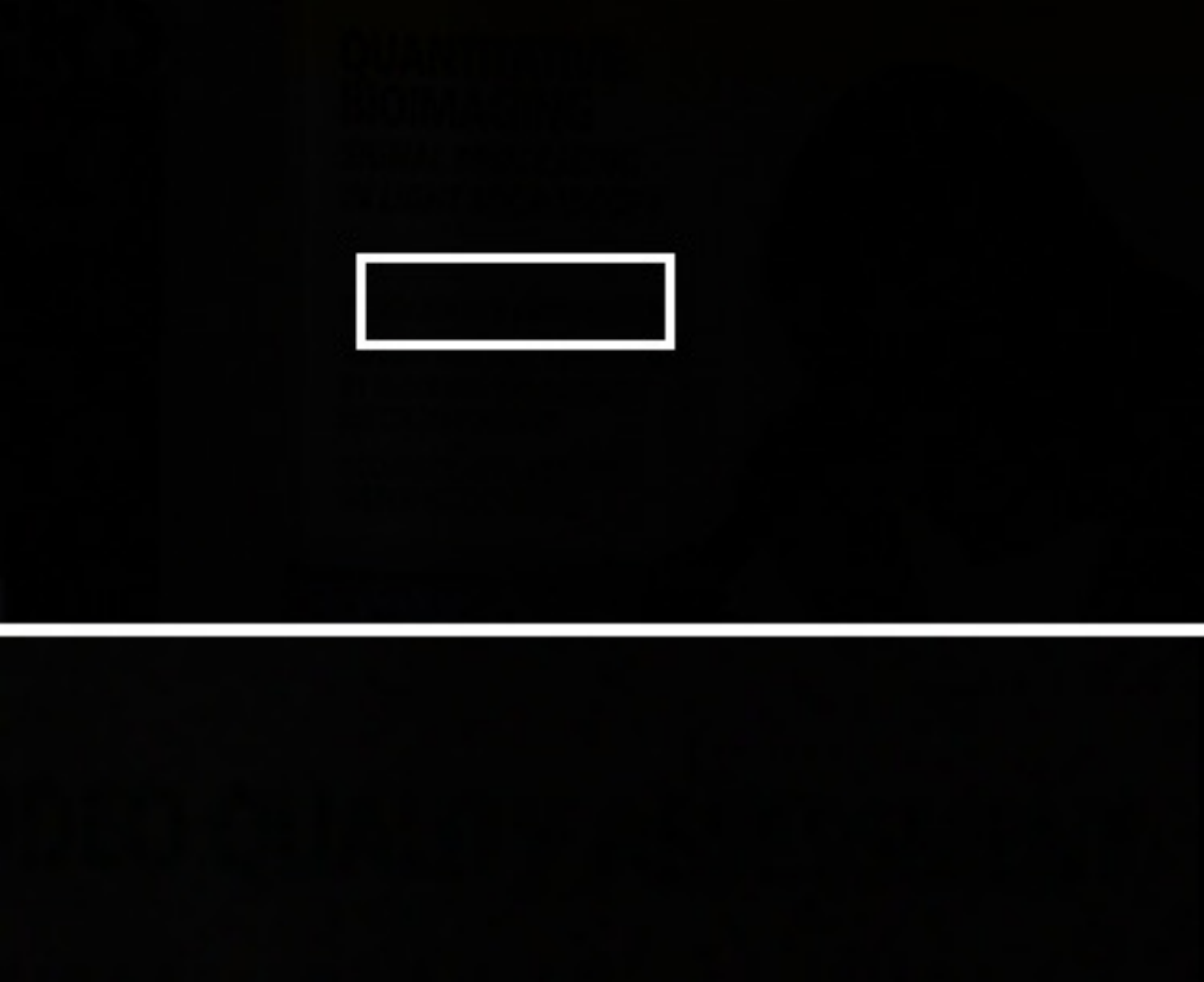}\hspace{-1mm}
		}\subfigure[LIME \cite{LIME} ]{
		\includegraphics[width=0.32\linewidth]{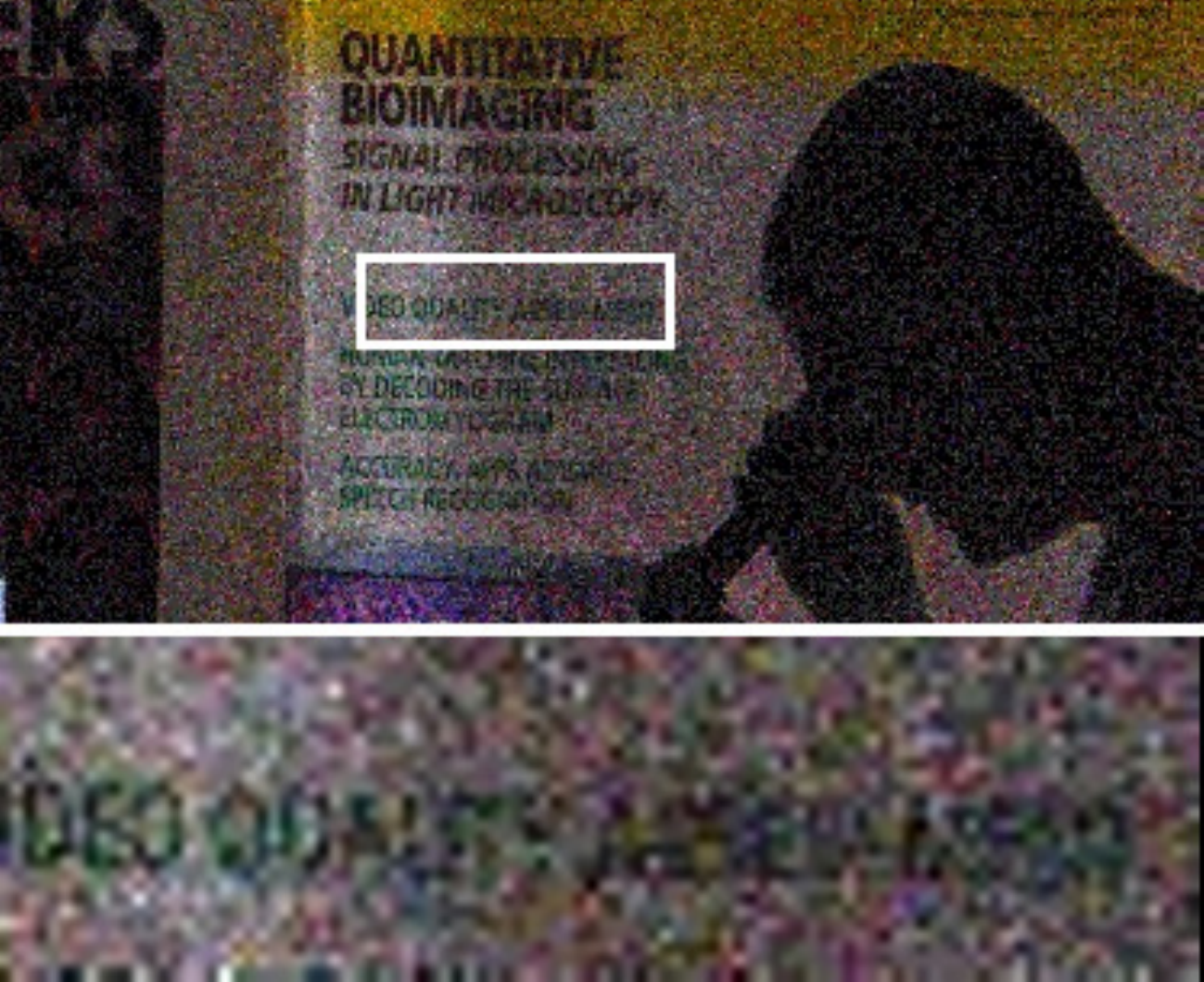}\hspace{-1mm}
		}\subfigure[LIME+NBNet \cite{nbn} ]{
		\includegraphics[width=0.32\linewidth]{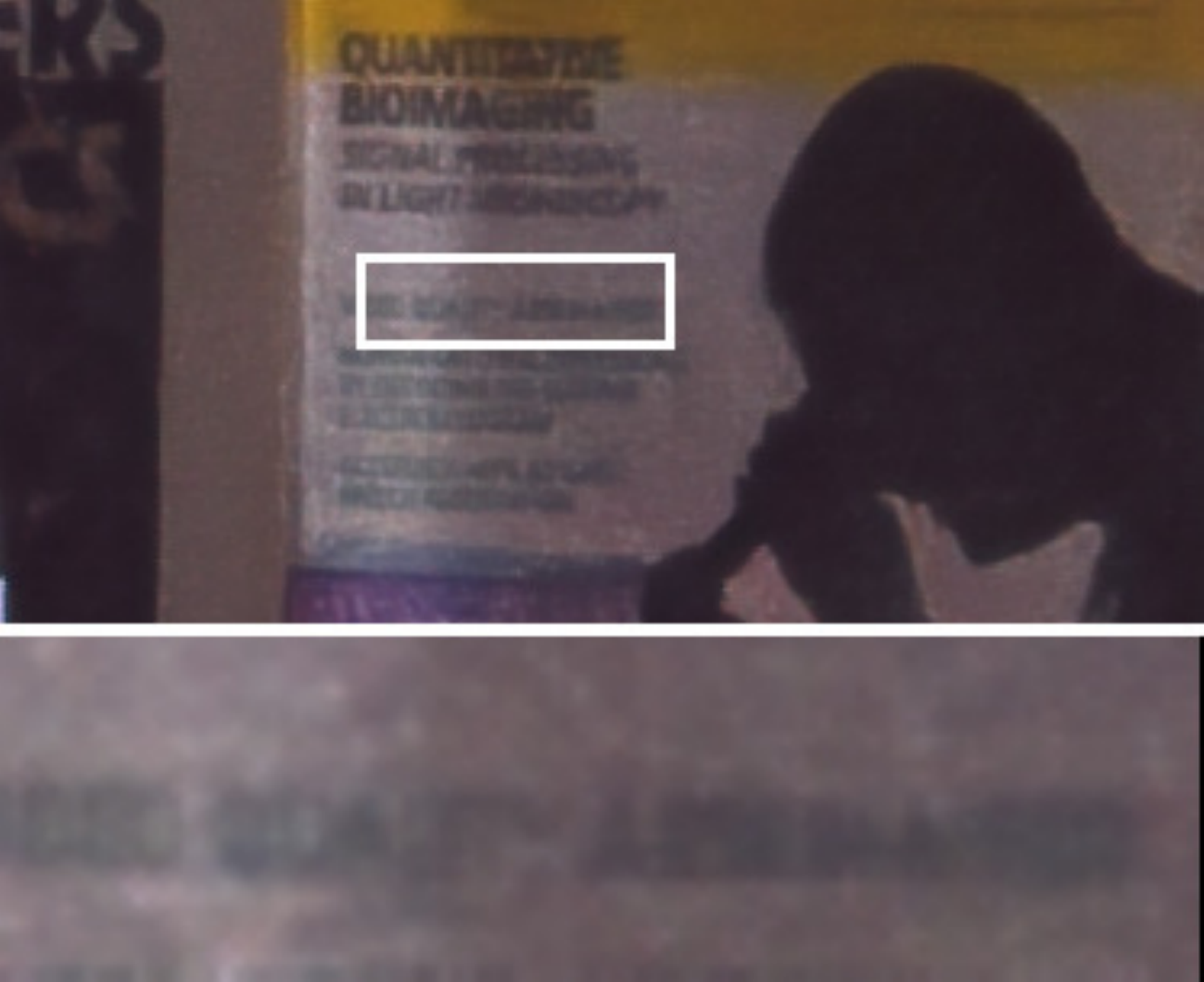}\hspace{-1mm}
		}\vspace{-0.3cm}
		
		\subfigure[NBNet+LIME ]{
		\includegraphics[width=0.32\linewidth]{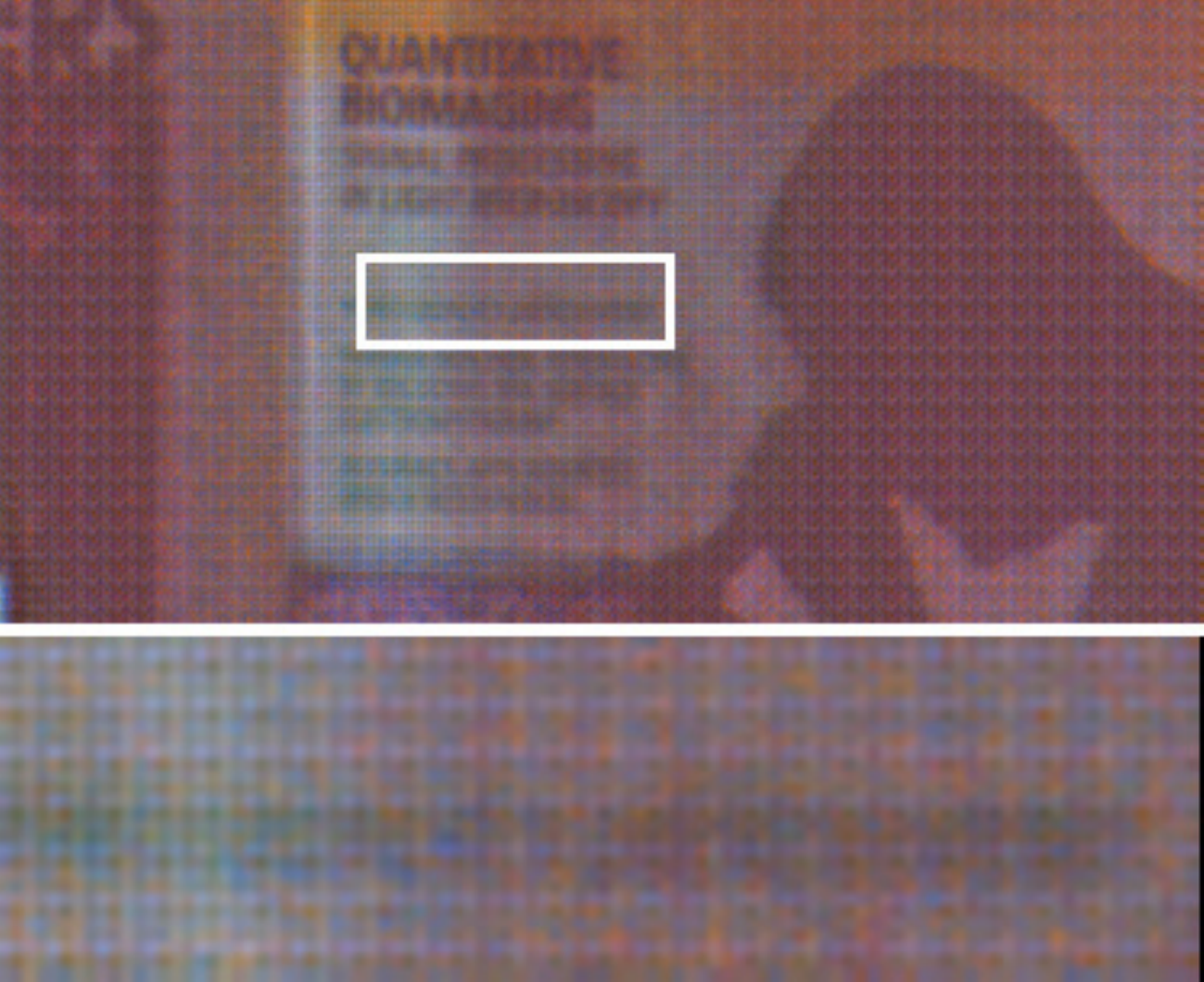}\hspace{-1mm}
		}\subfigure[AGLLNet \cite{atl}]{
		\includegraphics[width=0.32\linewidth]{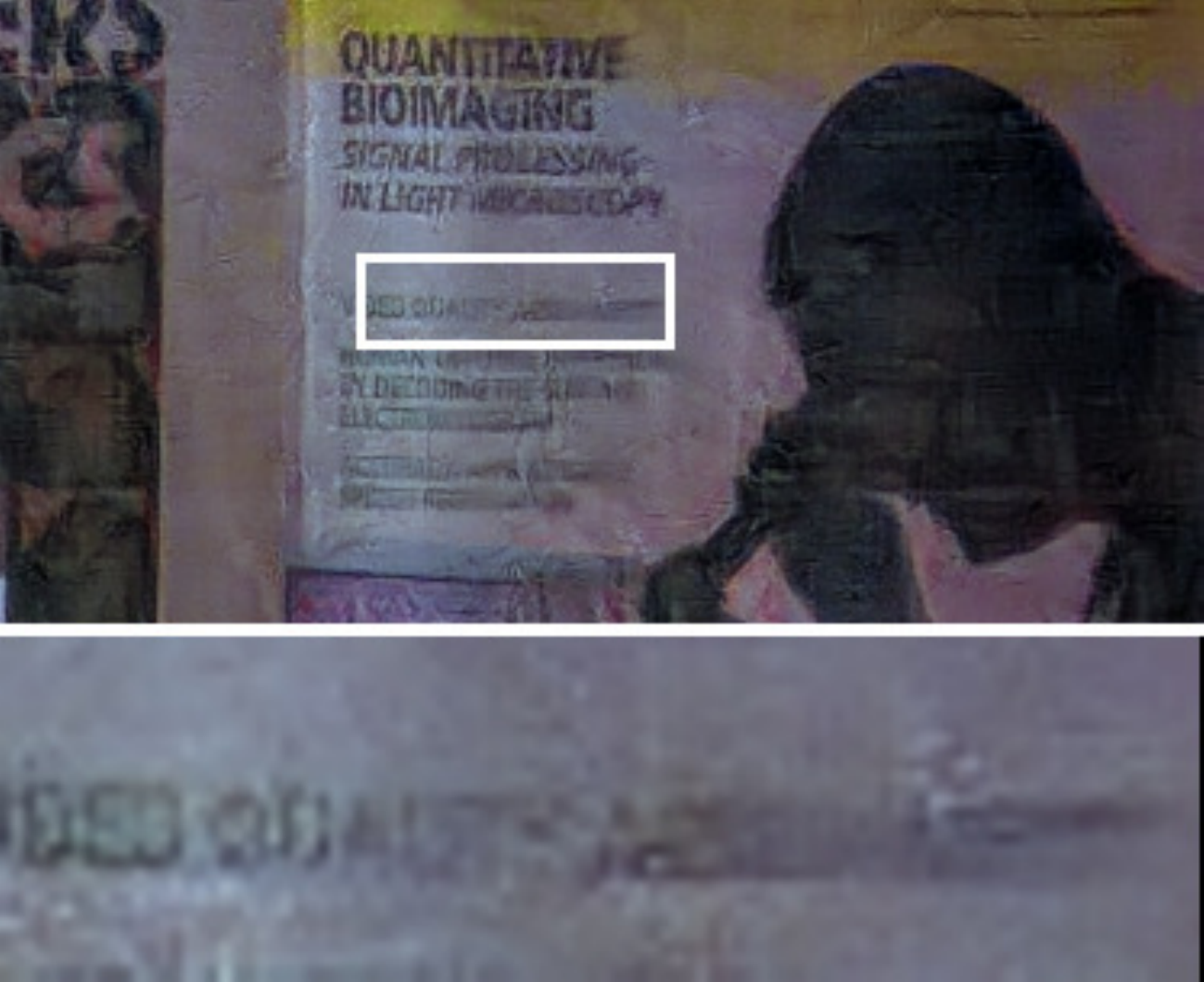}\hspace{-1mm}
		}\subfigure[D\&E \cite{LDE} ]{
		\includegraphics[width=0.32\linewidth]{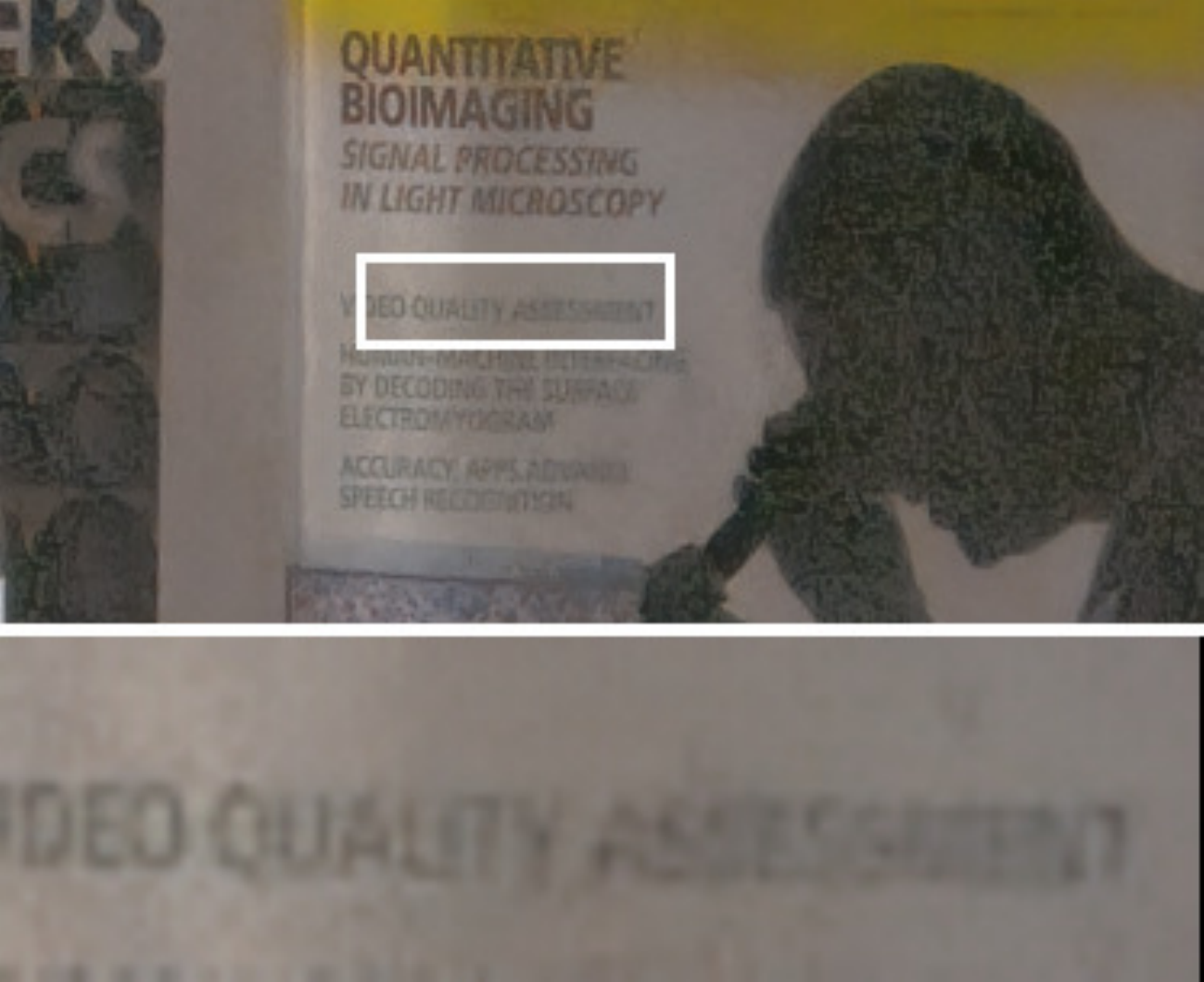}\hspace{-1mm}
		}\vspace{-0.3cm}		
		
		\subfigure[MIRNet \cite{MIRNet} ]{
		\includegraphics[width=0.32\linewidth]{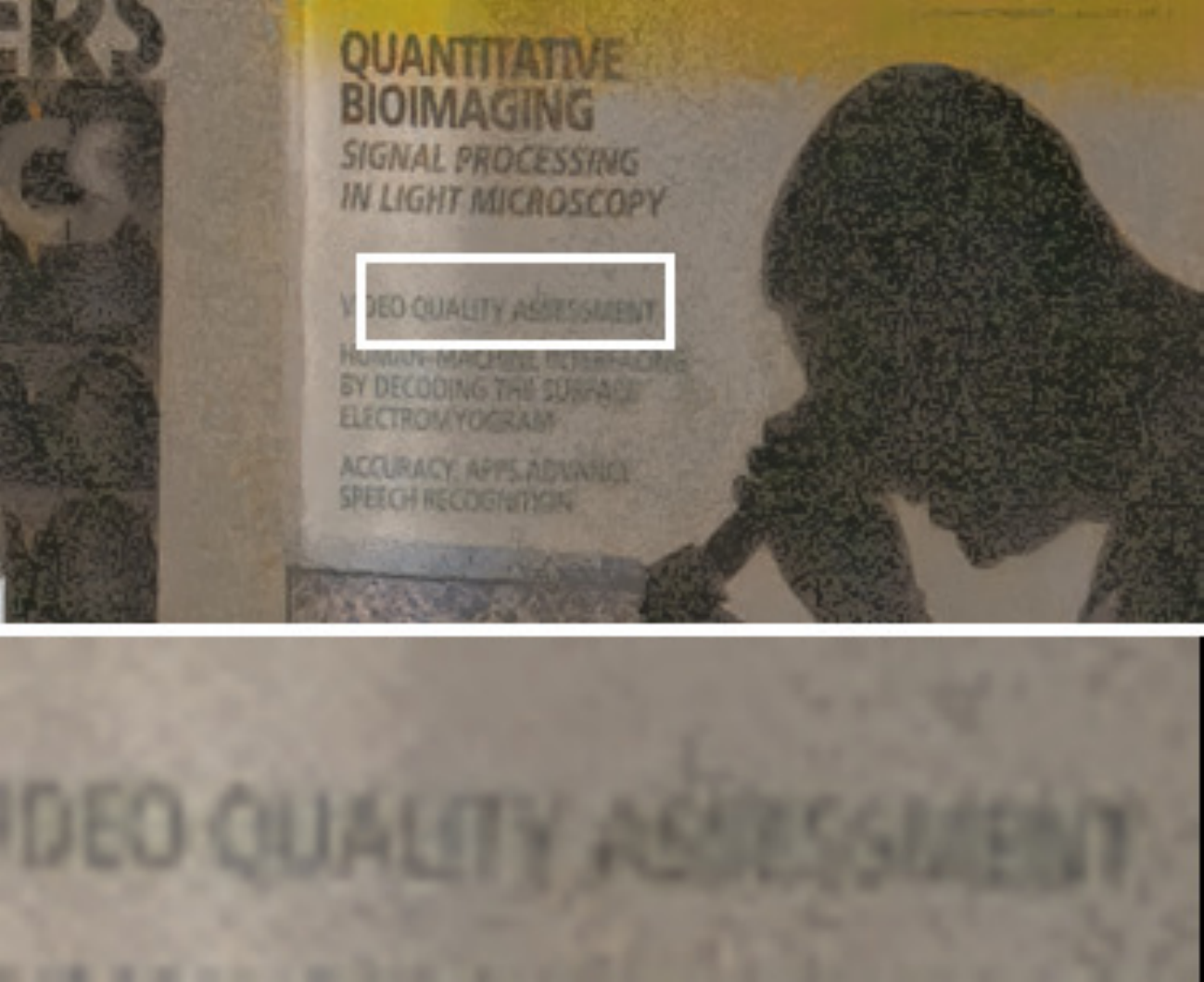}\hspace{-1mm}
		}\subfigure[ \textbf{UTVNet (Ours)} ]{
		\includegraphics[width=0.32\linewidth]{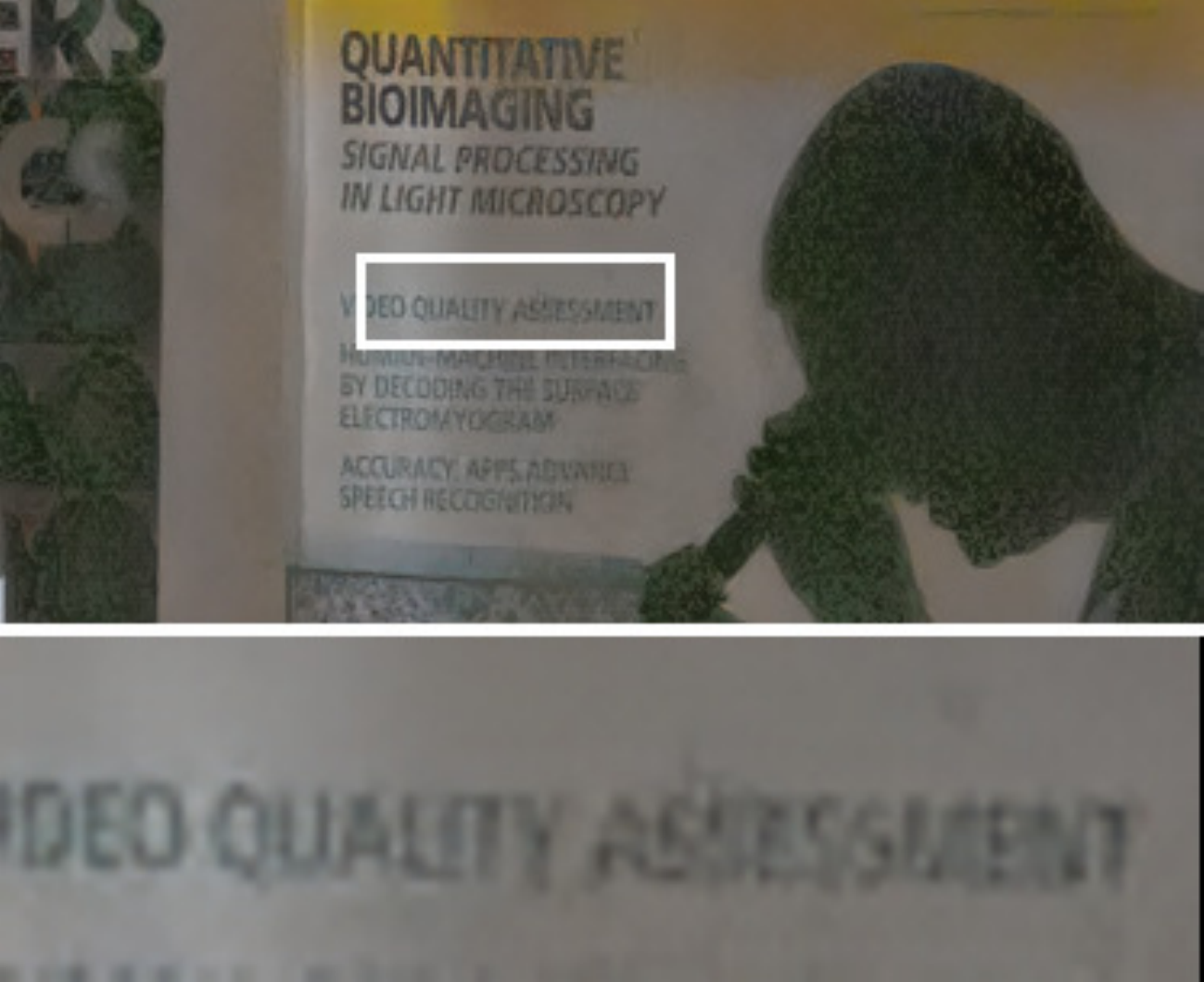}\hspace{-1mm}
		}\subfigure[Ground Truth ]{
		\includegraphics[width=0.32\linewidth]{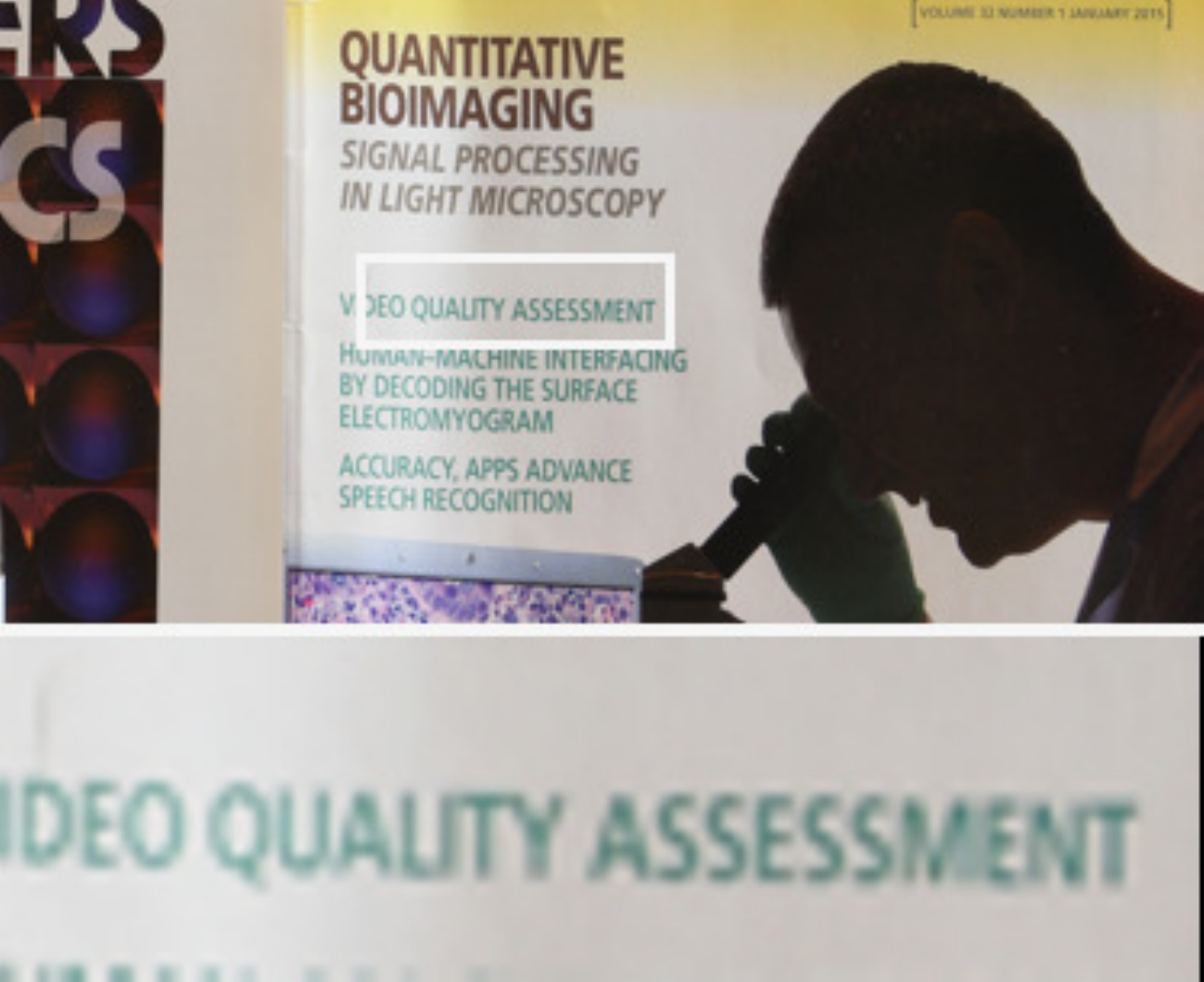}\hspace{-1mm}
		}\vspace{-0.3cm}

	\end{center}
	\caption{Performance in real captured low-light scene. (a)  a sRGB image captured by a NikonD850 camera from ELD dataset \cite{rewd}. (b-d) results of the luminance correction and denoising approaches. (f) and (g) are outputs of two recent methods without noise estimation. (e) the result of   \cite{atl} trained on the synthetic low-light dataset to predict noise maps. (h) the result of proposed  UTVNet.  \textit{All images were cropped to make their details more visible.
}}
\end{figure}
\section{Introduction}
The insufficient light provided by low-light scenarios is the main limitation of capturing high-quality photographs.    Besides enlarging the aperture and prolonging the exposure time to gather more light when taking images,  another flexible solution is to increase the ISO setting of camera to adjust the analog gain or both analog and digital gains as the compensation for the brightness. However, a higher ISO setting will decrease the signal-to-noise ratio (SNR). The SNR should be carefully managed to control the noise  for providing higher possibilities of post-processing \cite{googlo}. Unfortunately, such a manipulation is difficult for non-expert users who lack the appropriate photography skills. 

Restoration of real-world low-light images is a challenging process in computer vision, which focuses on recovering underlying clean images from observed noisy ones and correcting the luminance without amplifying the noise. Most existing methods \cite{LIME,whitebox, upe} focus on enhancing noise-free underexposed images. Such methods may solve the low visibility problem, but the noise will amplified when adjusting the exposure as shown in Fig.\;1 (b).  Meanwhile,  existing denoising approaches \cite{FFDnet,cbdnet,nbn} may fail to suppress noise for low-light images, since most of them focus on images taken under normal luminance, the small magnitude between pixels in extremely low-light images hinder the detection of noise from details as shown in Fig.\;1 (c, d).
 
 As real-world low-light images exhibit different noise levels, recent low-light image restoration methods \cite{LDE,MIRNet} may lack of robustness without noise estimation as shown in Fig.\;1 (f, g). In raw Bayer space, a physics-based noise formation model \cite{rewd} has been proposed for low-light raw images. When converting  raw data to sRGB, the distribution of noise is affected by the image processing pipeline (ISP). Existing method \cite{atl} learned noise maps by synthetic low-light image dataset, but the efficiency is degraded by the inaccurate noise model and the gap between synthetic images and real photographs as shown in Fig.\;1 (e). 
 
In this paper, an adaptive unfolding total variation network (UTVNet) is presented to restore the extremely low-light images in sRGB space. To boost the robustness in real captured scenes, we approximate the noise level inspired by the relationship between the noise level and the balancing parameter in model-based image denoising method. Specifically, we formulate an adaptive total variation (TV) regularization by introducing a learnable noise level map that serves as the balancing parameters for each pixel to control the trade-off between noise reduction and detail preservation. To learn the noise level map from the real low-light dataset, designing regularizers in loss function is infeasible. Therefore, we unfold the TV minimization process via the Alternating Direction Method of Multiplier (ADMM) \cite{adam} to separate the minimization problem into three sub-problems with closed-form solutions, which provides the inferences of fidelity and smoothness constraints for learning the noise level map. Subsequently, a noise-free low-frequency layer is automatically generated along with the noise level map. Since the noise is smoothed out by the TV regularization in the low-frequency layer, the luminance correction is applied in this layer to avoid noise amplification. Meanwhile, the noise suppression process is guided by the approximated noise level map in the high-frequency layer. The main contributions of this work are as follows:\begin{figure*}
\begin{center}
 \includegraphics[width=0.97\linewidth,height=6.6cm]{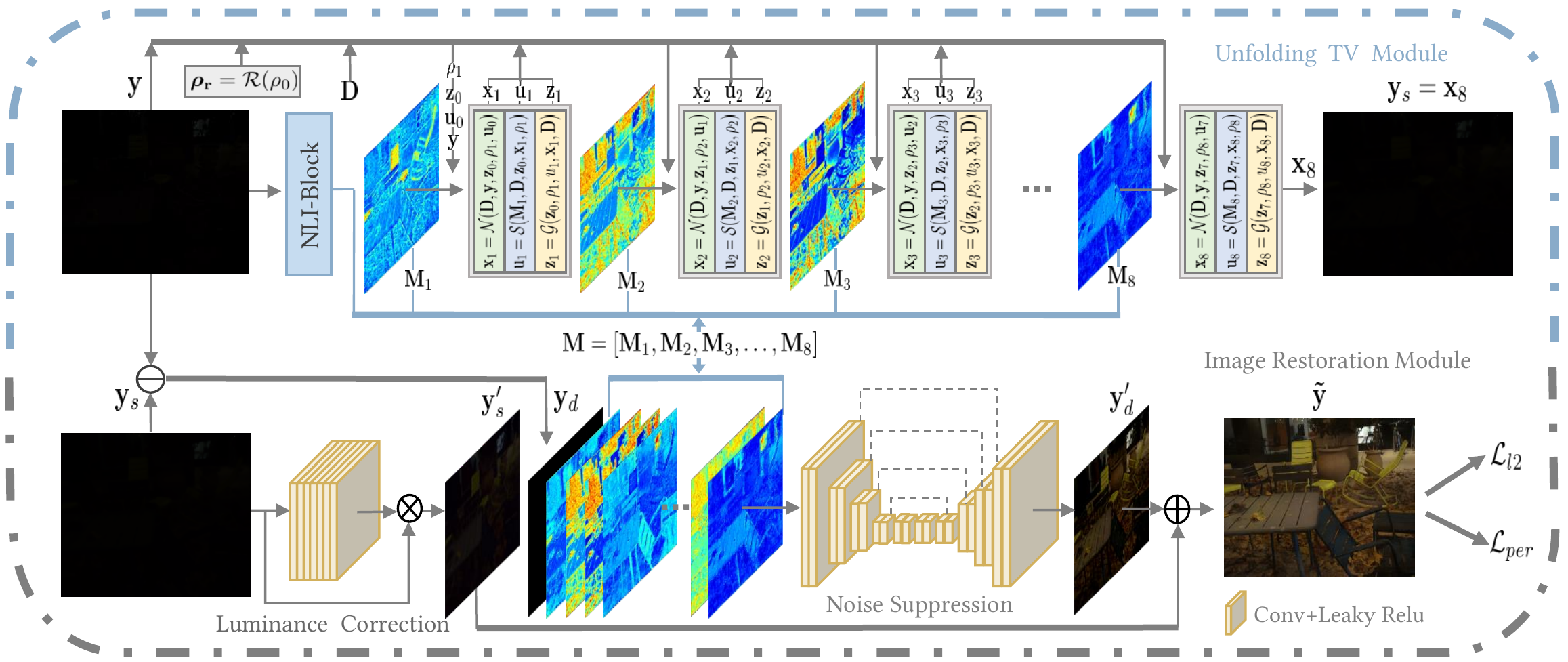}
\end{center}
\vspace{-0.2cm}
   \caption{The proposed UTVNet  consists of an unfolding TV module and an image restoration module. First, the given image $\mathbf{y}$ is  fed into unfolding TV module  to estimate noise level map $\mathbf{M}$ and generate noise-free smooth layer $\mathbf{y_{s}}$. Then, in image restoration module, $\mathbf{y_{s}}$ is used to correct the luminance, and the noise suppression in detail layer $\mathbf{y_{d}}$ is guided by the approximated noise level map $\mathbf{M}$. We describe the details of the two modules in Section 3. }
\label{fig:short}
\end{figure*}
\begin{itemize}

\item 
 By integrating a learnable nois level map, we formulate an  adaptive TV regularization  and propose a new noise level approximation method  for real low-light photographs in sRGB  color space.

\item We propose an UTVNet that unfolds the total variation minimization algorithms to provide fidelity and smoothness constraints to learn the noise level map without designing special loss function.

\item UTVNet is robust in restoring the  real captured low-light images with various noise levels. 
\end{itemize}

\section{Related Work}
  \textbf{Low-light image enhancement.} Most low-light image enhancement approaches only point to underexposure without considering noise suppression, like learning-based methods \cite{upe,csrnet, hdrnet}, and retinex-based methods \cite{LIME,crm}, in which most of the cases are noise-free, therefore, these methods handle the low-visibility problem only. Some other methods take noise suppression into account in raw Bayer space.\;Chen\;\emph{et al.}\;\cite{learningtosee} proposed a raw low-light image dataset to learning the mapping from raw short-exposure images to the long-exposure references. Wei \emph{et al.} \cite{rewd} proposed a physics-based noise formation model for low-light raw image denoising. However, as the noise is affected by the ISP during the conversion from raw to sRGB, modeling noise is more difficult in the sRGB color space.\;Lv \emph{et al.}  \cite{atl} constructed a large sRGB synthetic dataset to learn noise maps for low-light images.\;The performance is related to the gap between synthetic images and real photographs. Xu \emph{et al.}\;\cite{LDE} suppressed the noise for low-light images by learning a low-frequency layer of the input image using the reference generated by a guided filter \cite{guided}.  Zamir \emph{et al.}\;\cite{MIRNet} proposed a multi-scale residual block that selects useful sets of kernels from each branch using a self-attention approach. Without noise estimation, such noise suppression networks may lack robustness in real captured scenes.

\textbf{Image denoising.}
 Some blind denoising methods can handle the real-world image denoising tasks, these methods often combine noise estimation with non-blind denoising.  Xu \emph{et al.}\;\cite{nlic} proposed a multi-channel weighted unclear norm minimization model for real color image denoising. Nam\;\emph{et al.}\;\cite{nan} proposed a cross-channel noise model in sRGB space which took the correlation between the color channels into consideration.\;Among learning-based methods,  Zhang\;\emph{et al.}\;\cite{FFDnet} stretched the noise level as a noise-level map to deal with spatially variant noise.\;Cheng\;\emph{et al.}\;\cite{nbn}  proposed a network that separated signal and noise by learning  a set of reconstruction basis in the feature space. Guo\;\emph{et al.}\;\cite{cbdnet} generated synthetic images and adopted a sub-network to learn the mapping from estimated noise level to ground truth. Instead of generating such a dataset, our UTVNet aims to learn the noise level approximation directly from real low-light images.

\textbf{Deep unrolling architecture.}
Model-based denoising methods are flexible in handling noise with various levels. Most model-based denoising methods can be formulated as\begin{small}
\begin{equation}
\hat{\mathbf{x}}=\arg \min _{\mathbf{x}} \frac{1}{2 \sigma^{2}}\|\mathbf{x}-\mathbf{y}\|^{2}+\lambda \mathbf{\Phi}(\mathbf{x}),
\end{equation}
\end{small}where $\frac{1}{2 \sigma^{2}}\|\mathbf{x}-\mathbf{y}\|^{2}$ is the data fidelity term, $\sigma$ is the noise level , $\lambda$ is the balancing parameters, $\mathbf{\Phi}(\boldsymbol{x})$ is the regularization term, $\hat{\mathbf{x}}$ is the solution to the problem, $\mathbf{x}$ and $\mathbf{y}$ are clean image and observed image,\;respectively.\;The performance of these methods is associated with image priors, such as self-similarity \cite{nonelocal}, sparsity  \cite{sparsity}, and low-rank \cite{rank}. However, most model-based methods focus only on how to carefully design the denoising priors, the parameter $\lambda$ is neglected. It is often set as an empirical value. Recently, deep unrolling methods have been proposed that explicitly unroll iterative optimization algorithms into learnable deep architectures combined with convolutional neural networks (CNNs). ADMM \cite{admm} has been used in \cite{rui} to unroll the minimization process for image layer separation, and the convergence has been analysed in \cite{adcov}. Meanwhile, the half-quadratic splitting algorithm \cite{hqsp} has been utilized in \cite{dnopi,FFDnet,bygu,USRnet} to unfold the minimization problems for image denoising and super-resolution. Different from those methods that utilize CNNs to fit the physical priors in minimization problems, our UTVNet adopts the TV regularization and unfolds the minimization process for the fidelity and smoothness inferences.

\section{Methodology}
In this section, we introduce the proposed UTVNet. An overview of the process is shown in Fig.\;2.\;The network consists of two modules, including an unfolding TV module and an image restoration module. To approximate the noise level, we start with the analysis of the noise level and the balancing parameter in model-based denoising methods. Then we formulate an adaptive TV regularization and describe each module of the network in detail below.

\subsection{Noise Level vs. Balancing Parameter }

As described in previous work \cite{FFDnet}, when $\lambda$  is combined into $\sigma$ in Eq.\;(1), setting noise level $\sigma$  also plays the role of setting $\lambda$  to control the trade-off between noise reduction and detail preservation. Inspired by this relationship, as for certain regularization terms, if we combine $\lambda$  and $\sigma$ as the balancing parameter in Eq.\;(1), the noise level can be approximated by learning such balancing parameter via unrolling architecture. In our model, we adopt the TV  regularization term for three reasons. First, TV regularization can better preserve sharp edges \cite{tved,newyv} and also be used in enhancing images with poor visibility \cite{tipbio}. Second, there are no extra parameters of the original TV regularization, accordingly, the trade-off is controlled only by the balancing parameter, which is easy and accurate for approximation. Third, instead of using CNNs to learn the explicit mapping of solutions to the minimization problem, one can easily find closed-form solutions for the TV regularization term, which provide the fidelity and smoothness constraints without using special loss functions.

\subsection{Adaptive Total Variation Regularization }

We first define operator  $\mathbf{D}$ as a collection of two sub-operators $\mathbf{D}=[\mathbf{D}_{x}^{T}\;\mathbf{D}_{y}^{T} ]^{T}$, where $ \mathbf{D}_{x}, \mathbf{D}_{y}$ are the first-order forward finite-difference operators along the horizontal and vertical directions respectively. Thus, the anisotropic TV regularization term can be written as $\|\mathbf{Dx}\|_{\mathrm{1}}$. Meanwhile, the balancing parameter $\lambda$ can be combined into the regularization term, written as $\|\lambda\mathbf{Dx}\|_{\mathrm{1}}$. Since the real-world noise exhibits different patterns, following \cite{FFDnet,cbdnet}, we stretch $\lambda$  into a map $\mathbf{\hat{M}}=[\mathbf{M}_{x}^{T}\;\mathbf{M}_{y}^{T}]^{T}$, where the size of $\mathbf{M}_{x}$ and $\mathbf{M}_{y}$ are the same as $\mathbf{x}$. Therefore, the adaptive TV regularization minimization problem is formulated as\begin{small}
 \begin{equation}
\underset{\mathbf{x}}{\operatorname{minimize}} \quad \frac{1}{2}\| \mathbf{x}-\mathbf{y}\|^{2}+\|\mathbf{\hat{M}}\circ\left(\mathbf{Dx}\right)\|_{\mathrm{1}},
\end{equation}
\end{small}where ``$\circ $'' denotes component-wise multiplication. In our UTVNet, we set $\mathbf{M}_{x}=\mathbf{M}_{y}=\mathbf{M}$ as the noise level map that we need to approximate in the network.
By introducing intermediate variable $\mathbf{u}=\mathbf{Dx}$, we have the augmented Lagrangian function of Eq. (2) \begin{small}  
\begin{equation}
L(\mathbf{x}, \mathbf{u}, \mathbf{z})\!=\!\frac{1}{2}\|\mathbf{ x}-\mathbf{y}\|^{2}\!+ \|\mathbf{\hat{M}}\circ\mathbf{u}\|_{1}\!-\mathbf{z}^{T}\!(\mathbf{u}-\mathbf{D} \mathbf{x})+\frac{\rho_{r}}{2}\|\mathbf{u}-\mathbf{D} \mathbf{x}\|^{2},
\end{equation}
\end{small}where $\mathbf{z}$ is the Lagrange multiplier, $\rho_{r}$ is a regularization parameter. Through ADMM, the problem can be solved by iteratively updating $\mathbf{x}$, $\mathbf{u}$, and  $\mathbf{z}$,  other variables treated as constants. Consequently, the unfolding inference can be obtained by solving the below sub-problems \begin{small}
\begin{equation}
\left\{\begin{array}{l}
\mathbf{x}_{k+1}\!=\!\underset{\mathbf{x}}{\arg \min } \frac{1}{2}\| \mathbf{x}\!-\mathbf{y}\|^{2}\!\!-\!\mathbf{z}_{k}^{T}\left(\mathbf{u}_{k}\!-\!\mathbf{D} \mathbf{x}\right)\!+ \!\frac{\rho_{r}}{2}\left\|\mathbf{u}_{k}\!-\!\mathbf{D} \mathbf{x}\right\|^{2}, \\
\mathbf{u}_{k+1}\!\!=\!\underset{\mathbf{u}}{\arg \min }\|\mathbf{\hat{M}}\circ\mathbf{u}\|_{1}\!-\!\mathbf{z}_{k}^{T}\!\!\left(\!\mathbf{u}\!-\!\mathbf{D} \mathbf{x}_{k+1}\!\right)\!+\!\!\frac{\rho_{r}}{2}\!\!\left\|\mathbf{u}\!\!-\!\!\mathbf{D} \mathbf{x}_{k+1}\right\|^{2},\\
\mathbf{z}_{k+1}= \mathbf{z}_{k}-\rho_{r}\left(\mathbf{u}_{k+1}-\mathbf{D} \mathbf{x}_{k+1}\right),
\end{array}\right.
\end{equation}
\end{small}where $k$ denotes the $k$-th iteration.\;Next, the fast Fourier transform (FFT) and the shrinkage function can be utilized to solve x-subproblem and u-subproblem respectively \cite{ladmmtv,newyv}. Hence, the closed-form solutions to the sub-problems in Eq.\;(4)\;is
\begin{equation}
\left\{\begin{array}{l}
\mathbf{x}_{k+1}=\mathcal{F}^{-1}\left[\frac{\mathcal{F}\left[ \mathbf{y}+\rho_{r} \mathbf{D}^{T} \mathbf{u}_{k}-\mathbf{D}^{T} \mathbf{z}_{k}\right]}{1+\rho_{r}\left(\left|\mathcal{F}\left[\mathbf{D}_{x}\right]\right|^{2}+\left|\mathcal{F}\left[\mathbf{D}_{y}\right]\right|^{2}\right)}\right],\vspace{0.1cm} \\ 
\mathbf{u}_{k+1}=\max \left\{\left|\mathbf{v}_{k+1}\right|-\frac{\mathbf{\hat{M}}}{\rho_{r}}, 0\right\} \cdot \operatorname{sign}\left(\mathbf{v}_{k+1}\right),\\
\mathbf{z}_{k+1}= \mathbf{z}_{k}-\rho_{r}\left(\mathbf{u}_{k+1}-\mathbf{D} \mathbf{x}_{k+1}\right),
\end{array}\right.
\end{equation} where $\mathcal{F}\left(\cdot\right)$ and $\mathcal{F}^{-1}\left(\cdot\right)$ denote FFT and inverse FFT, respectively, $sign \left(\cdot\right)$ denotes the sign function, $\mathbf{v}$ is defined as $\mathbf{v}_{k+1}=\mathbf{D} \mathbf{x}_{k+1}+\left(1/ \rho_{r}\right) \mathbf{z}_{k} $. By using the above three closed-form solutions iteratively, the minimization problem in Eq.\;(2) can be solved. We then analyze the relationship between three closed-from solutions and design unfolding TV module to learn the noise level map $\mathbf{M}$.

\subsection {Unfolding Total Variation Module} 
Based on three closed-form solutions, we design an unfolding TV Module as shown in Fig.\;2. First, image $\mathbf{y}$ is fed into noise level initialization block (NLI-block) to predict noise level map $\mathbf{M}$. Then, we unfold the TV minimization problem and iteratively utilize Eq.\;(5) to guarantee the smoothness and fidelity constraints for learning $\mathbf{M}$ end-to-end and generating noise-free smooth layer $\mathbf{y_{s}}$.

\textbf{Unfolding TV Architecture.} In Eq.\;(4), it can be observed that the data fidelity term and the TV regularization term are decoupled into the x- and u- subproblems. The  parameter $\mathbf{M}$ only exists in the u-subproblem. Because $\mathbf{M}$ is also the hyper-parameter of the iterative algorithm, instead of learning a fixed $\mathbf{M}$ for all iterations, here we follow  \cite{USRnet}, in which  learning a series of hyper-parameters varies in different iterations. Hence, we set $\boldsymbol{\mathbf{M}}=\left[\mathbf{M}_{1}, \mathbf{M}_{2},\mathbf{M}_{3}, \ldots, \mathbf{M}_{K}\right]$, $\boldsymbol{\mathbf{\rho_{r}}}=\left[\rho_{1}, \rho_{2},\rho_{3}, \ldots, \rho_{K}\right]$, where $K$ is the number of iteration times. For speed-accuracy trade-off, $K$ is set to $8$. Following the unfolding optimization, the unfolding TV architecture alternatively solves the  $\mathbf{x}$, $\mathbf{u}$, $\mathbf{z}$ sub-problems by Eq.\;(5), they  are written as below and illustrated in Fig.\;2
\begin{equation}
\left\{\begin{array}{l}
\mathbf{x}_{k}=\mathcal{N}(\mathbf{y},\mathbf{z}_{k-1},\rho_{k},\mathbf{D},\mathbf{u}_{k-1}),\\
\mathbf{u}_{k}=\mathcal{S}(\mathbf{M}_{k},\mathbf{D},\mathbf{z}_{k-1},\mathbf{x}_{k},\rho_{k}), \\
\mathbf{z}_{k}=\mathcal{G}(\mathbf{D},\mathbf{z}_{k-1},\rho_{k},\mathbf{u}_{k},\mathbf{x}_{k}).
\end{array}\right.
\end{equation}In the following descriptions, we will analyze the importance of three functions for learning the noise level map $\mathbf{M}$.

Function  $\mathcal{N}\left(\cdot\right)$ and function  $\mathcal{G}\left(\cdot\right)$ only contain the hyper-parameter $\rho_{k}$, but such two function are essential for the UTVNet. Function $\mathcal{N}\left(\cdot\right)$ aims to solve the fidelity term sub-problem, and guarantees the inference between the noise-free smooth layer $\mathbf{y}_{s}$ and the original image $\mathbf{y}$. Such inference provides the fidelity prior that speeds up the convergence of the whole network when training our model. Function $\mathcal{G}\left(\cdot\right)$  is associated with constraint $\mathbf{u}=\mathbf{Dx}$, as iteration goes, the Lagrange multiplier $\mathbf{z}$  should be updated. In PyTorch, torch.fft.fftn and torch.fft.ifftn are used to implement $\mathcal{F}\left(\cdot\right)$ and $\mathcal{F}^{-1}\left(\cdot\right)$.

Function $\mathcal{S}\left(\cdot\right)$  contains the approximated noise level map $\mathbf{M}$. The function can be regarded as a special smoothness constraint for the low-frequency layer $\mathbf{y_{s}}$, where it smooths out details and noise guided by the magnitude of noise level map $\mathbf{M}$ for each pixel. Using such a smoothness inference between the original image and the smoothed one, we can learn the parameter $\mathbf{M}$ for input images without designing a special loss function. As shown in Fig.\;2, the parameter $\mathbf{M}$ varies in each iteration to control the trade-off between noise reduction and detail preservation.

There are two learning parameters in this architecture, $\boldsymbol{\mathbf{\rho_{r}}}$ and $\mathbf{M}$. $\boldsymbol{\mathbf{\rho_{r}}}$ is generated by $\boldsymbol{\mathbf{\rho_{r}}}=\mathcal{R}(\rho_{0})$ in Fig. 2, where we adopt the hyper-parameter module in \cite{USRnet} that uses three $1\times1$ convolutions with two ReLU and one Softplus activation functions. To avoid dividing by extremely small $\rho_{k}$, we add $1e-6$ to the final output. The prediction of  parameter $\mathbf{M}$  will be described in detail below.

\textbf{NLI-Block.} 
Before iteration, a coarse-to-fine noise level initialization block  is proposed in Fig.\;2 to predict the noise level in a coarse-to-fine way. The details of the NLI-block can be seen in Fig.\;3. We first estimate global noise variation ${\boldsymbol{\mathbf{\varepsilon}}}$  for three channels ${\boldsymbol{\mathbf{\varepsilon}}}=\left[\varepsilon^{r},\varepsilon^{g},\varepsilon^{b}\right]$ as coarse approximated values according to \cite{nes} \begin{small}
\begin{equation}
\mathbf{\varepsilon}^{c} =\sqrt{\frac{\pi}{2}} \frac{1}{6(w-2)(h-2)} \sum_{(i, j)}\left|\left(\mathbf{\mathbf{y}^{c}} \otimes\mathbf{ N}\right)(i, j)\right|,
\end{equation}
\end{small}
where $c\in \{r,g,b\}$ denotes the RGB channels, ``$\otimes$'' denotes the convolution operator. $(i,j)$ denotes indices of the pixel position. $h$ and $w$ are the  height and width of $\mathbf{y}^{c}$, respectively. $\mathbf{N}$ is defined as

\begin{small}
\begin{equation}
\mathbf{N}=\left[\begin{array}{ccc}
1 & -2 & 1 \\
-2 & 4 & -2 \\
1 & -2 & 1
\end{array}\right].
\end{equation}
\end{small}Then, we adopt a seven dilated convolution layers with LeakyReLU activation function for the first six layers to predict the residual $\mathbf{R}^{\prime}=\left[\mathbf{R}_{1}^{\prime}, \mathbf{R}_{2}^{\prime}, \ldots, \mathbf{R}_{K}^{\prime}\right]$, which can be used to modify  the global noise  variation ${\boldsymbol{\mathbf{\varepsilon}}}$ and generate  precise noise level values for each pixel. Hence, the noise level map $\mathbf{M}^{\prime}=\left[\mathbf{M}_{1}^{\prime}, \mathbf{M}_{2}^{\prime}, \ldots, \mathbf{M}_{K}^{\prime}\right]$ is formulated as \begin{small}
\begin{equation}
\boldsymbol{\mathbf{M}^{\prime}}=\left[\mathbf{R}_{1}^{\prime}+\boldsymbol{\mathbf{\varepsilon}} , \mathbf{R}_{2}^{\prime}+\boldsymbol{\mathbf{\varepsilon}}, \ldots, \mathbf{R}_{K}^{\prime}+\boldsymbol{\mathbf{\varepsilon}}\right].
\end{equation}
\end{small}We define such methods for two reasons. First, the small pixel values make it harder to  learn the noise level map $\mathbf{M}$ directly  in extremely low-light images. The modification of the global noise variation provides a better initialization and accelerate the speed of convergence for the unfolding TV module. Second, as indicated in \cite{FFDnet}, a larger noise level may guarantee the role of $\mathbf{M}$ in controlling the trade-off between noise reduction and detail preservation. The global noise variation also provides a basic magnitude for learning the parameter $\mathbf{M}$, which ensures an accurate approximation of the noise level. As the values in the noise level map should be positive, the final output $\mathbf{M}_{k}^{c}(i,j)$ is defined as \begin{small}
\begin{equation}
\mathbf{M}_{k}^{c}(i,j)=\left\{\begin{array}{cc}
\mathbf{M}_{k}^{\prime c}(i,j) & \mathbf{M}_{k}^{\prime c}(i,j)>0\vspace{0.1cm}, \\
{ \varepsilon^{c}} &\mathbf{M}_{k}^{\prime c}(i,j) \leq 0,
\end{array}\right.
\end{equation}
\end{small} $k\in\{1,2,3,\ldots,K\}$ denotes $k$-th iteration. For brevity, in Fig.\;3, Eq.\;(7) and Eq.\;(9) are  denoted as Eq.\;(11a), meanwhile,  Eq.\;(10) is denoted as Eq. (11b)
\begin{subnumcases}{}
\{\mathbf{M}^{\prime},\boldsymbol{\mathbf{\varepsilon}}\}=\mathcal{H}\left(\mathbf{R}^{\prime},\mathbf{y},\mathbf{N},h,w\right),\\
\mathbf{M}=\mathcal{A}\left(\mathbf{M}^{\prime},\boldsymbol{\mathbf{\varepsilon}}\right).
\end{subnumcases}$\mathcal{A}(\cdot)$   is similar to ReLU  but contains an important modification. Different from ReLU that totally sets all negative values as zero, we set all negative values to the global noise variation ${\boldsymbol{\mathbf{\varepsilon}}}$. Such a treatment also provides an adaptive higher noise level value that guarantees the role of $\mathbf{M}$ as the balancing parameter $\lambda$. 
It is notable that the smoothed-out detail can be recovered when suppressing the noise guided by $\mathbf{M}$ in the image restoration module. The effectiveness of the NLI-block will be discussed in Section 4.3.

\begin{figure}
\begin{center}
   \includegraphics[width=0.95\linewidth]{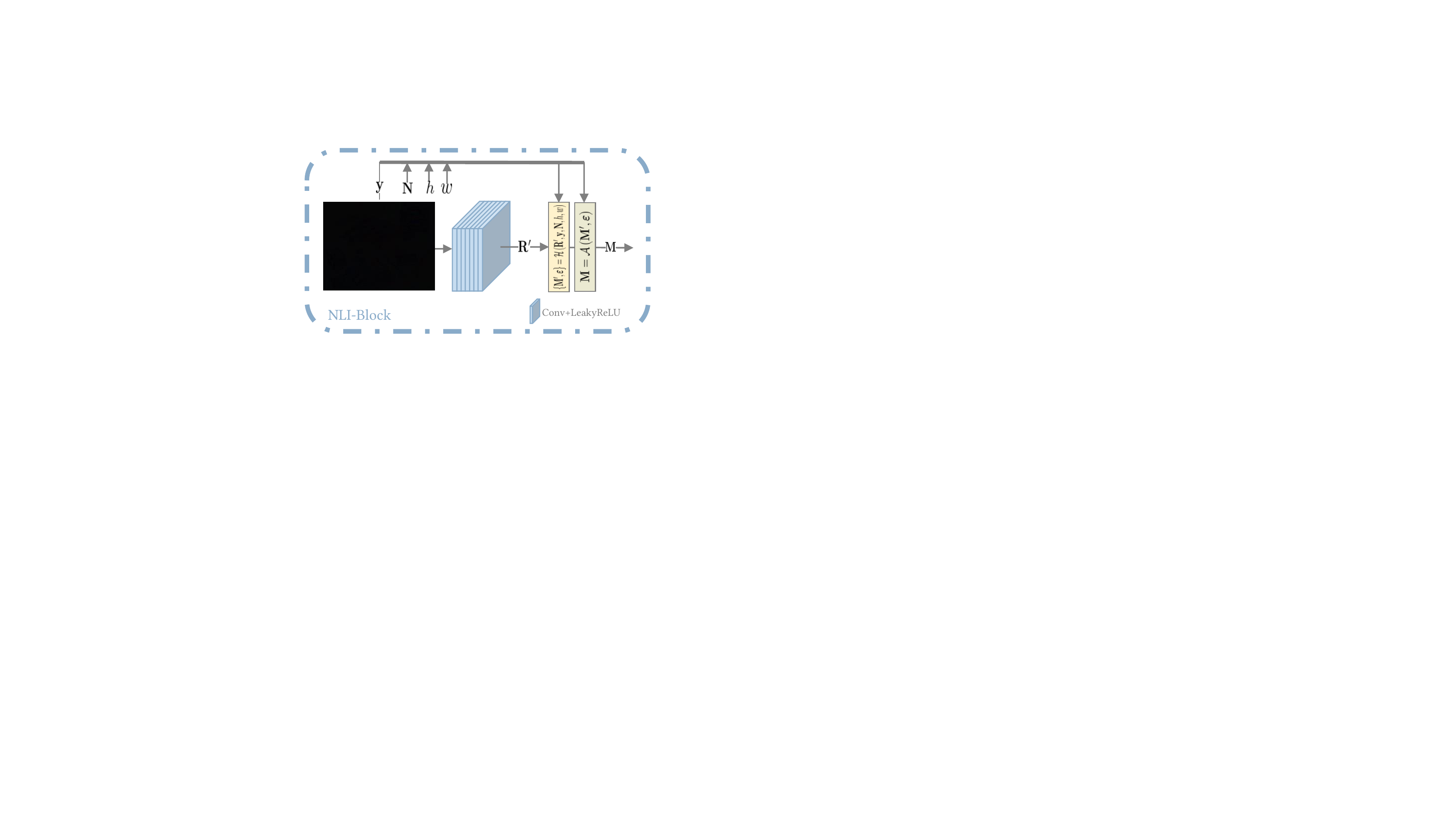}
\end{center}\vspace{-0.2cm}
   \caption{ Illustration of the NLI-Block. The details of function $\mathcal{H}(\cdot)$ and function $\mathcal{A(\cdot)}$ are described in Section 3.3.}
\end{figure}
\vspace{-0.1cm}

\begin{figure*}
 \hsize=\textwidth
  \subfigcapskip=-3pt
 \subfigbottomskip=-3pt
 \begin{center}
        \subfigure 
    { 
         \centering
        \begin{minipage}[b]{0.14\textwidth}
            \centering
            \includegraphics[width=1\linewidth,height=1.8cm]{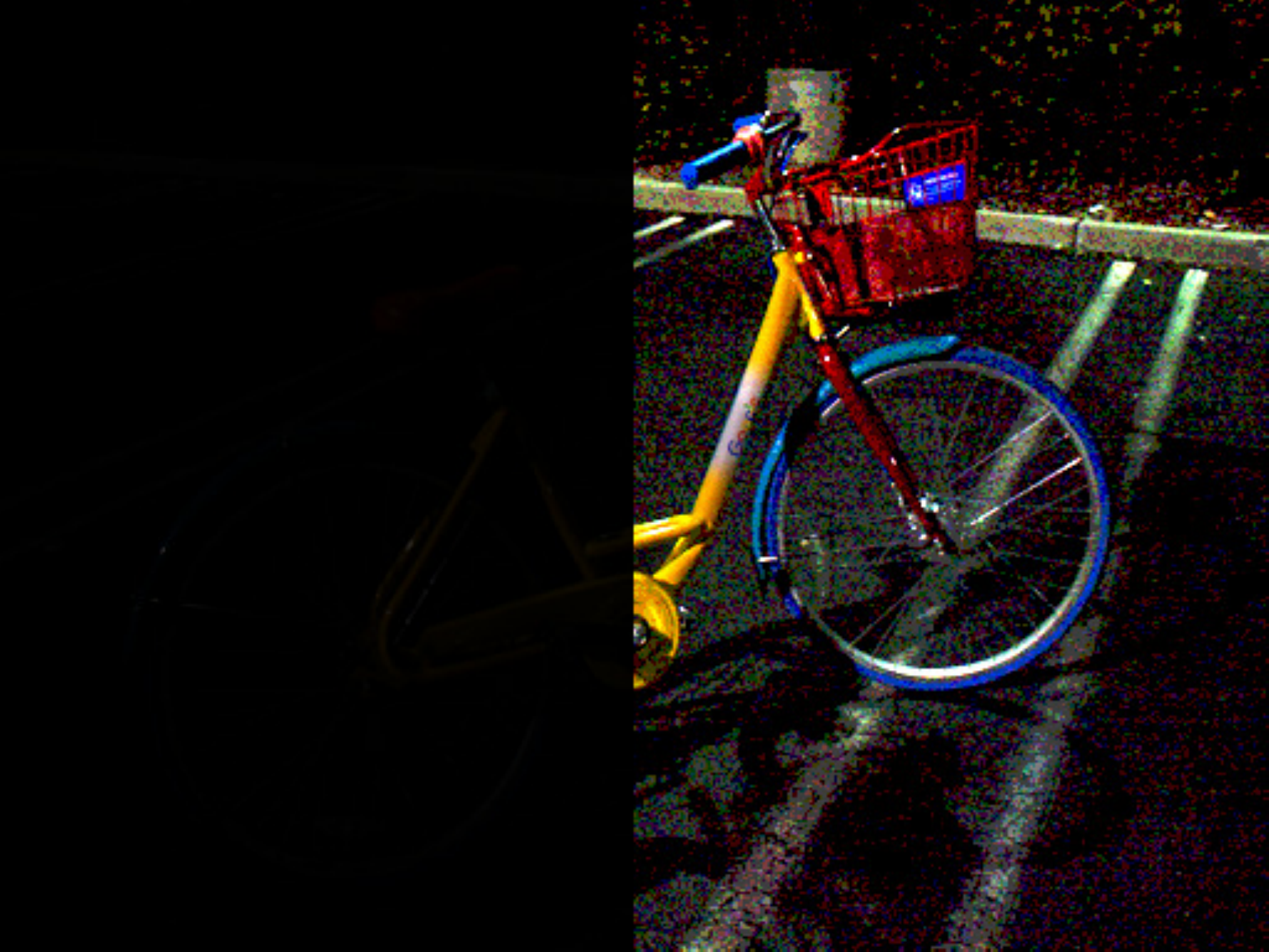}\vspace{0.3cm}
        \end{minipage}\hspace{-0.3mm}
   
        \begin{minipage}[b]{0.14\textwidth}
            \centering
            \includegraphics[width=1\linewidth,height=1.8cm]{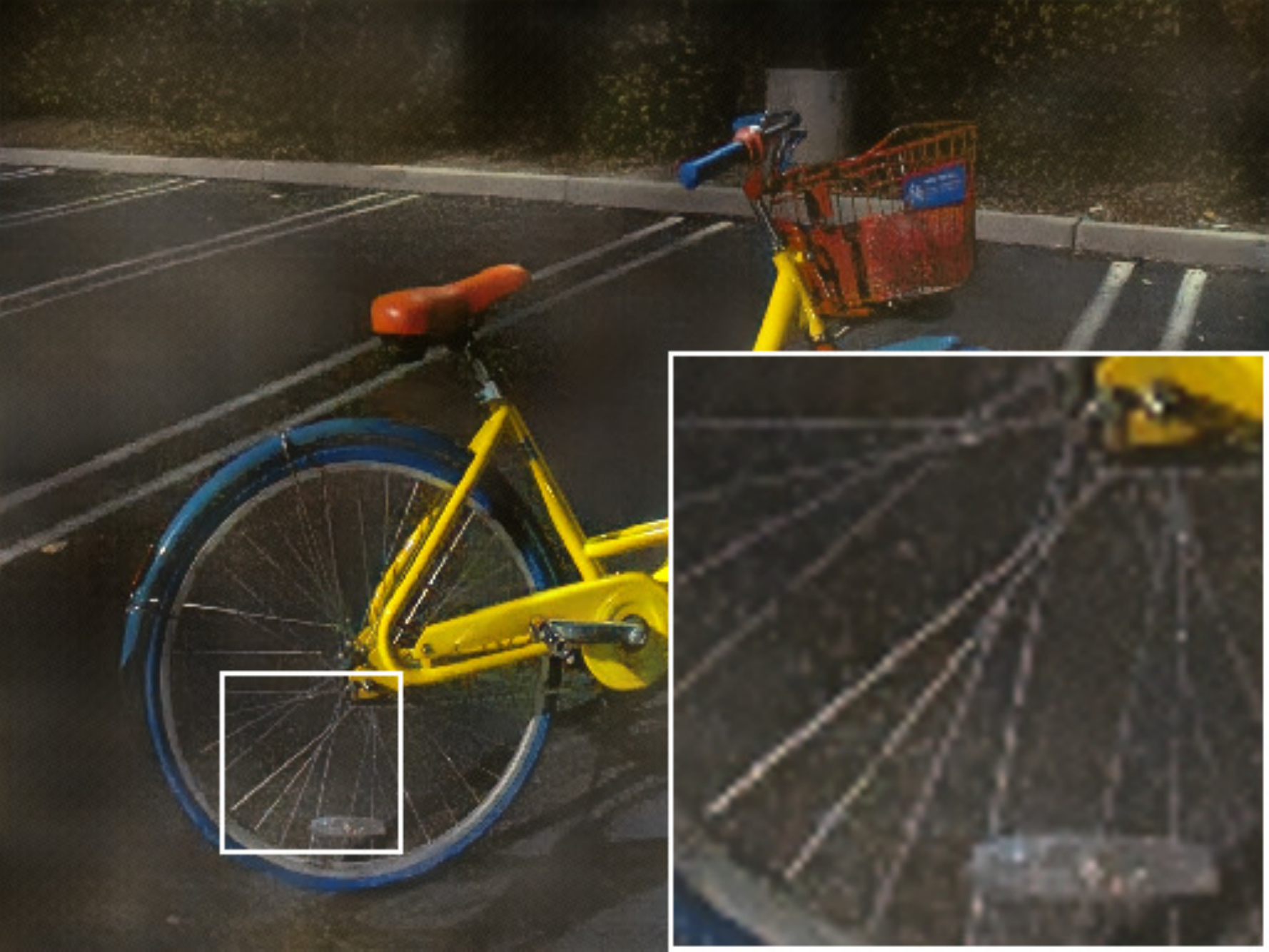}\vspace{0.3cm}
        \end{minipage}\hspace{-0.3mm}
  
         \begin{minipage}[b]{0.14\textwidth}
            \centering
            \includegraphics[width=1\linewidth,height=1.8cm]{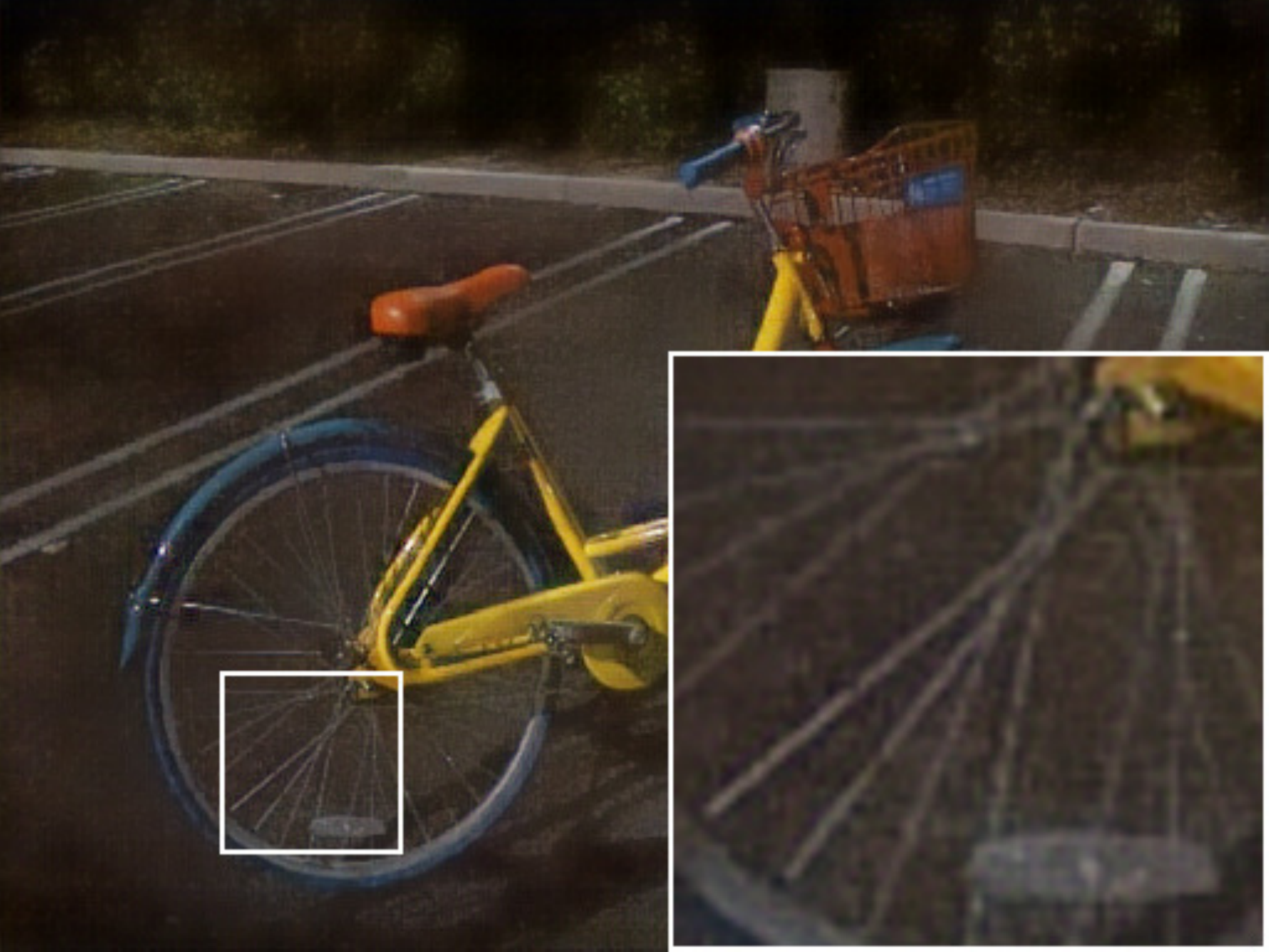}\vspace{0.3cm}
        \end{minipage}\hspace{-0.3mm}
  
        \begin{minipage}[b]{0.14\textwidth}
            \centering
            \includegraphics[width=1\linewidth,height=1.8cm]{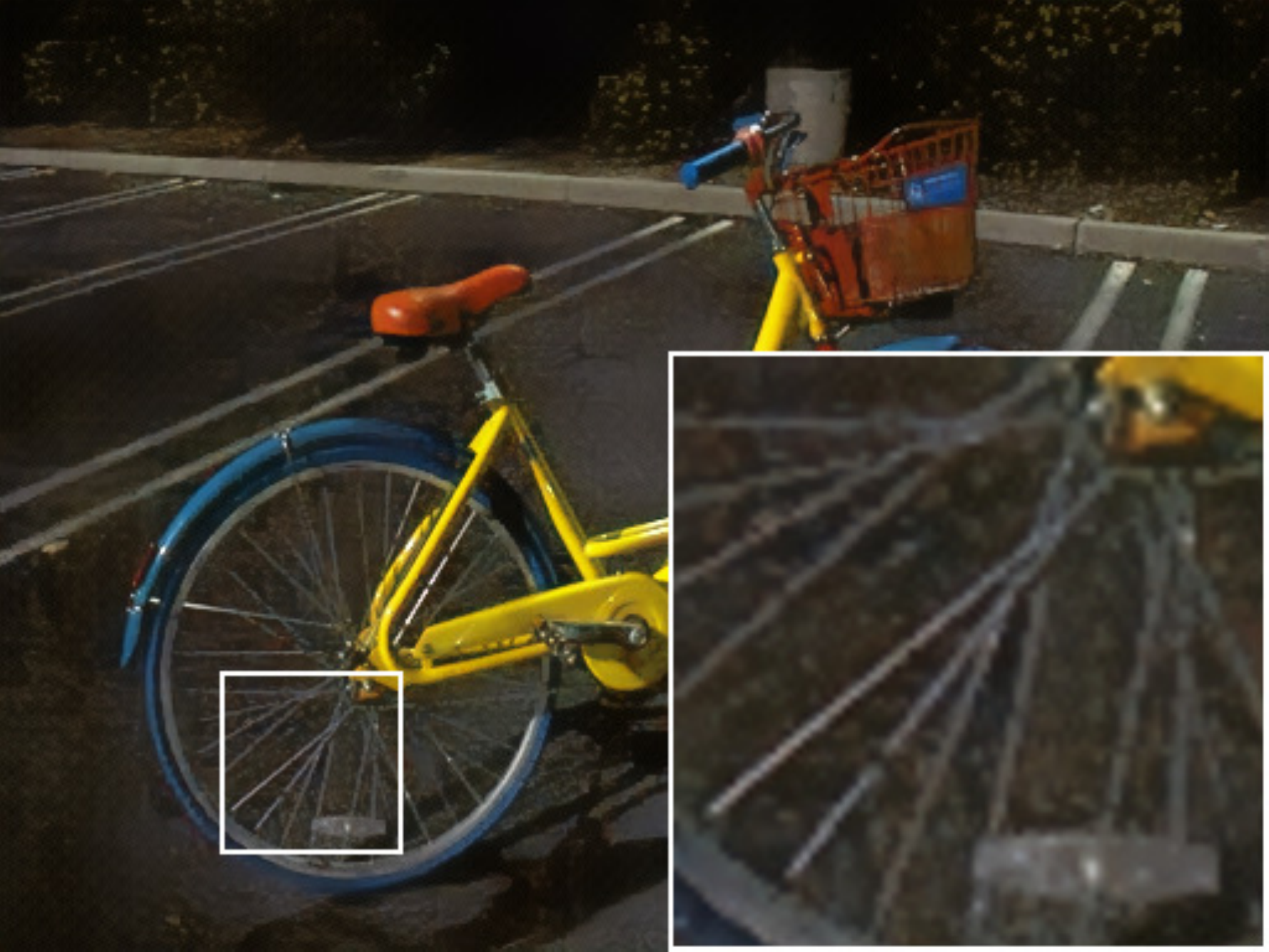}\vspace{0.3cm}
        \end{minipage}\hspace{-0.3mm}

      \begin{minipage}[b]{0.14\textwidth}
            \centering
            \includegraphics[width=1\linewidth,height=1.8cm]{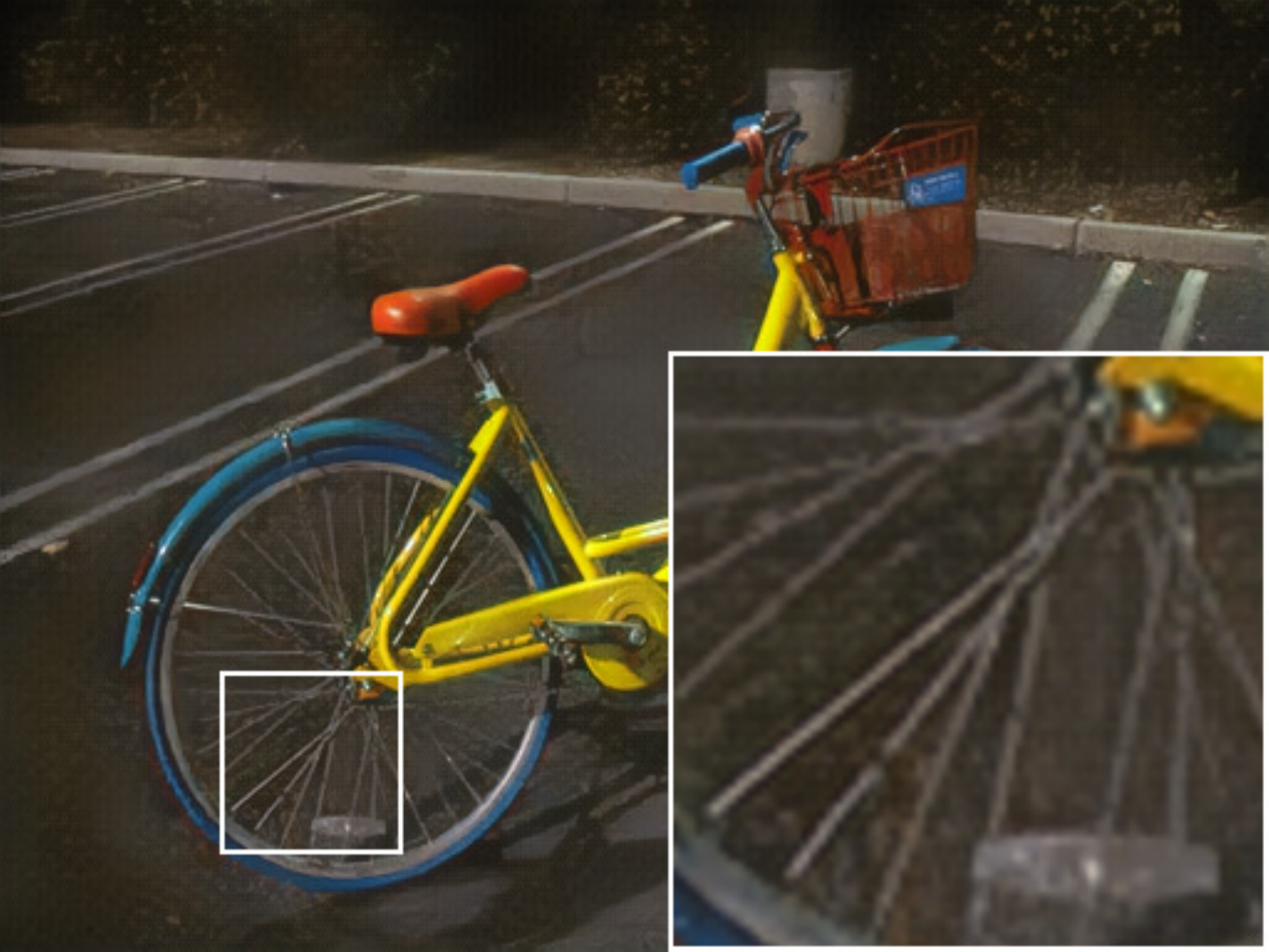}\vspace{0.3cm}
        \end{minipage}\hspace{-0.3mm}

         \begin{minipage}[b]{0.14\textwidth}
            \centering
            \includegraphics[width=1\linewidth,height=1.8cm]{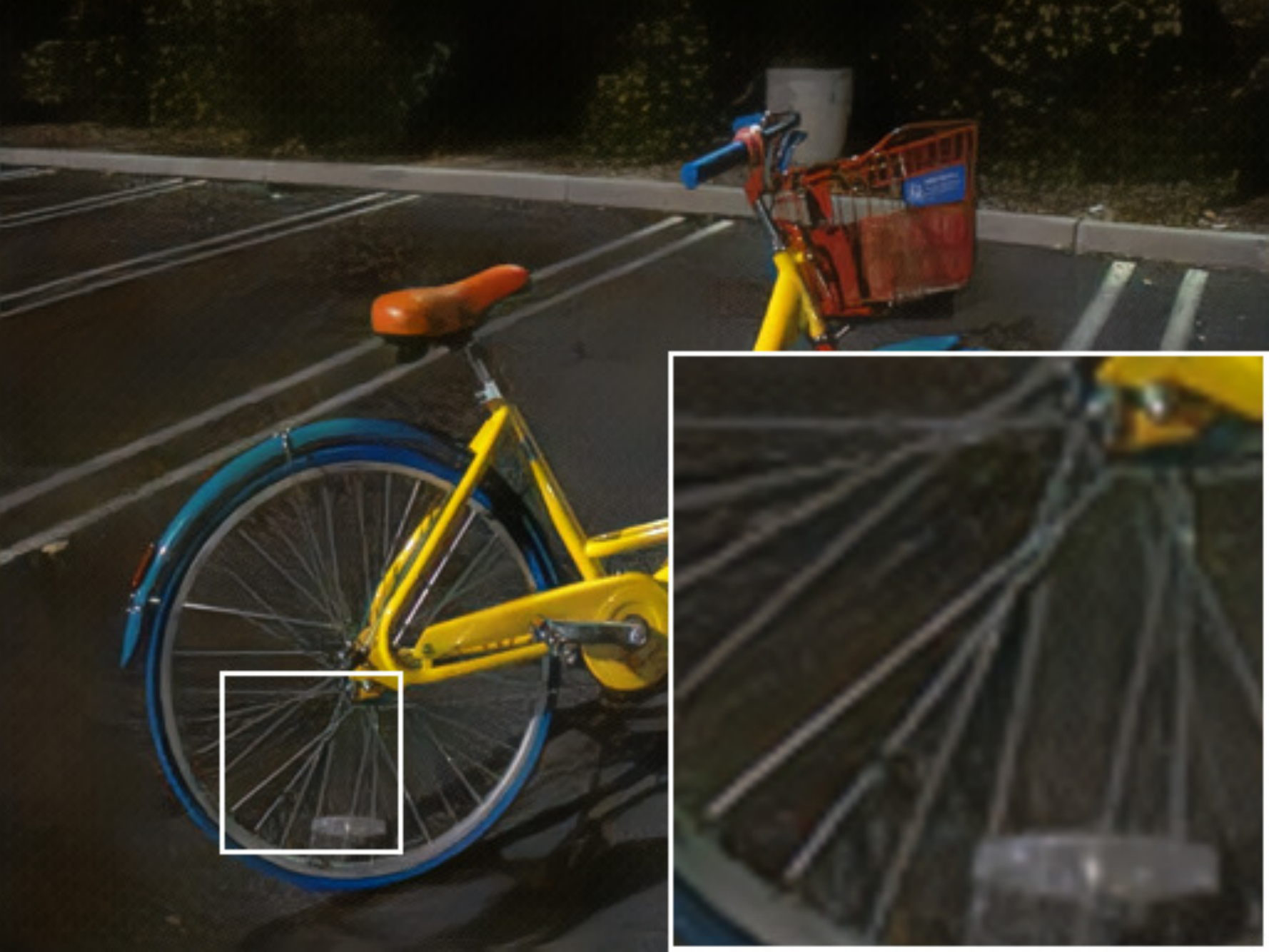}\vspace{0.3cm}
        \end{minipage}\hspace{-0.3mm}
        
         \begin{minipage}[b]{0.14\textwidth}
            \centering
            \includegraphics[width=1\linewidth,height=1.8cm]{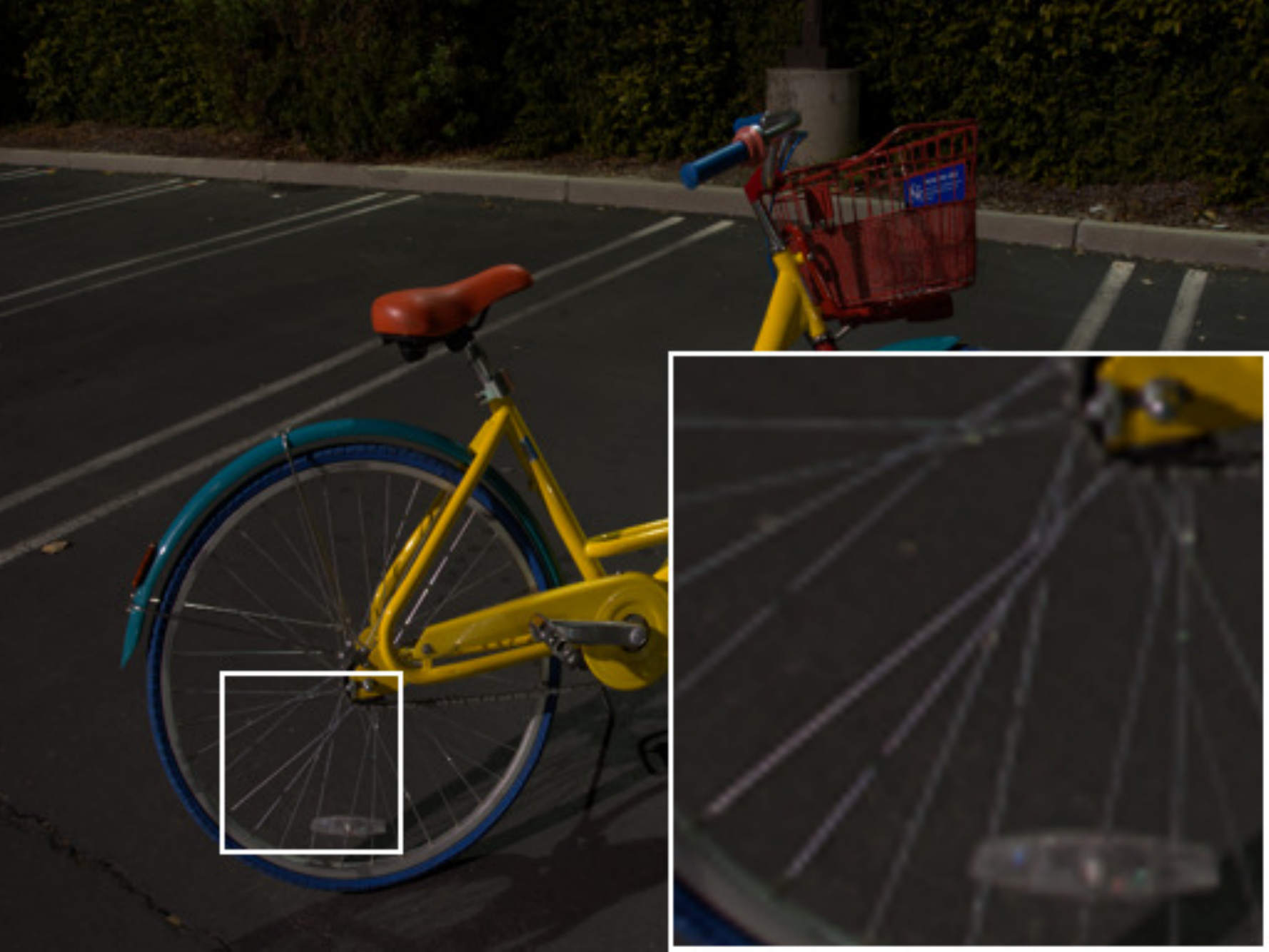}\vspace{0.3cm}
        \end{minipage}\hspace{-0.3mm}
       }

\vspace{-0.45cm}

              \subfigure 
    {
         \centering
        \begin{minipage}[b]{0.14\textwidth}
            \centering
            \includegraphics[width=1\linewidth,height=1.8cm]{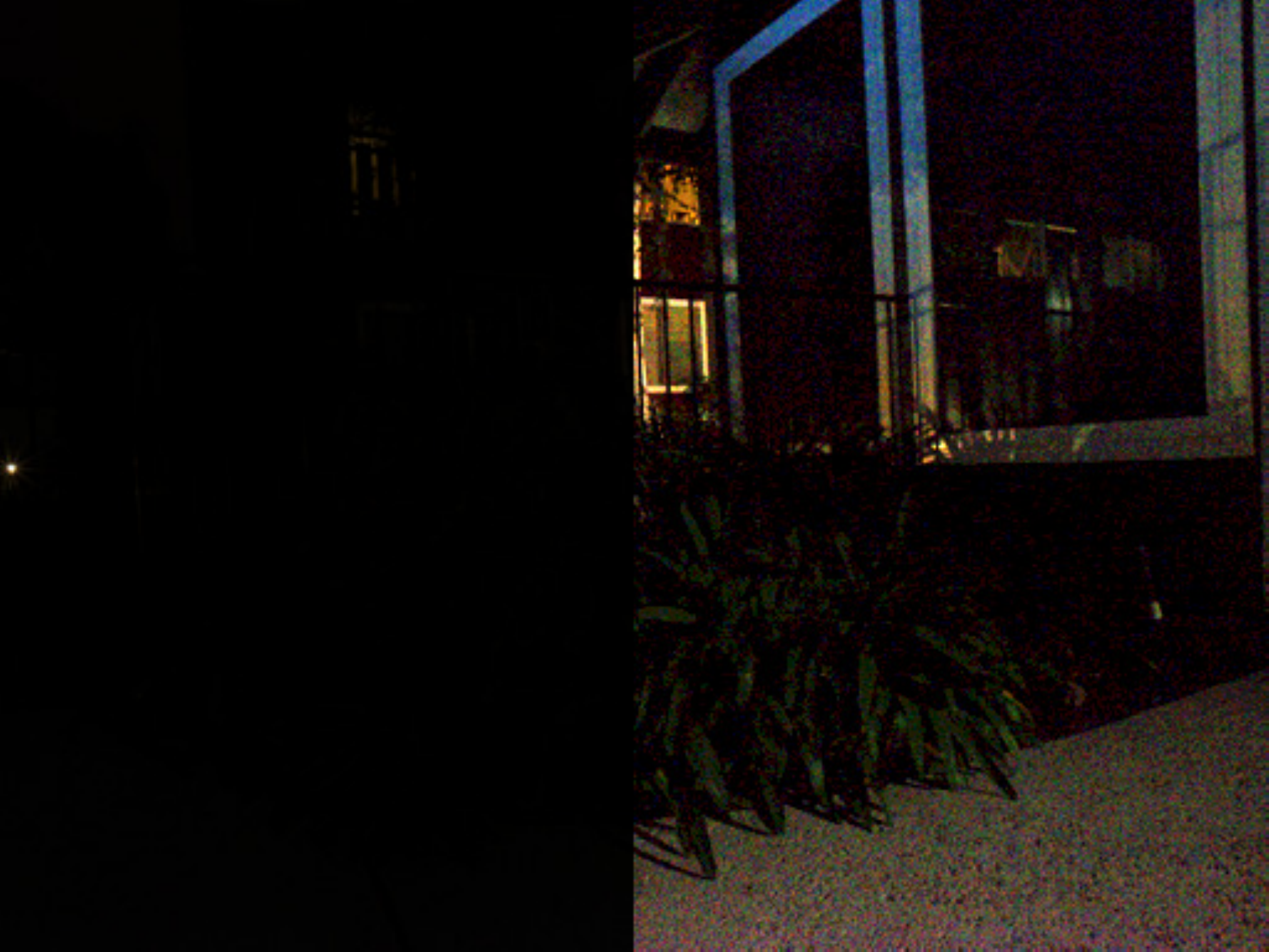}\vspace{0.1cm}
        \end{minipage}\hspace{-0.3mm}
   
         \begin{minipage}[b]{0.14\textwidth}
            \centering
            \includegraphics[width=1\linewidth,height=1.8cm]{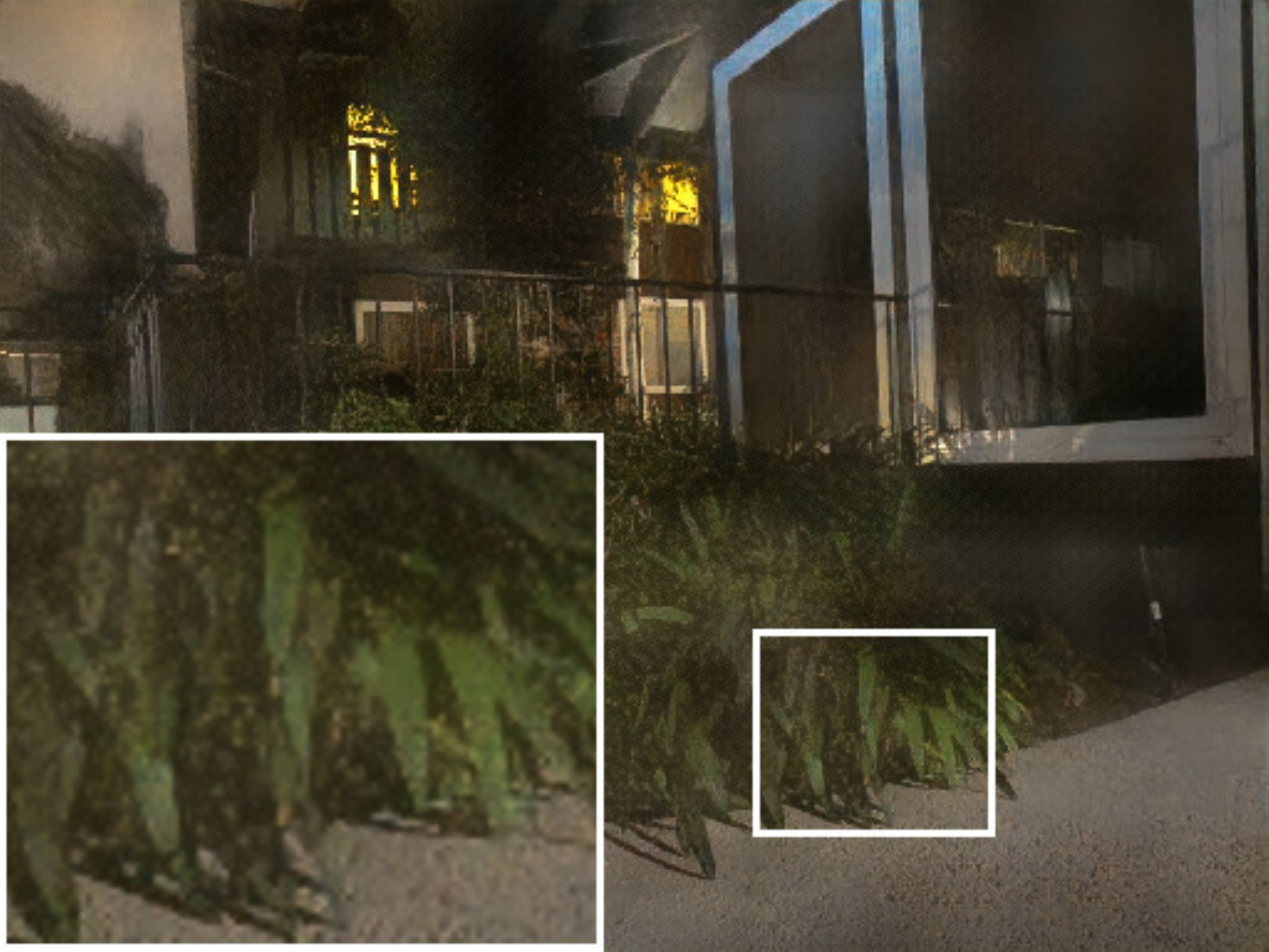}\vspace{0.1cm}
        \end{minipage}\hspace{-0.3mm}
   
        \begin{minipage}[b]{0.14\textwidth}
            \centering
            \includegraphics[width=1\linewidth,height=1.8cm]{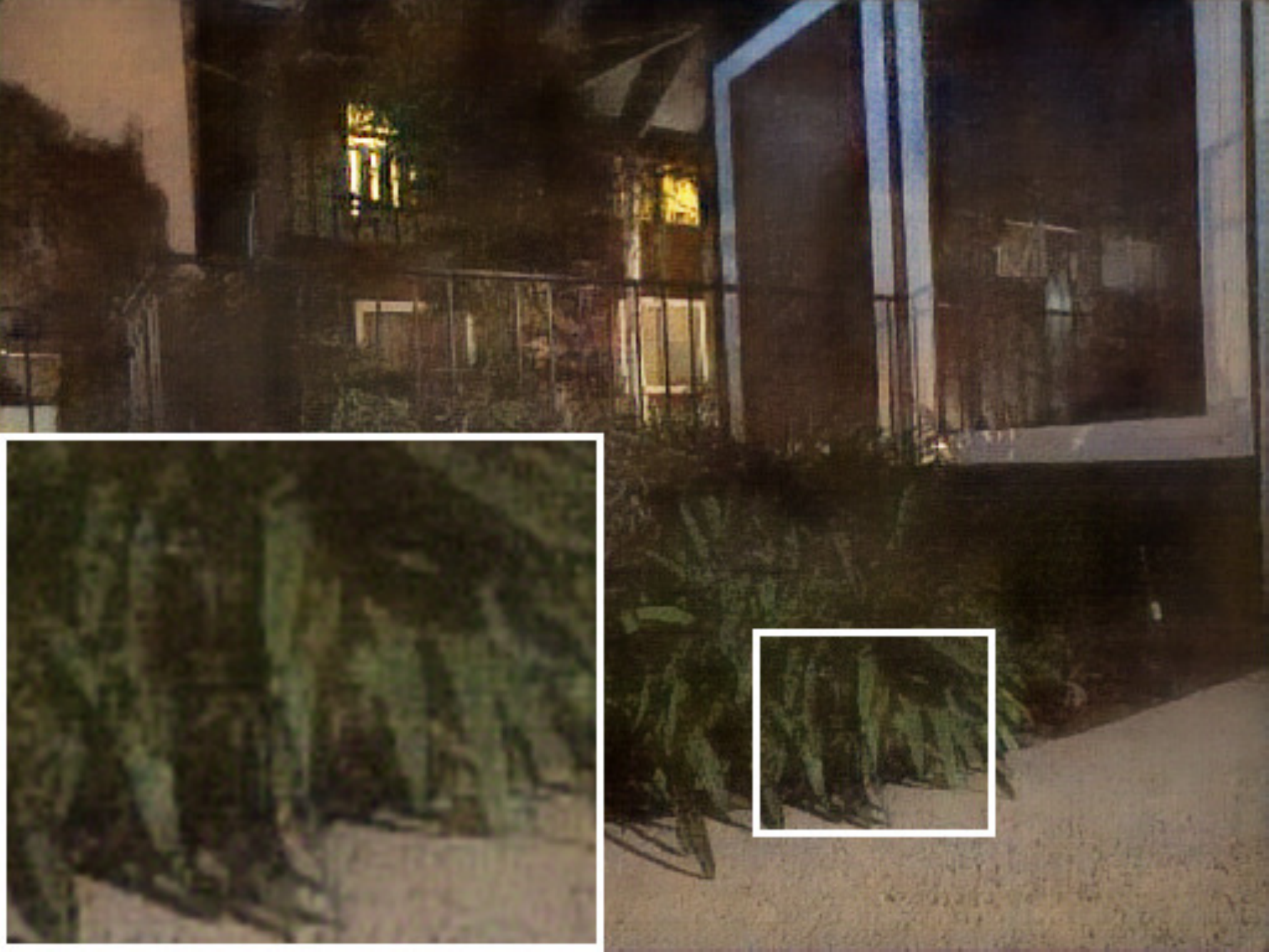}\vspace{0.1cm}
        \end{minipage}\hspace{-0.3mm}
   
        \begin{minipage}[b]{0.14\textwidth}
            \centering
            \includegraphics[width=1\linewidth,height=1.8cm]{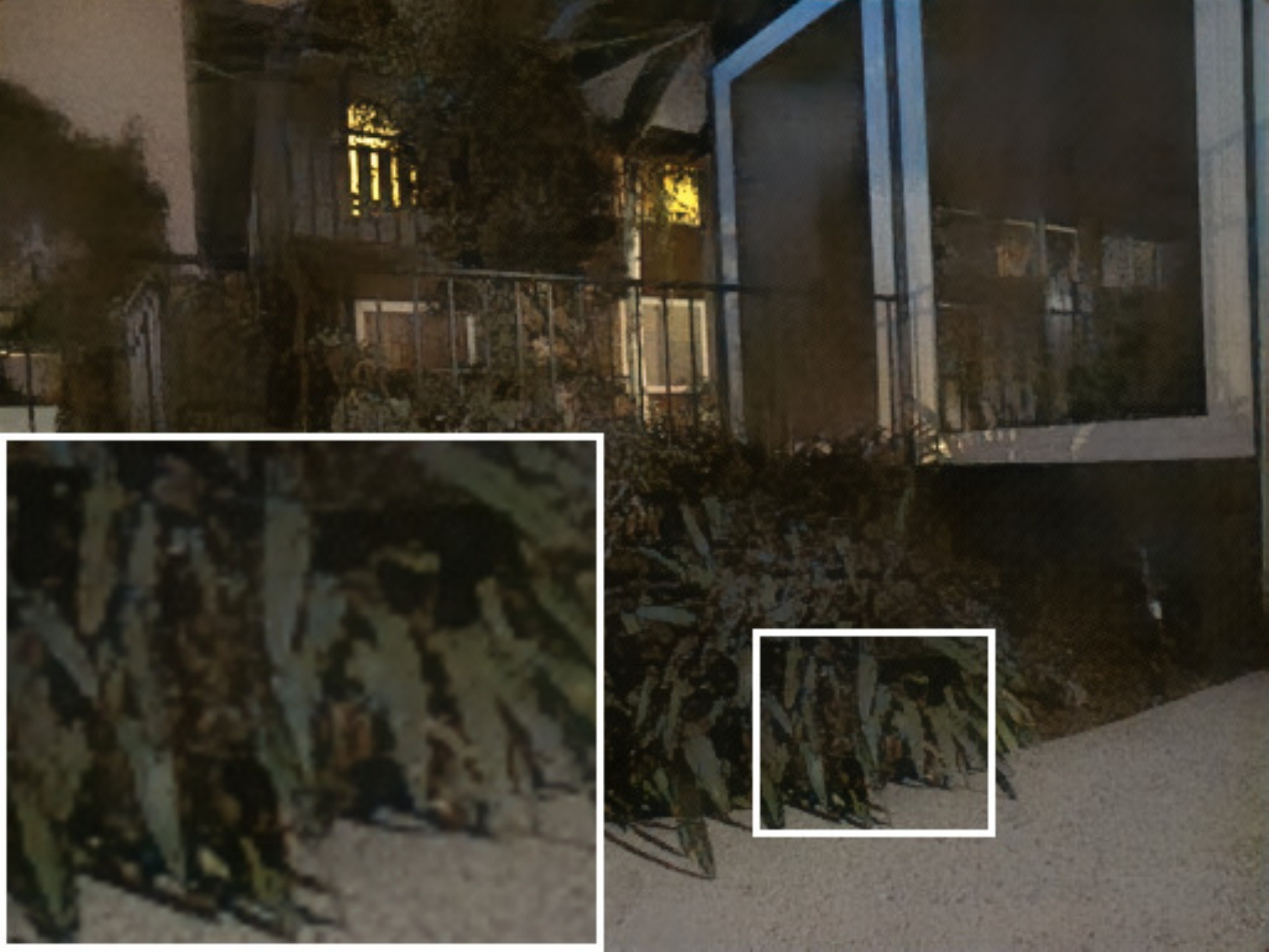}\vspace{0.1cm}
        \end{minipage}\hspace{-0.3mm}
 
         \begin{minipage}[b]{0.14\textwidth}
            \centering
            \includegraphics[width=1\linewidth,height=1.8cm]{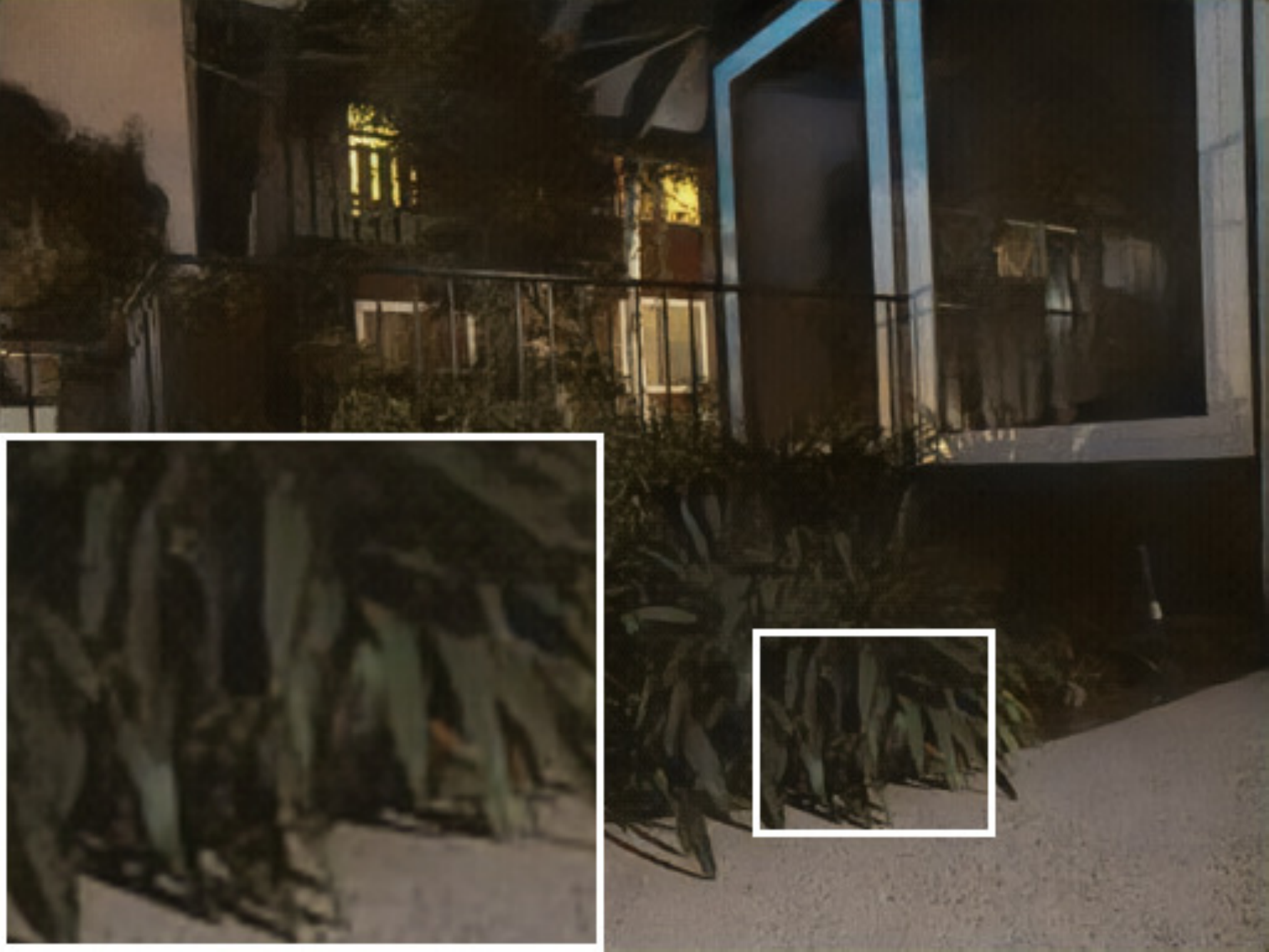}\vspace{0.1cm}
        \end{minipage}\hspace{-0.3mm}
  
         \begin{minipage}[b]{0.14\textwidth}
            \centering
            \includegraphics[width=1\linewidth,height=1.8cm]{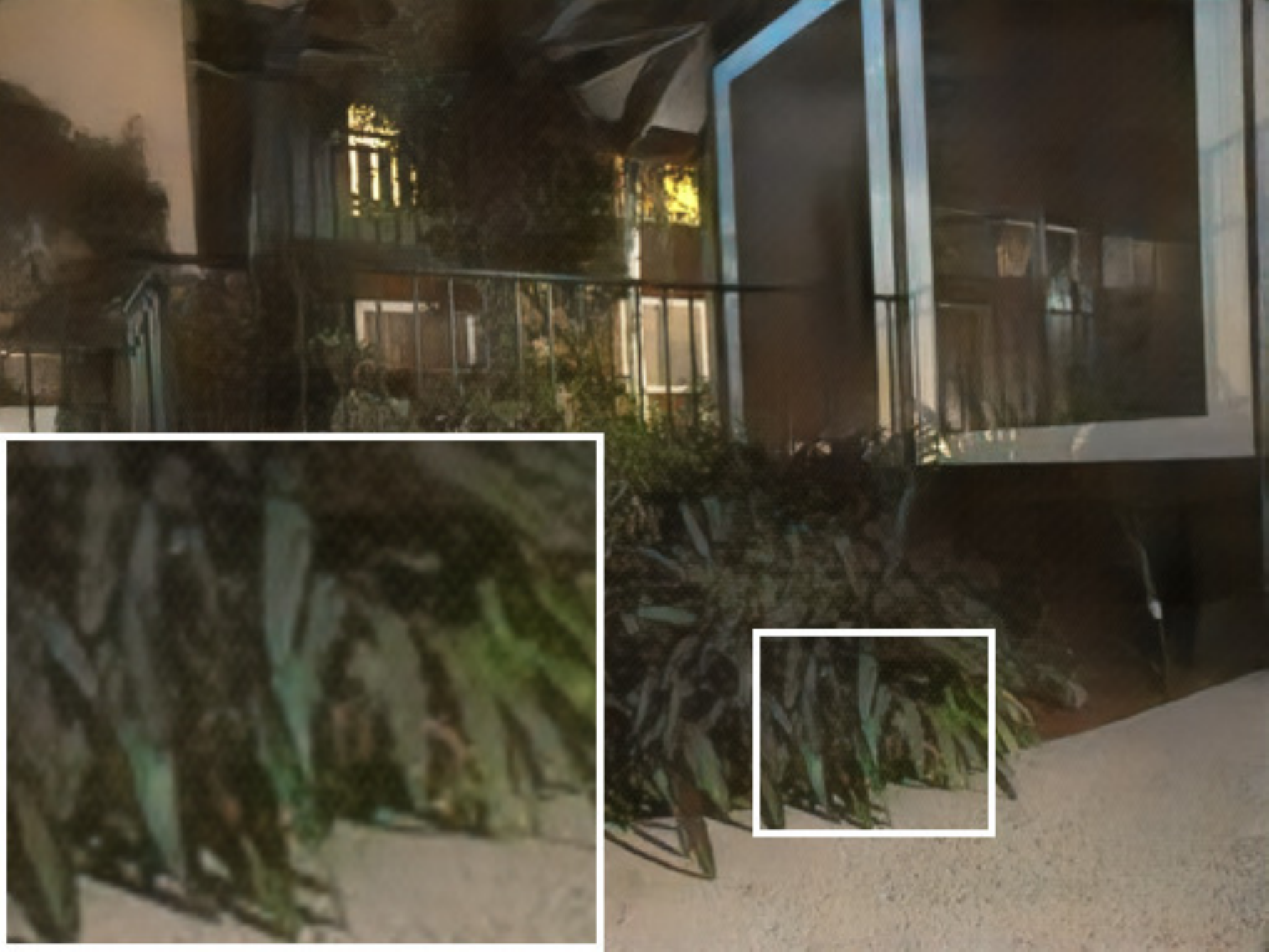}\vspace{0.1cm}
        \end{minipage}\hspace{-0.3mm}
  
         \begin{minipage}[b]{0.14\textwidth}
            \centering
            \includegraphics[width=1\linewidth,height=1.8cm]{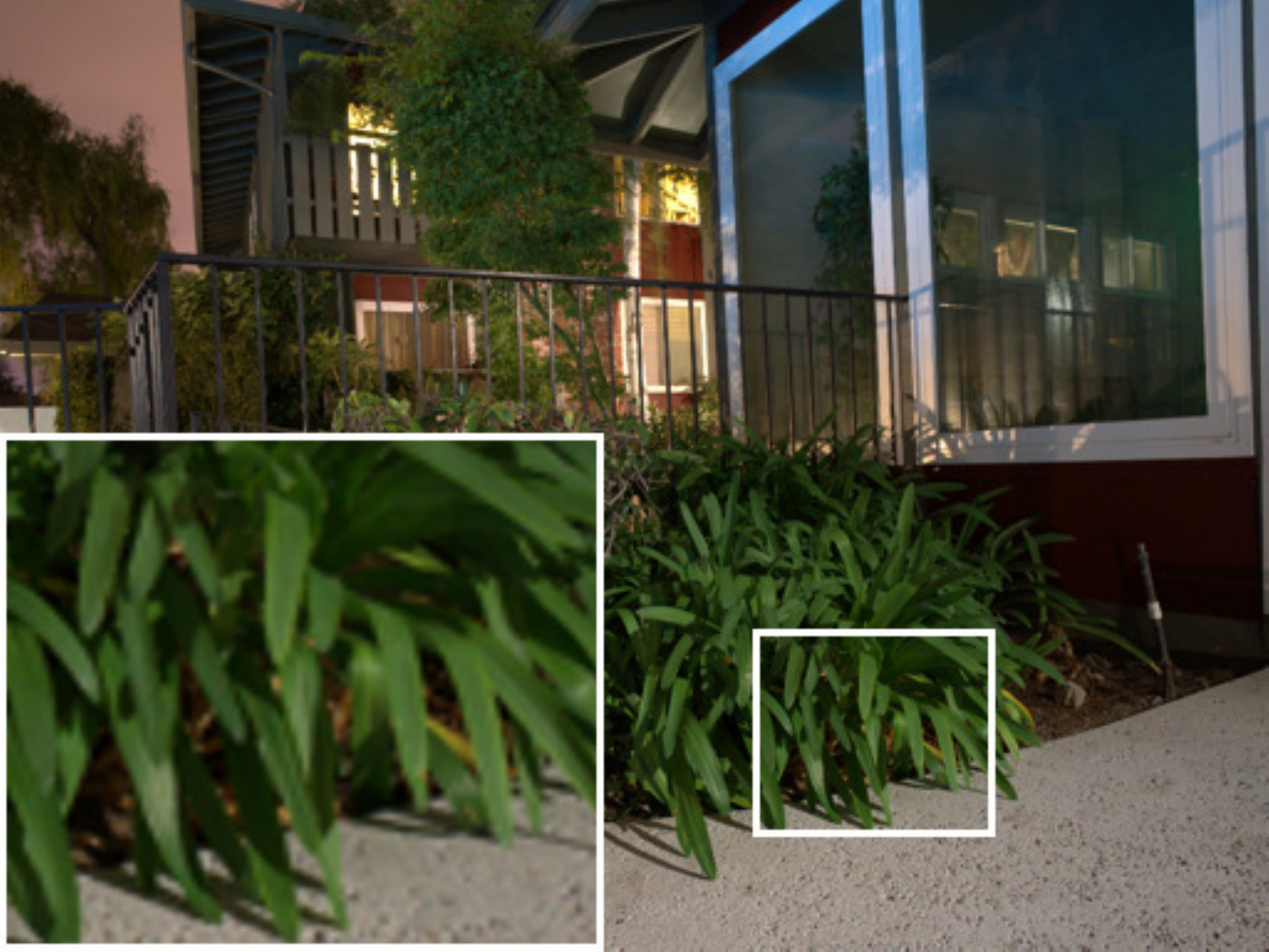}\vspace{0.1cm}
        \end{minipage}\hspace{-0.3mm}
       }

      \vspace{-0.25cm}

         \subfigure
    {
         \centering
        \begin{minipage}[b]{0.14\textwidth}
            \centering
            \includegraphics[width=1\linewidth,height=1.8cm]{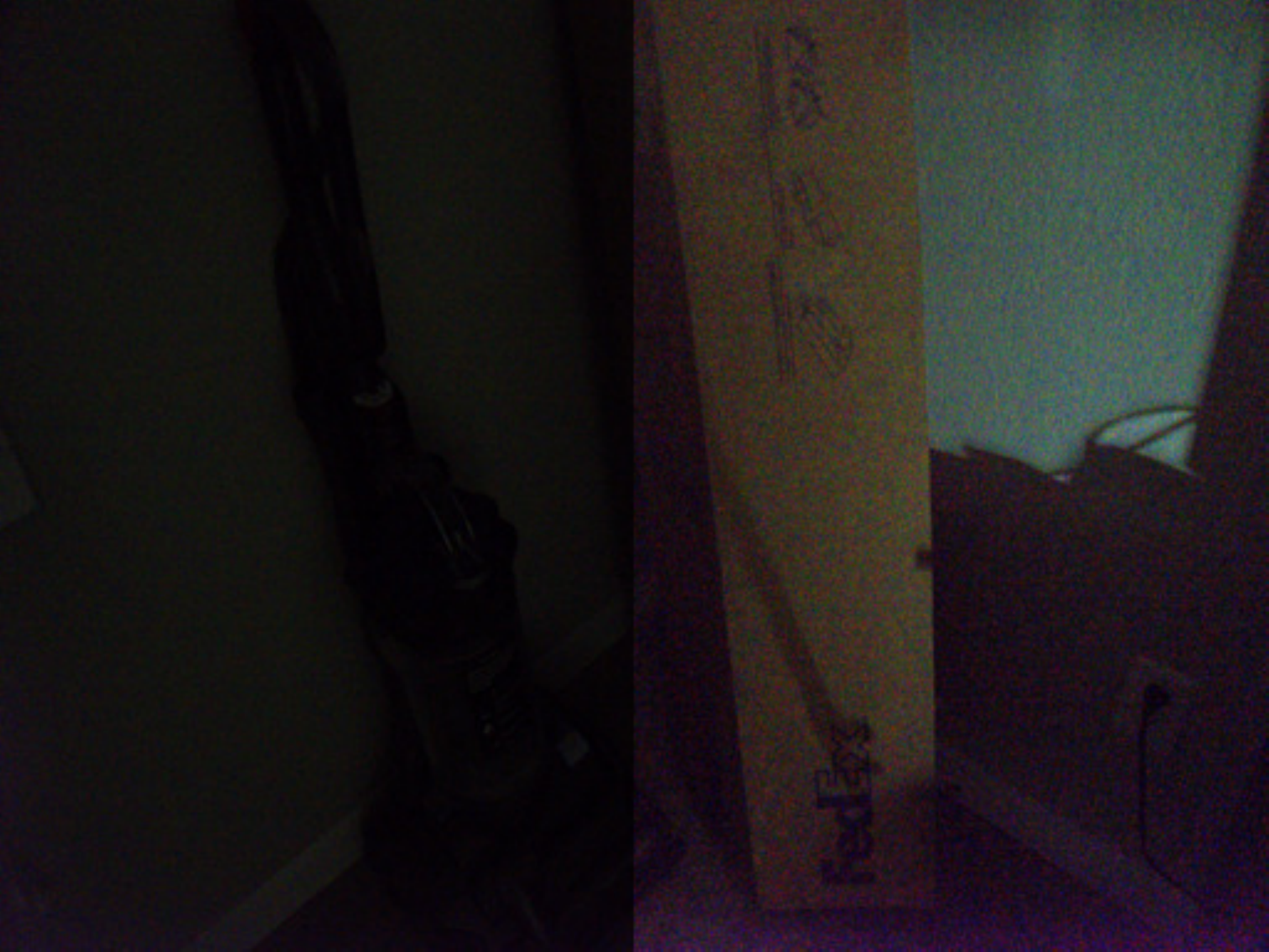}\vspace{0.1cm}
        \end{minipage}\hspace{-0.3mm}
    
         \begin{minipage}[b]{0.14\textwidth}
            \centering
            \includegraphics[width=1\linewidth,height=1.8cm]{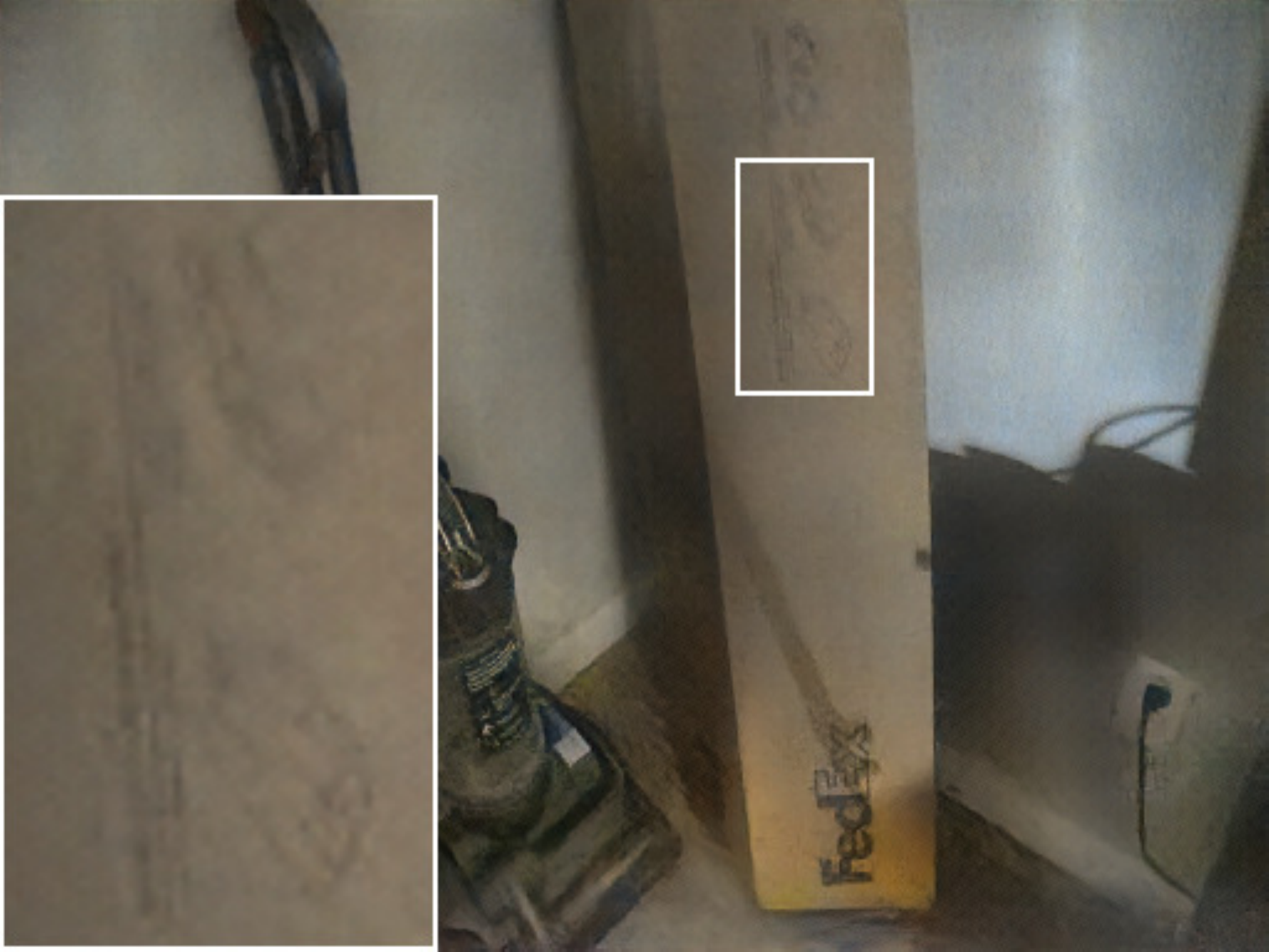}\vspace{0.1cm}
        \end{minipage}\hspace{-0.3mm}
    
        \begin{minipage}[b]{0.14\textwidth}
            \centering
            \includegraphics[width=1\linewidth,height=1.8cm]{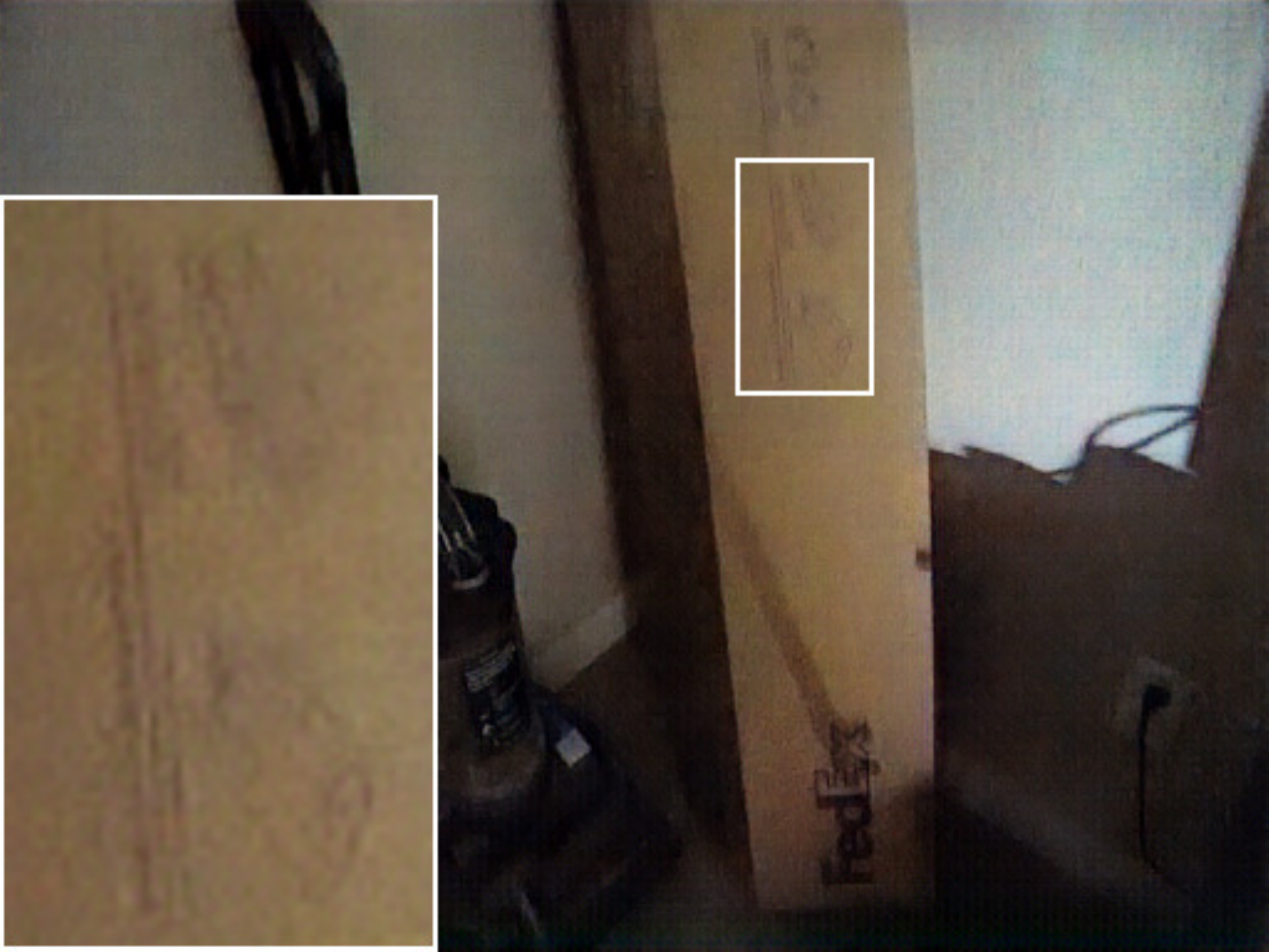}\vspace{0.1cm}
        \end{minipage}\hspace{-0.3mm}
    
        \begin{minipage}[b]{0.14\textwidth}
            \centering
            \includegraphics[width=1\linewidth,height=1.8cm]{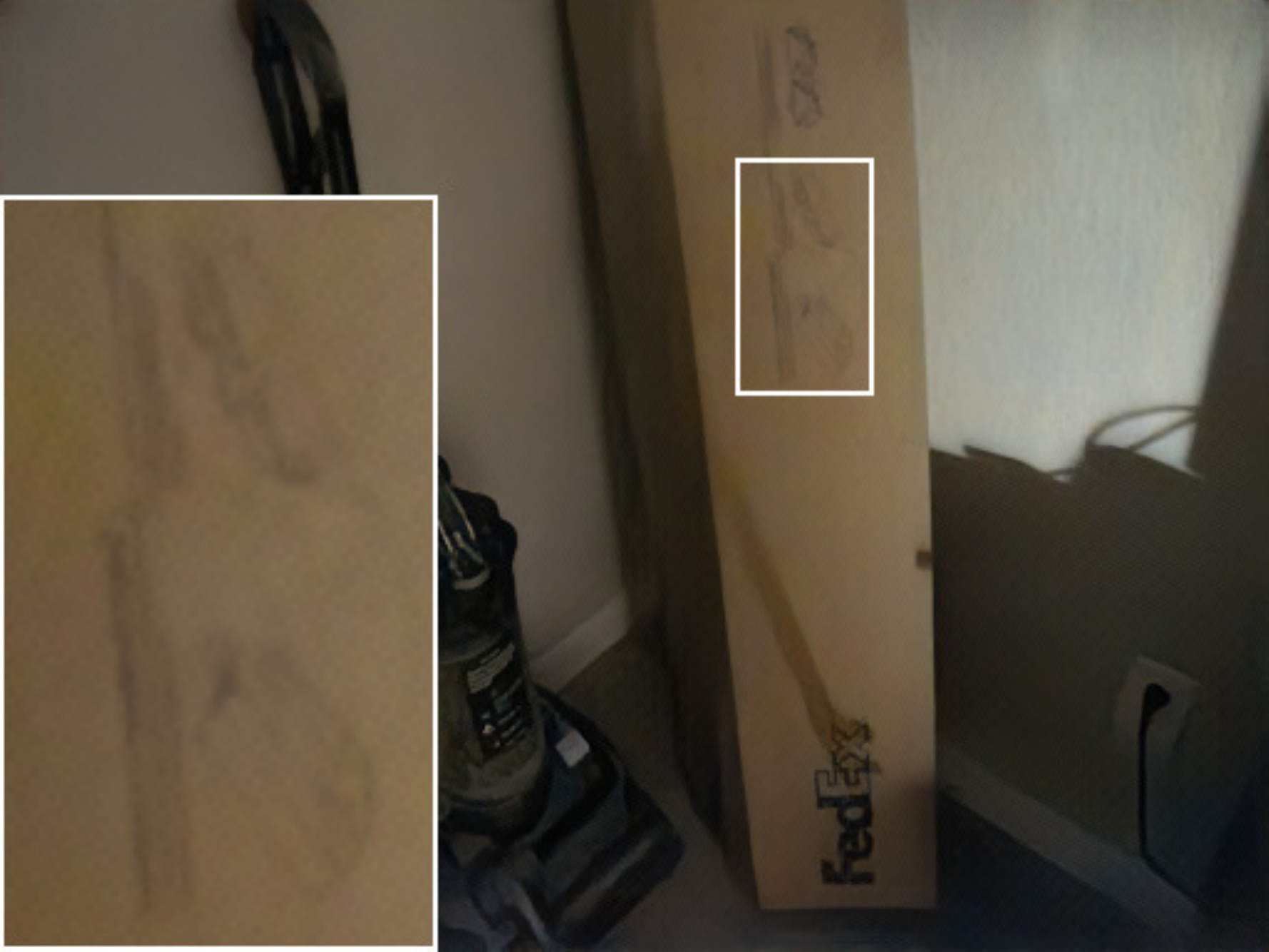}\vspace{0.1cm}
        \end{minipage}\hspace{-0.3mm}
   
         \begin{minipage}[b]{0.14\textwidth}
            \centering
            \includegraphics[width=1\linewidth,height=1.8cm]{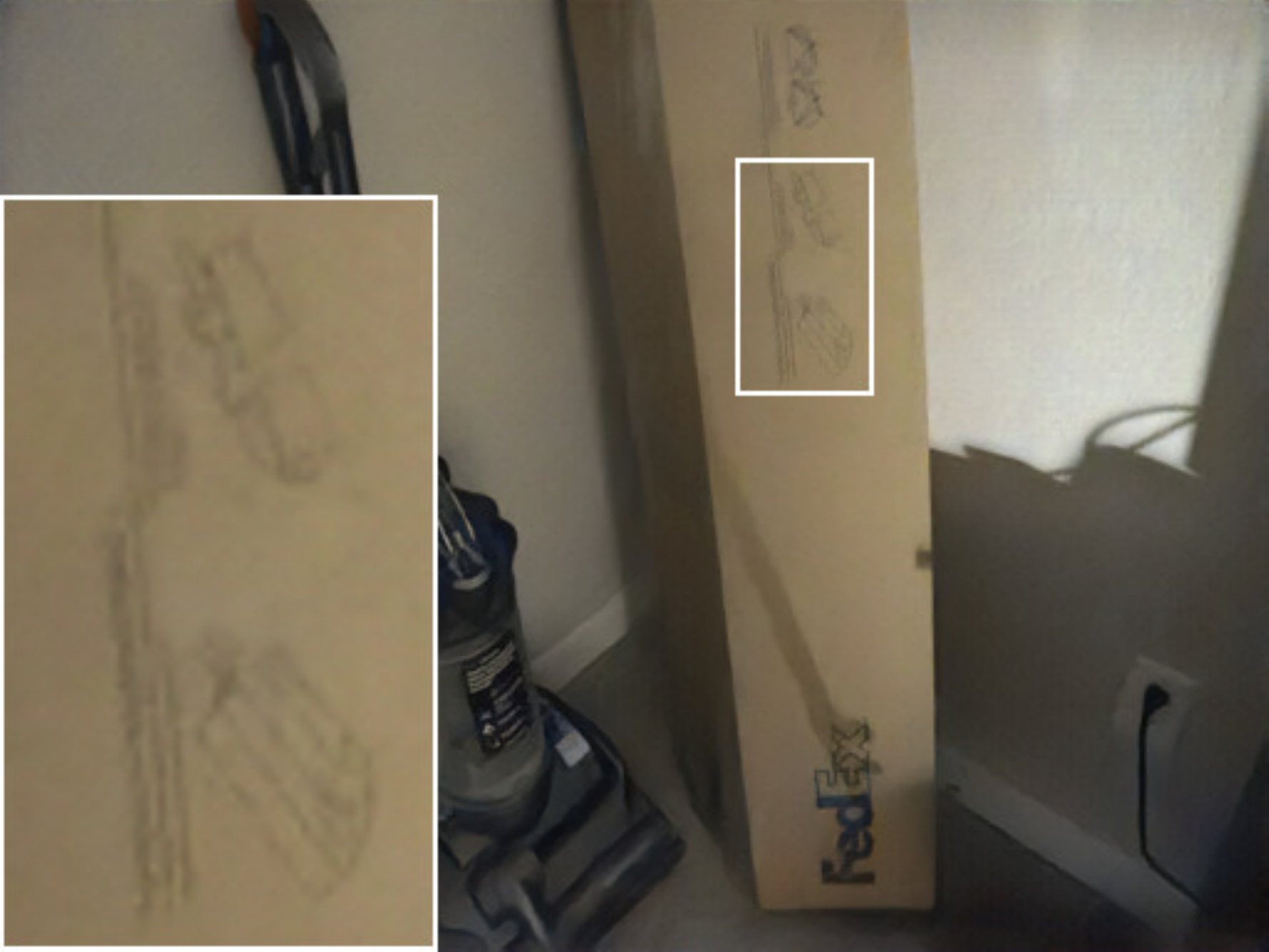}\vspace{0.1cm}
        \end{minipage}\hspace{-0.3mm}
   
         \begin{minipage}[b]{0.14\textwidth}
            \centering
            \includegraphics[width=1\linewidth,height=1.8cm]{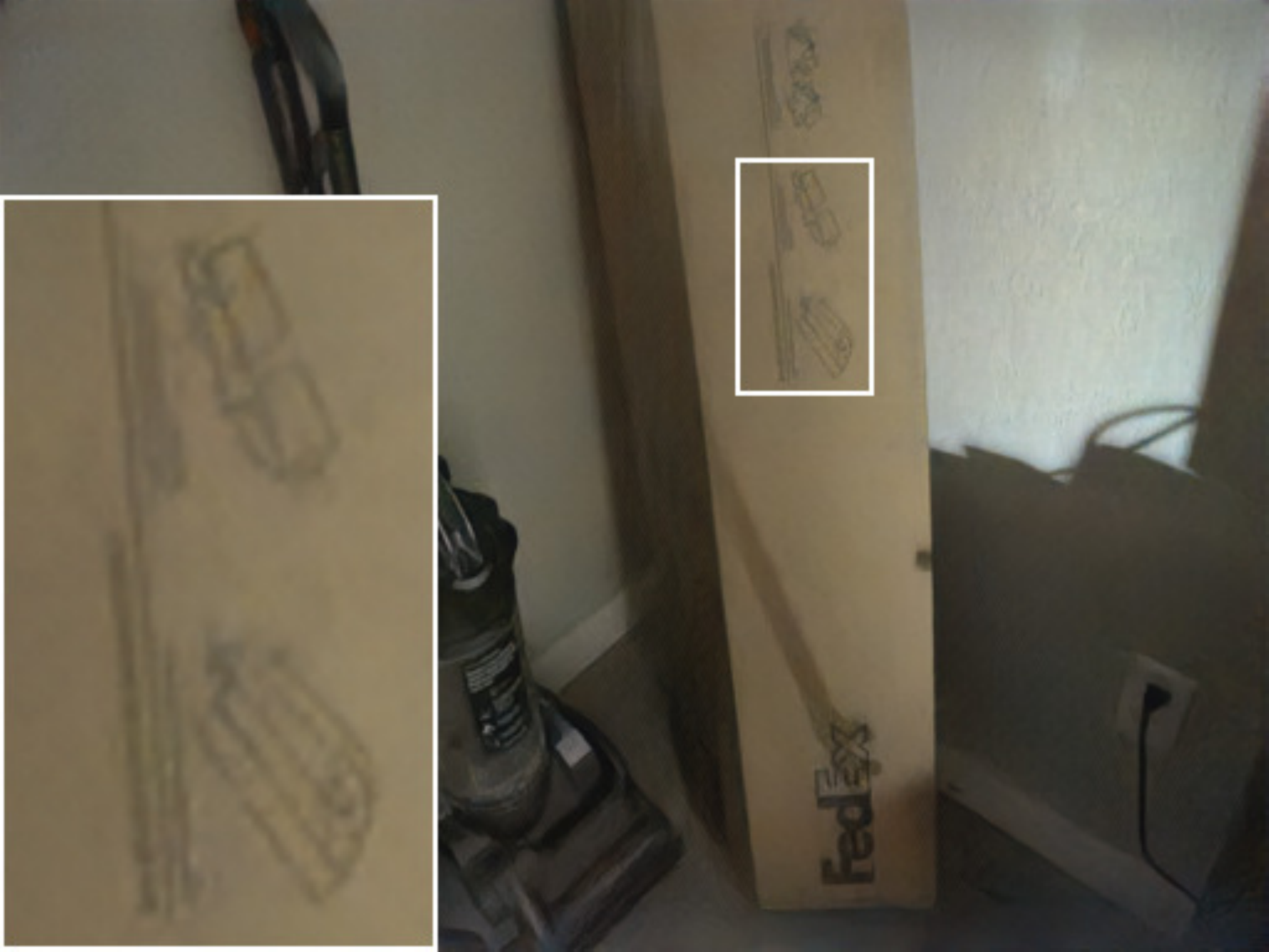}\vspace{0.1cm}
        \end{minipage}\hspace{-0.3mm}
  
         \begin{minipage}[b]{0.14\textwidth}
            \centering
            \includegraphics[width=1\linewidth,height=1.8cm]{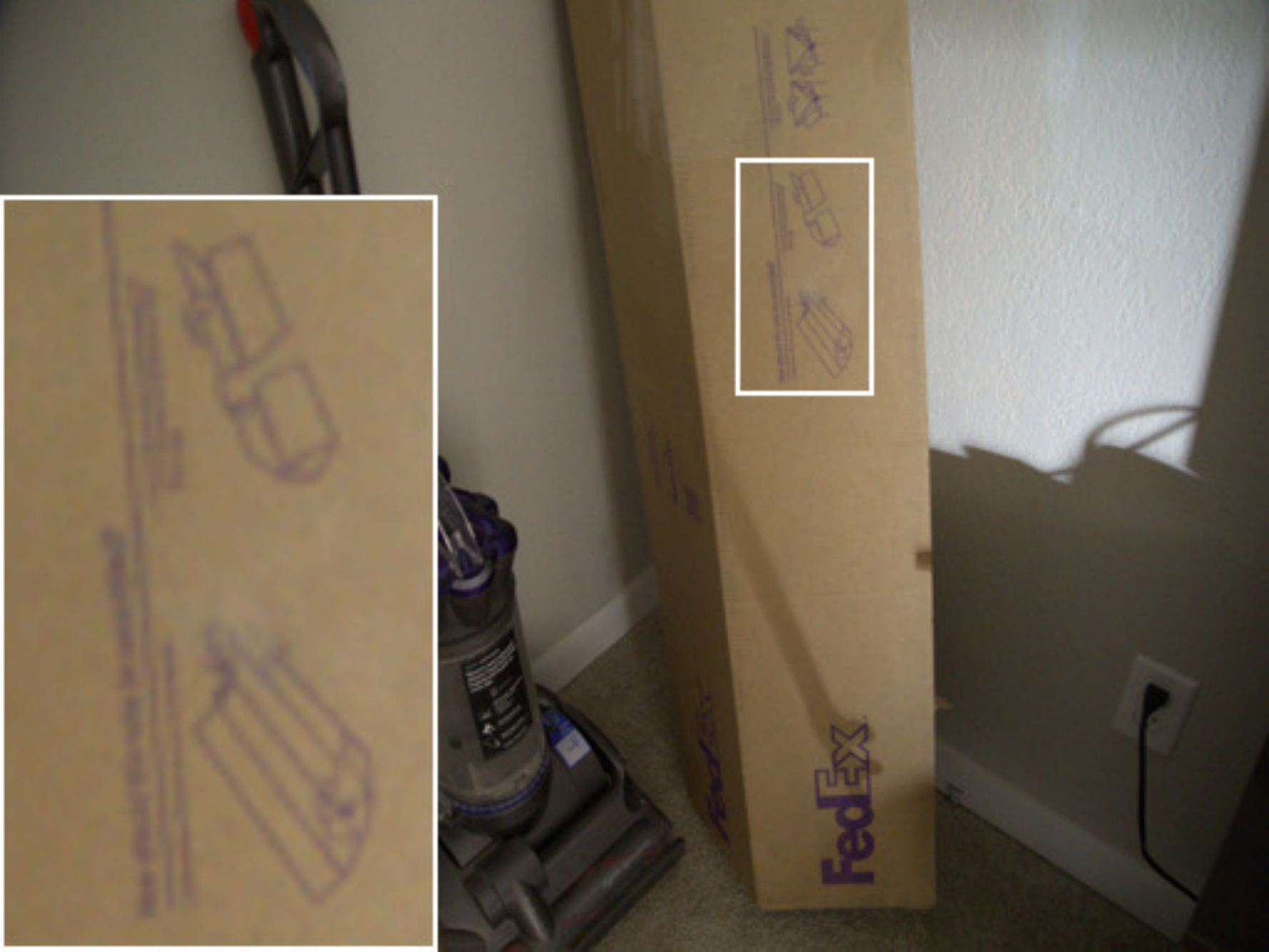}\vspace{0.1cm}
        \end{minipage}\hspace{-0.3mm}
       }
  
        \vspace{-0.25cm}
             
       \subfigure  
    { \footnotesize
         \centering
        \begin{minipage}[b]{0.14\textwidth}
            \centering

     \stackunder[5pt]{\includegraphics[width=1\linewidth,height=1.8cm]{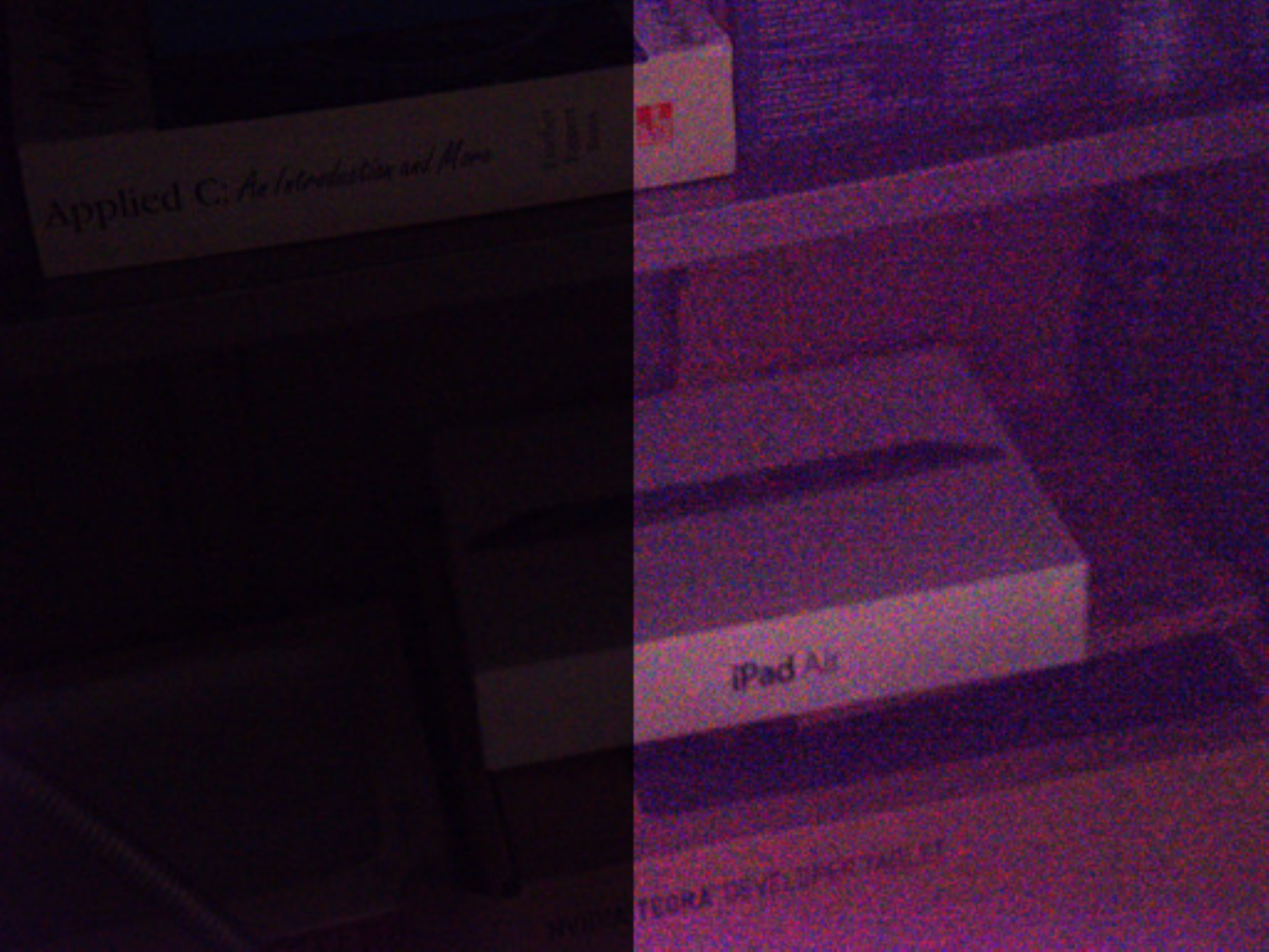}}{Input / LIME}
        \end{minipage}\hspace{-0.1mm}
  
         \begin{minipage}[b]{0.14\textwidth}
            \centering
       \stackunder[5pt]{\includegraphics[width=1\linewidth,height=1.8cm]{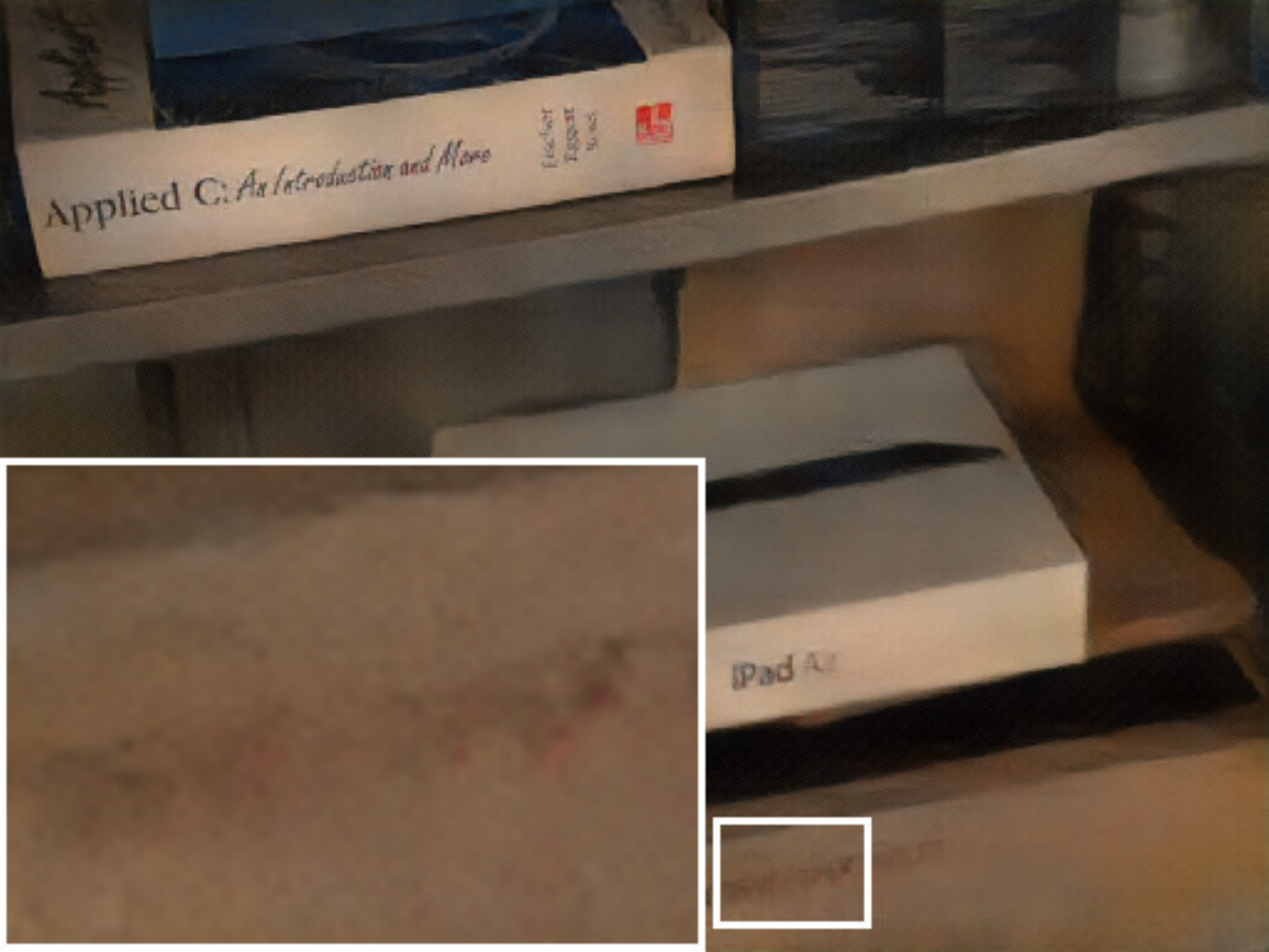}}{SID}      
        \end{minipage}\hspace{-0.1mm}
  
        \begin{minipage}[b]{0.14\textwidth}
            \centering
            \stackunder[5pt]{\includegraphics[width=1\linewidth,height=1.8cm]{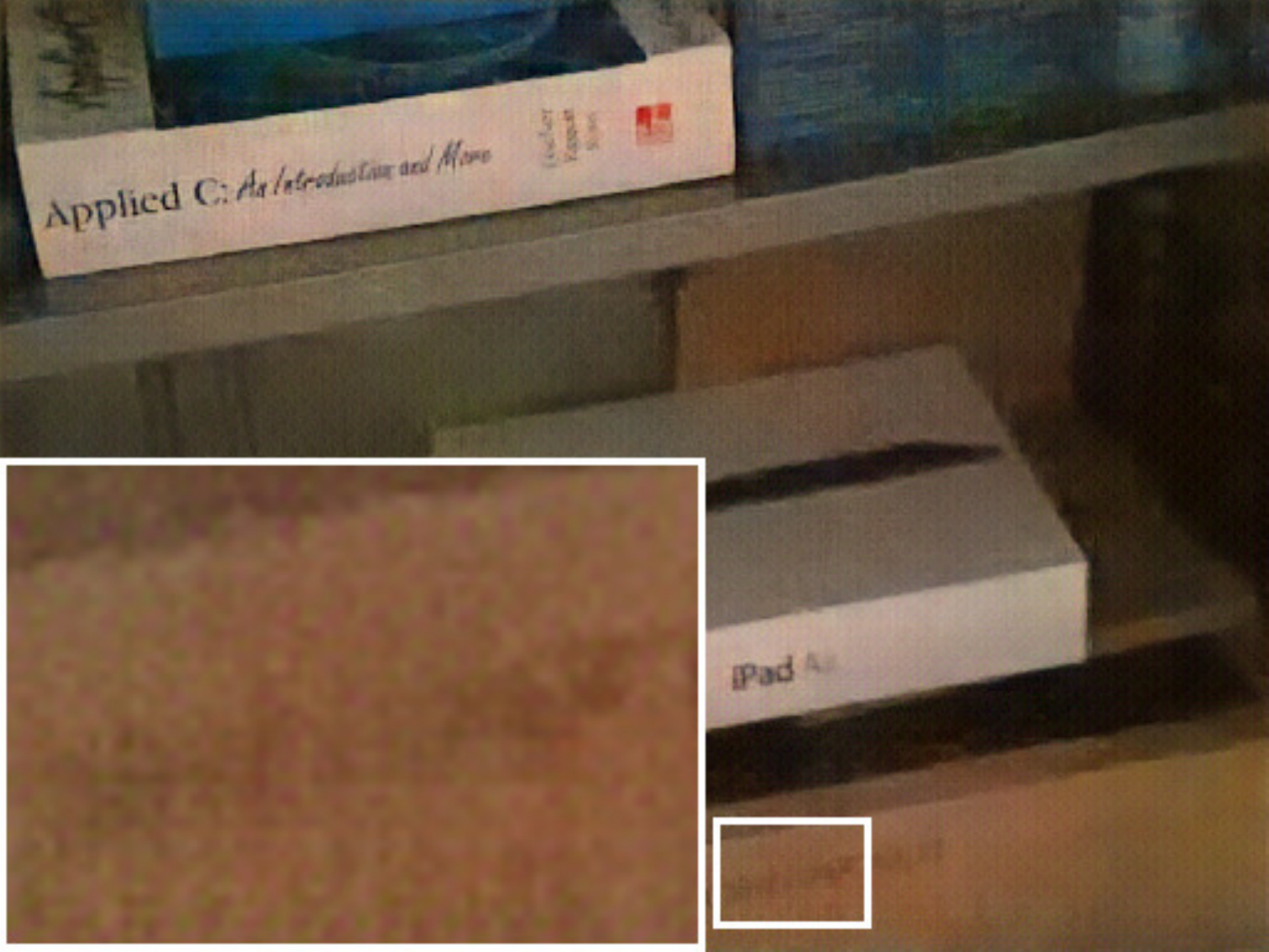}}{Pixel2Pixel}      
        \end{minipage}\hspace{-0.1mm}
  
        \begin{minipage}[b]{0.14\textwidth}
            \centering
            \stackunder[5pt]{\includegraphics[width=1\linewidth,height=1.8cm]{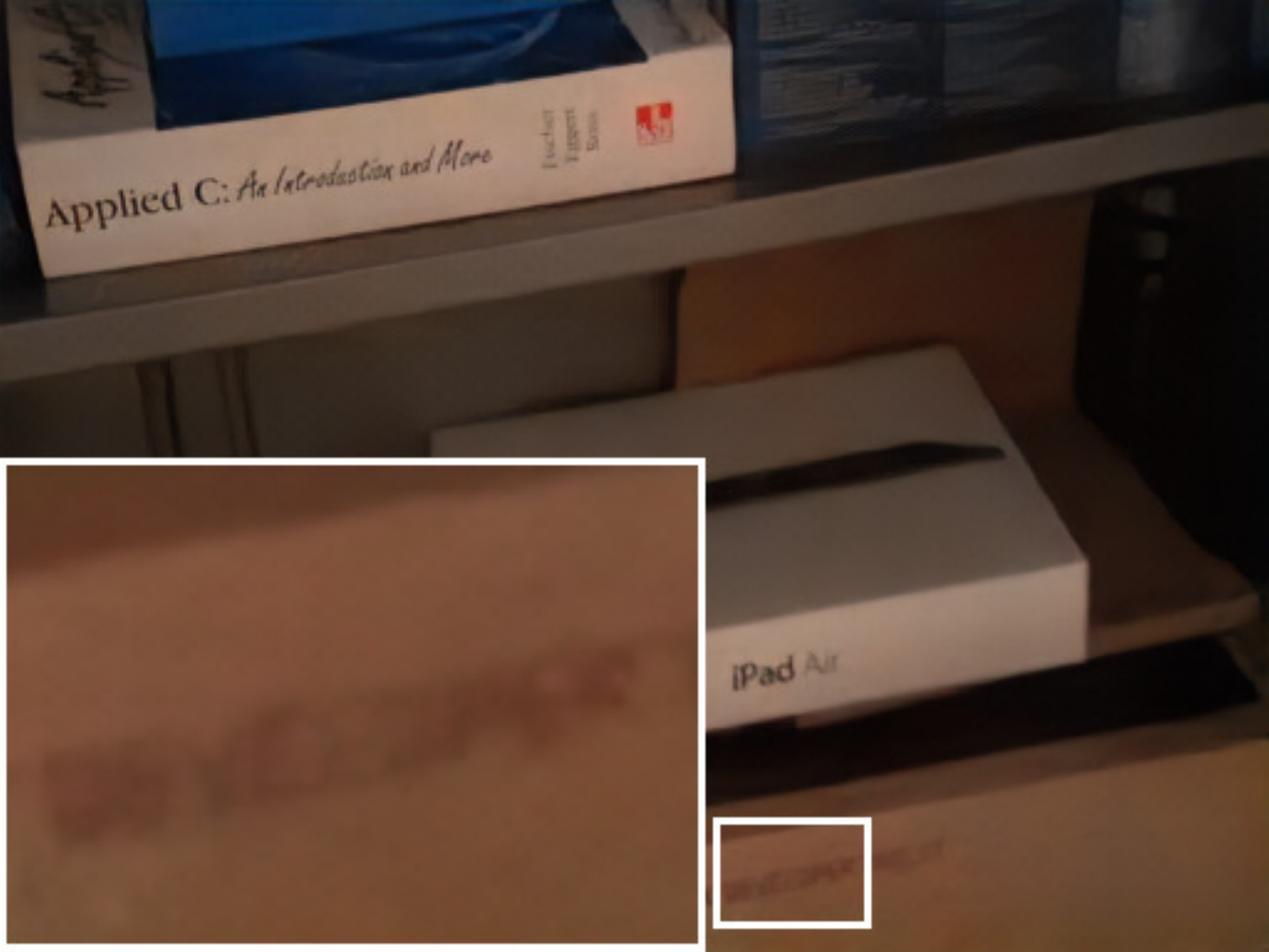}}{D\&E}      
        \end{minipage}\hspace{-0.1mm}
  
         \begin{minipage}[b]{0.14\textwidth}
            \centering
             \stackunder[5pt]{\includegraphics[width=1\linewidth,height=1.8cm]{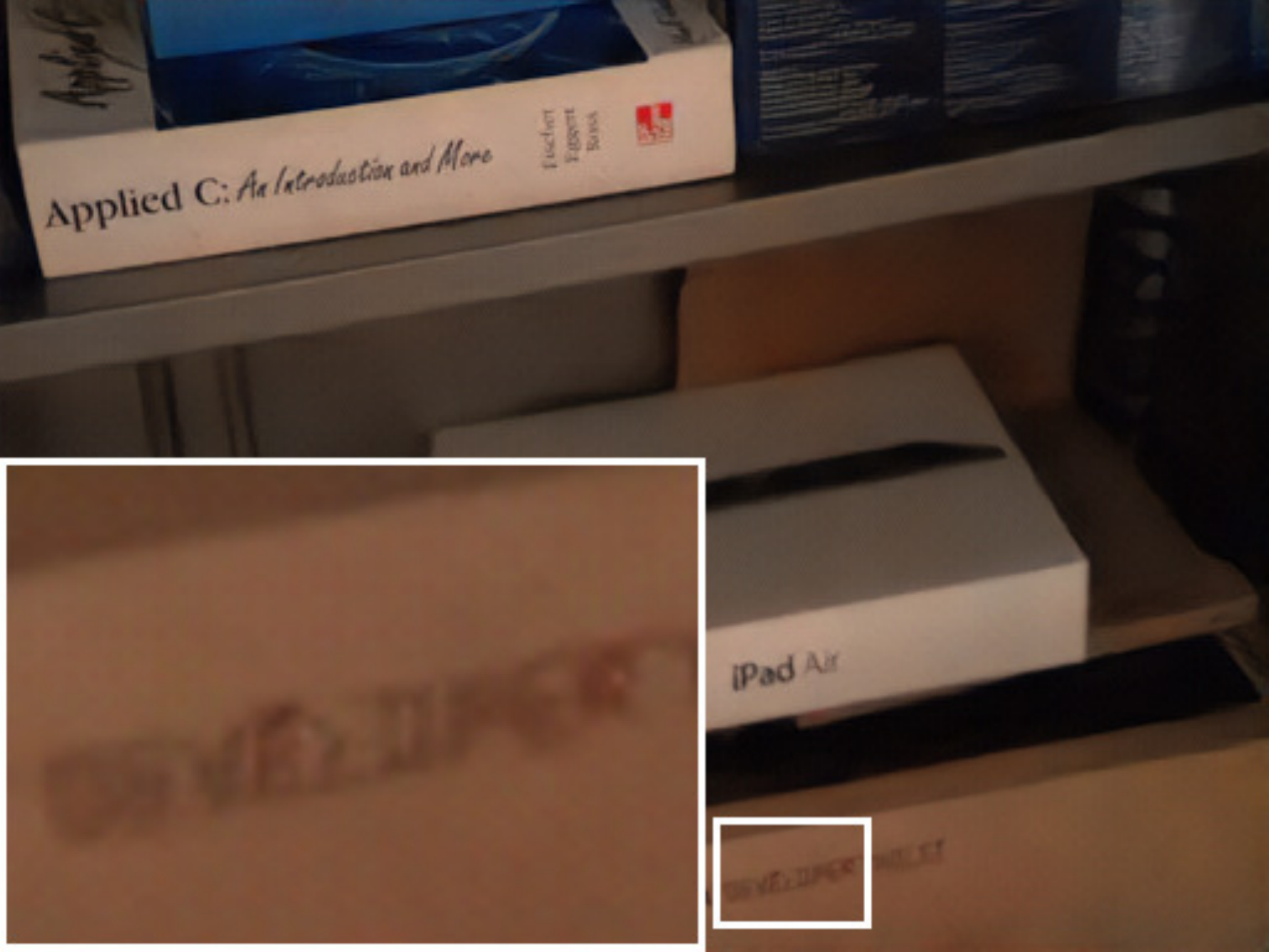}}{MIRNet}      
        \end{minipage}\hspace{-0.1mm}
  
         \begin{minipage}[b]{0.139\textwidth}
            \centering
             \stackunder[5pt]{\includegraphics[width=1\linewidth,height=1.8cm]{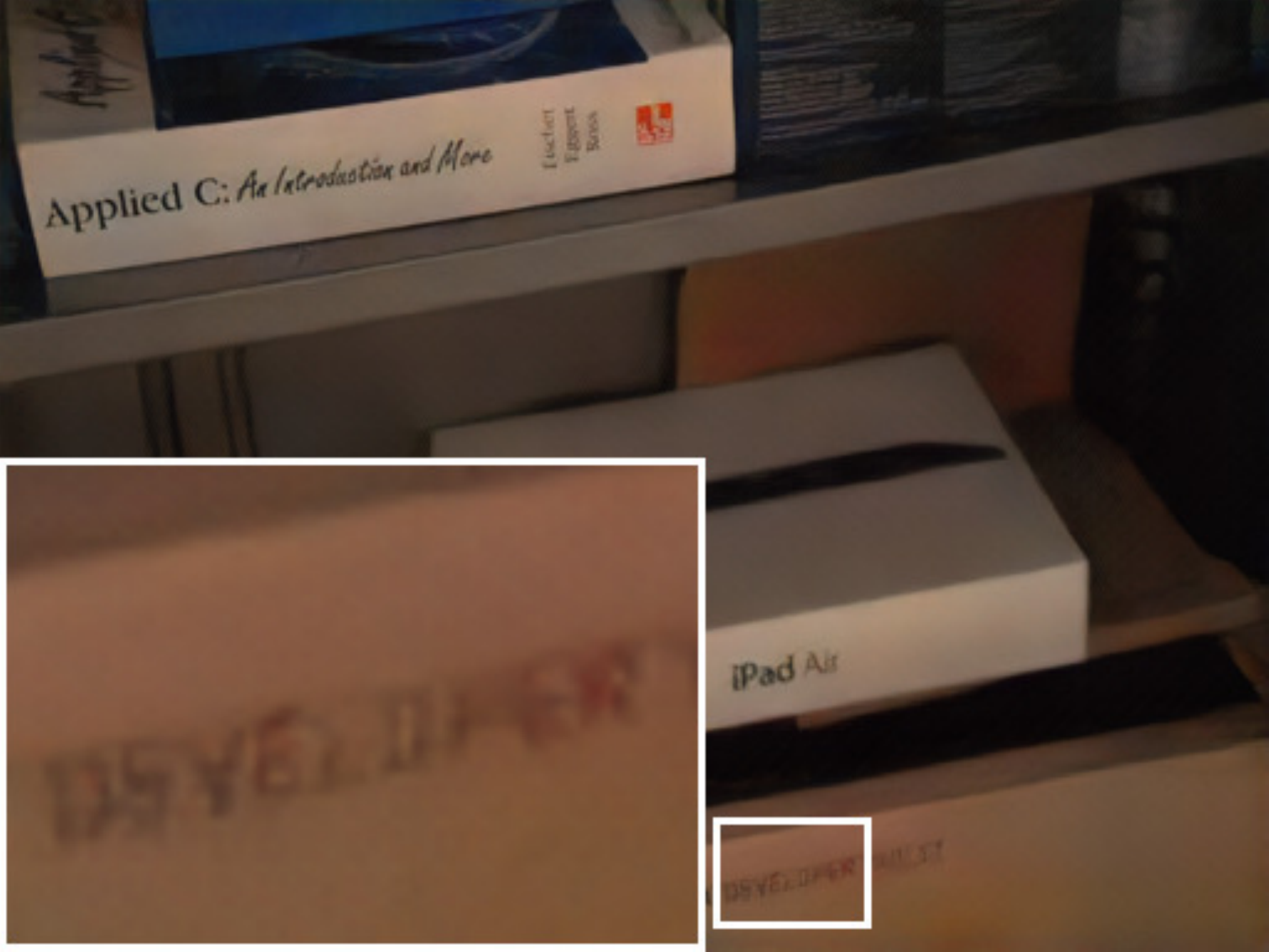}}{\textbf{UTVNet (Ours)}}      
        \end{minipage}\hspace{-0.1mm}
   
         \begin{minipage}[b]{0.14\textwidth}
            \centering
           \stackunder[5pt]{\includegraphics[width=1\linewidth,height=1.8cm]{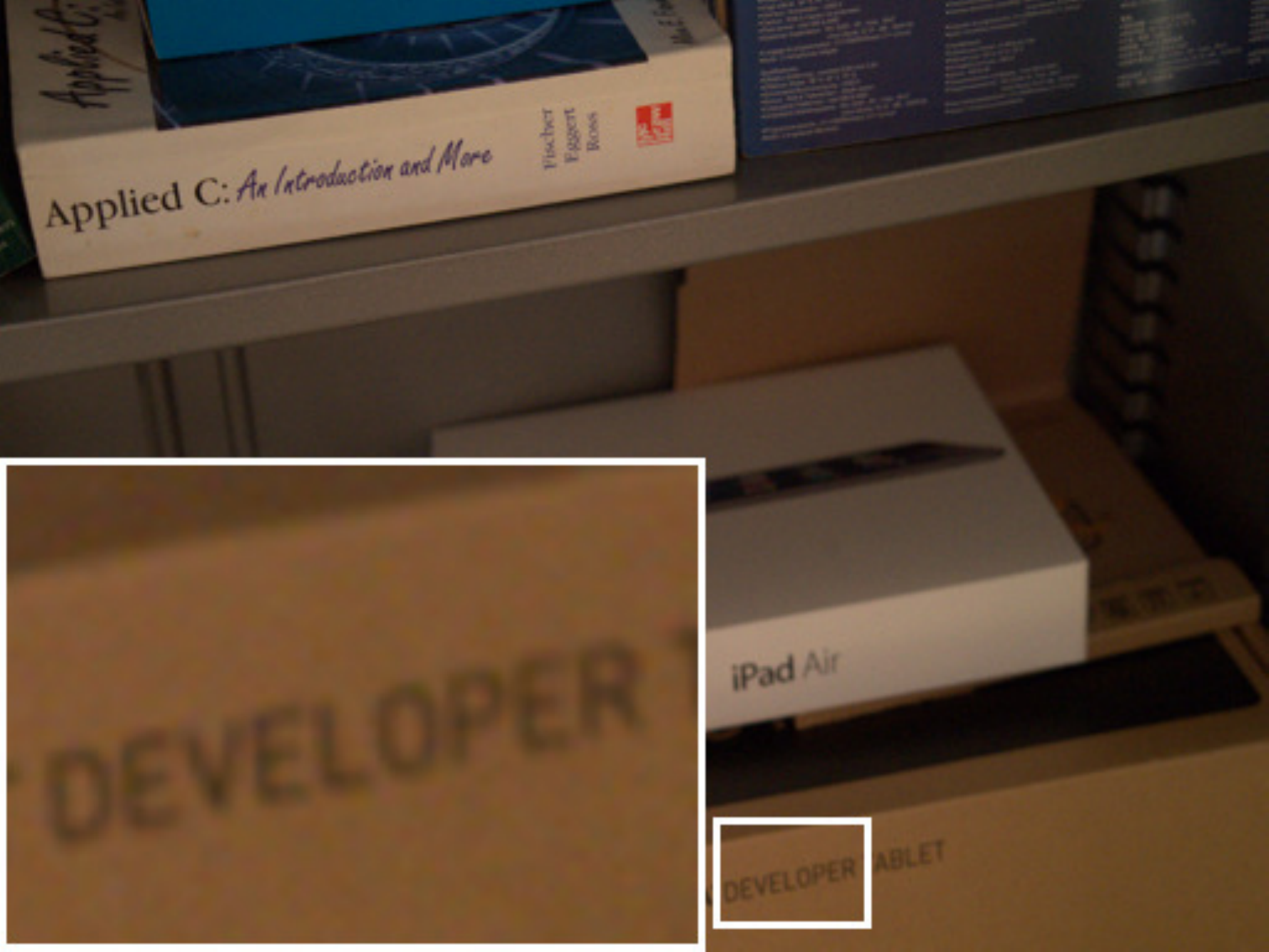}}{Ground Truth}  
        \end{minipage}
       }

 
\end{center}
\caption{ Visual results of different methods. UTVNet 
 outperforms other methods in finer details preservation. \textit{(Best viewed with zoom
)  } 
   }
  
\end{figure*}

\subsection{Low-light Image Restoration Module}

The image restoration module consists of a luminance correction sub-network and a noise suppression sub-network.\;Through unrolling the TV module, the noise-free smooth layer $\mathbf{y}_{s}$ and the approximated noise level map $\mathbf{M}$ are generated.\;As shown in Fig.\;2, $\mathbf{M}$ is utilized for noise suppression on the detail layer $\mathbf{y}_{d}=\mathbf{y}-\mathbf{y}_{s}$.\;Meanwhile, the smooth layer $\mathbf{y}_{s}$ is the input of the luminance correction sub-network. The details are described below. 

\textbf{Luminance correction sub-network.} 
The luminance correction sub-network generates luminance amplification for each pixel. By multiplying the luminance amplification with $\mathbf{y}_{s}$, we get the luminance corrected smooth layer $\mathbf{y}_{s}^{\prime}$. The sub-network adopts seven dilated convolution layers with LeakyReLU in the first six layers.

\textbf{Noise suppression sub-network.} 
As noted in \cite{bd,FFDnet}, taking both the noisy image and the noise level map as input helps generalize the learned model for blind denoising. For low-light image restoration, we also adopt a method that takes $\mathbf{y}_{d}$ and the approximated noise level map $\mathbf{M}$ as input for noise suppression and detail restoration of the original low-light image.\;As shown in Fig.\;2, we adopt U-Net \cite{unet} architecture with four scales, and the channels of the convolution layer in each scale are 32, 64, 128 and 256, respectively. Average pooling and fully-connected layers are used to process global features and connected them with former layers. $2\times2$ convolution and transposed convolution are adopted in downscaling and upscaling layers. For all convolution layers except the last, we use LeakyReLU as the activation function.

\subsection{End-to-End Training}
 As the TV minimization problem is solved in unfolding TV module when approximating the noise level map, the network can be trained end-to-end. We optimize weights and biases by minimizing the $L_{2}$ loss $\mathcal{L}_{l2}$ and the  perceptual loss $\mathcal{L}_{per} $ with a weight of 1 and 0.12, respectively.

\section{Experiment}
\textbf{Implementation detail.}  The proposed UTVNet\footnote{Code: \url{ https://github.com/CharlieZCJ/UTVNet }}  is implemented in PyTorch framework, before training, we initialize parameters for UTVNet. We set $\rho_{0}=2$, $\mathbf{x}_{0}=\mathbf{y}$, $\mathbf{u}_{0}=\mathbf{D}\mathbf{x}_{0}$, $\mathbf{z}_{0}=0$, and $K=8$.  To optimize parameters of UTVNet, we adopt the ADAM optimizer \cite{adam} for 320 epochs with an initial learning rate of $10^{-4}$.  The learning rate is reduced by half after 240 epochs. During training, the images are resized to have a long-edge of 512 pixels to avoid unnatural deformation. All the experiments are performed on a single NVIDIA TITAN Xp GPU.

\textbf{Dataset.}
To suppress noise and enhance real-world low-light images in the sRGB domain, we train the UTVNet with the sRGB-SID dataset proposed by \cite{LDE}. The sRGB-SID dataset is generated from SID \cite{learningtosee} raw dataset to model real-world noisy low-light sRGB images. It contains 4,198 image pairs for training and 1,196 image pairs for testing.   

\begin{table}
\footnotesize
\setlength{\extrarowheight}{2.5pt}
\setlength{\tabcolsep}{1.9mm}
\caption{Quantitative comparison of different methods on sRGB-SID dataset. * indicates the released pre-trained model on the same dataset. The best result is highlighted.}
\vspace{0.15cm}
\begin{tabular}{lccccc}
\toprule
\textbf{Method} & \textbf{PSNR } &  \textbf{SSIM } &  \textbf{$\mathbf{L2}_{Lab}$} &  \textbf{LPIPS} \\
\midrule

LIME \cite{LIME}       &15.515&0.3011&14.040&0.5230 \\

LIME \cite{LIME} + NBNet \cite{cbdnet}&16.280&0.3817&14.191&0.5087\\
NBNet \cite{nbn} + LIME \cite{LIME}&13.455&0.4416&15.920&0.6120\\

White-box \cite{whitebox}            &15.941&0.4149 &13.755&0.5569  \\
White-box \cite{whitebox} + NBNet \cite{cbdnet} &14.736&0.4938&13.624&0.5391 \\
NBNet \cite{nbn} + White-box \cite{whitebox}&13.875&0.4467&15.094&0.5964\\\hline

DPED \cite{dped}                     &16.528&0.5144&12.841&0.4924   \\
DeepUPE \cite{upe}          &13.412&0.2565&12.411&0.5188  \\

CSRNet \cite{csrnet}           &19.613&0.5696&12.183 &0.4779\\
DeepLPF \cite{lpf}          &19.645&0.6472&11.847&0.4207 \\

Pixel2Pixel \cite{p2p}     &21.182&0.6073&11.626&0.4959  \\
SID \cite{learningtosee}              &21.212&0.6601&10.998 &0.4026  \\
AGLLNet \cite{atl}   &18.641&0.6158&13.649 &0.4472\\

D\&E* \cite{LDE}                     &21.934&0.6950  &10.859 &0.3744\\
D\&E \cite{LDE}                     &22.157&0.6922  &10.751 &0.3794\\
MIRNet \cite{MIRNet}             &22.345&0.7031&10.750 &0.3562\\\hline

\textbf{UTVNet (Ours)} &\textbf{22.691}&\textbf{0.7179}&\textbf{10.581}&\textbf{0.3417}  \\

\bottomrule

\end{tabular}
\vspace{-0.3cm}
\end{table}

\subsection{Results on sRGB-SID dataset}
We compare the proposed UTVNet with 11 recent methods, including LIME \cite{LIME}, DPED \cite{dped}, D\&E \cite{LDE},  CSRNet \cite{csrnet}, SID \cite{learningtosee},   DeepUPE \cite{upe},  DeepLPF \cite{lpf},  White-Box \cite{whitebox}, Pixel2Pixel \cite{p2p}, AGLLNet \cite{atl},  MIRNet \cite{MIRNet}.\;For Pixel2Pixel, SID, DeepLPF, CSRNet, D\&E, MIRNet, we re-train their models for comparison. Because of the limitation on memory capacity, the number of features in the first layer of MIRNet is limited at 20. Meanwhile, the perceptual loss were adopted when training all the methods.

\begin{figure*}
 \hsize=\textwidth
  \subfigcapskip=-8pt
 \subfigbottomskip=-3pt

    \begin{minipage}{0.79\linewidth}
    \subfigure 
    {
        \begin{minipage}[b]{0.16\linewidth}
            \centering
            \includegraphics[width=1\linewidth,height=1.8cm]{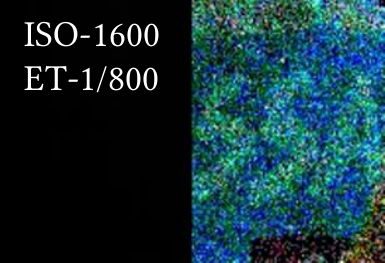}\vspace{0.1cm}
        \end{minipage}\hspace{-0.3mm}
   
        \begin{minipage}[b]{0.16\linewidth}
            \centering
            \includegraphics[width=1\linewidth,height=1.8cm]{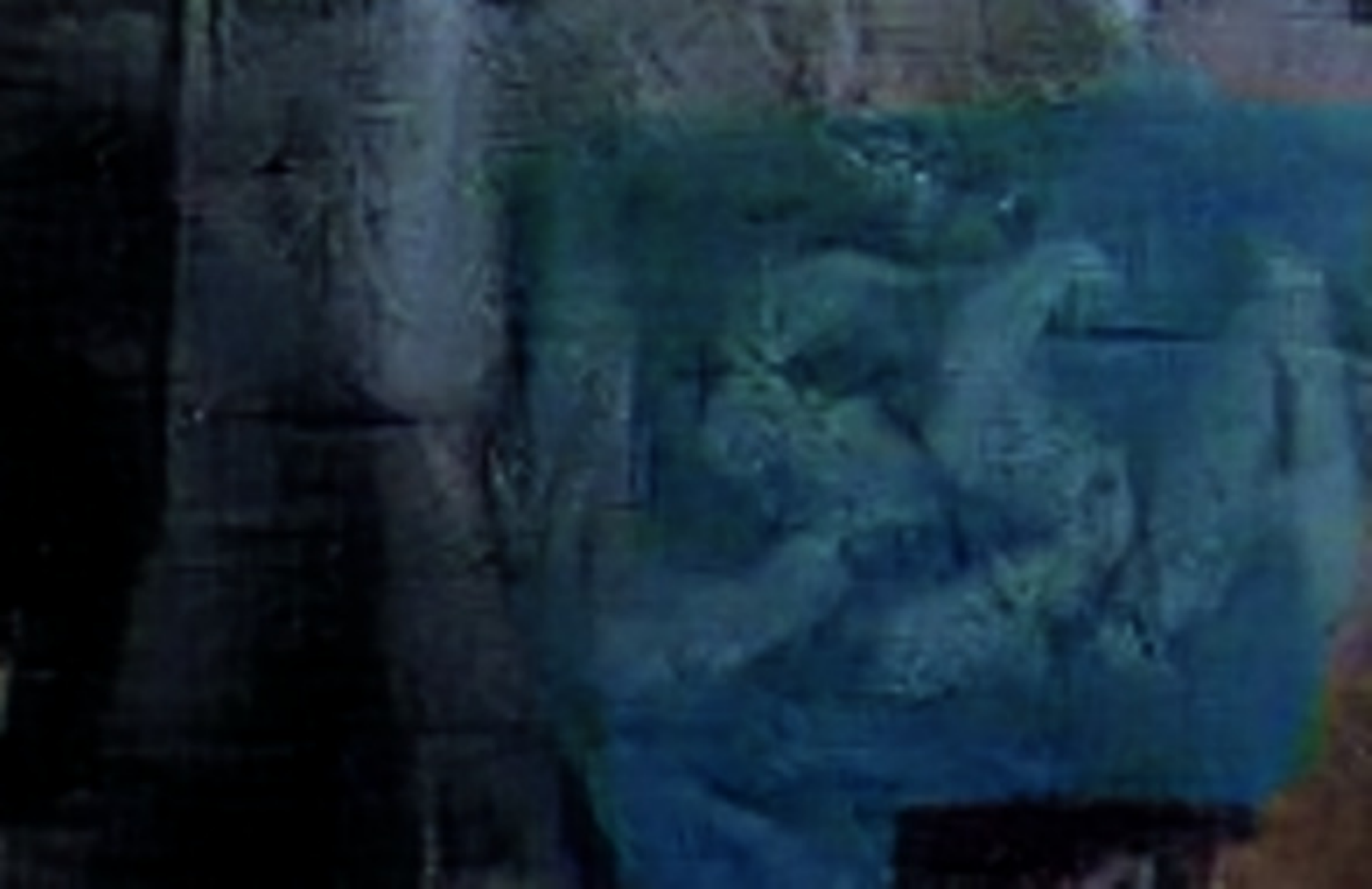}\vspace{0.1cm}
        \end{minipage}\hspace{-0.3mm}
   
         \begin{minipage}[b]{0.16\linewidth}
            \centering
            \includegraphics[width=1\linewidth,height=1.8cm]{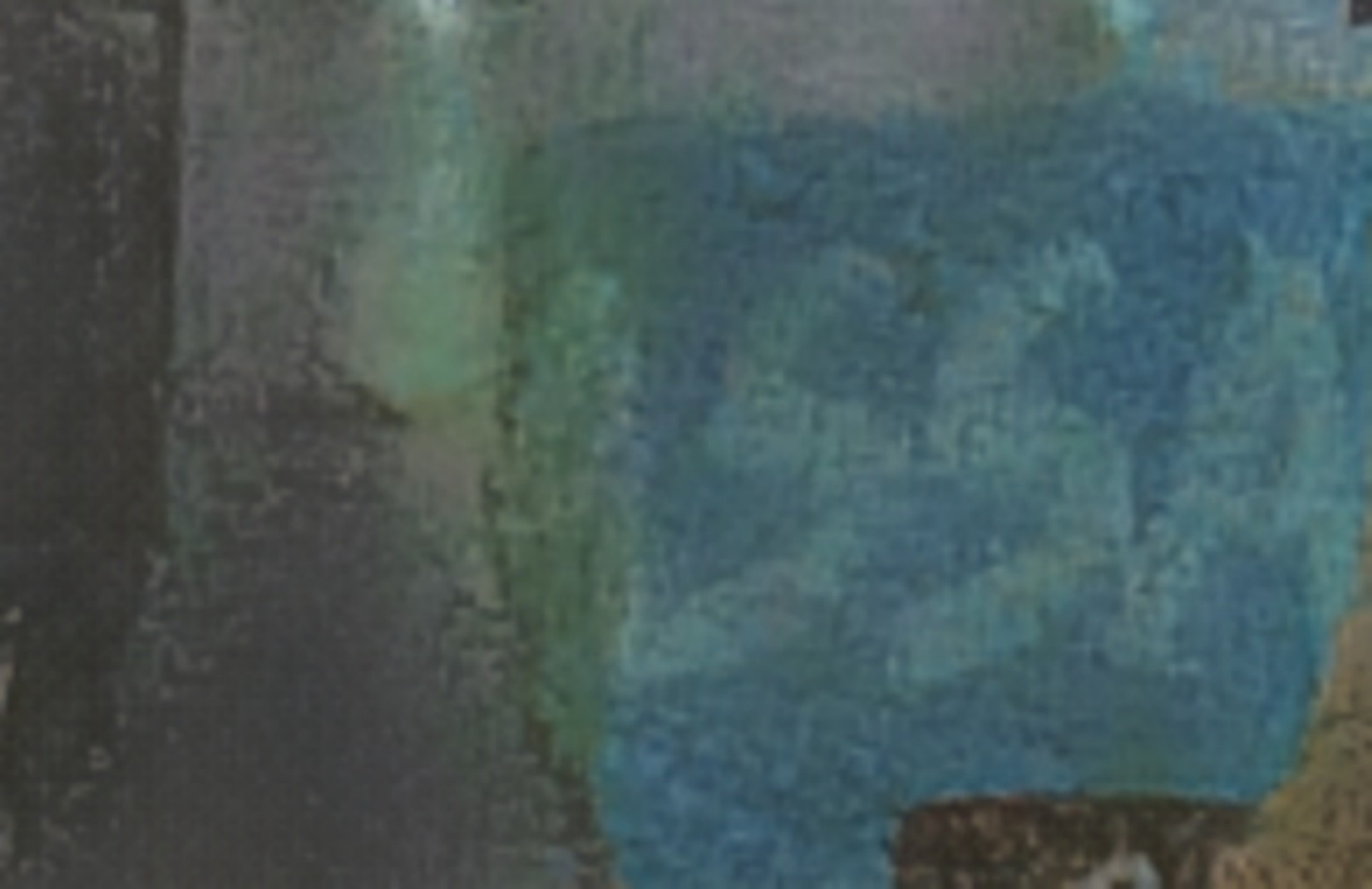}\vspace{0.1cm}
        \end{minipage}\hspace{-0.3mm}
   
         \begin{minipage}[b]{0.16\linewidth}
            \centering
            \includegraphics[width=1\linewidth,height=1.8cm]{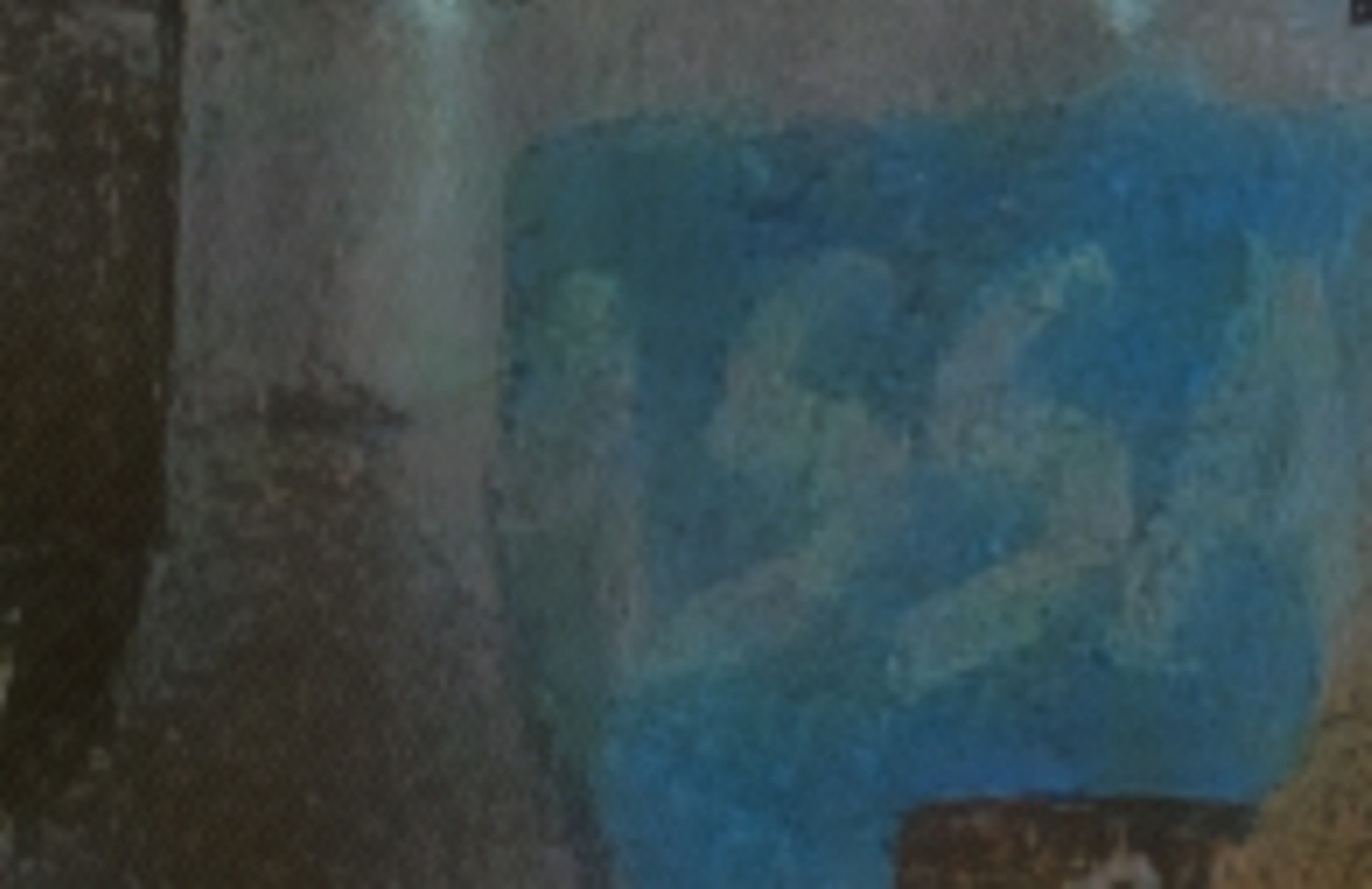}\vspace{0.1cm}
        \end{minipage}\hspace{-0.3mm}
   
         \begin{minipage}[b]{0.16\linewidth}
            \centering
            \includegraphics[width=1\linewidth,height=1.8cm]{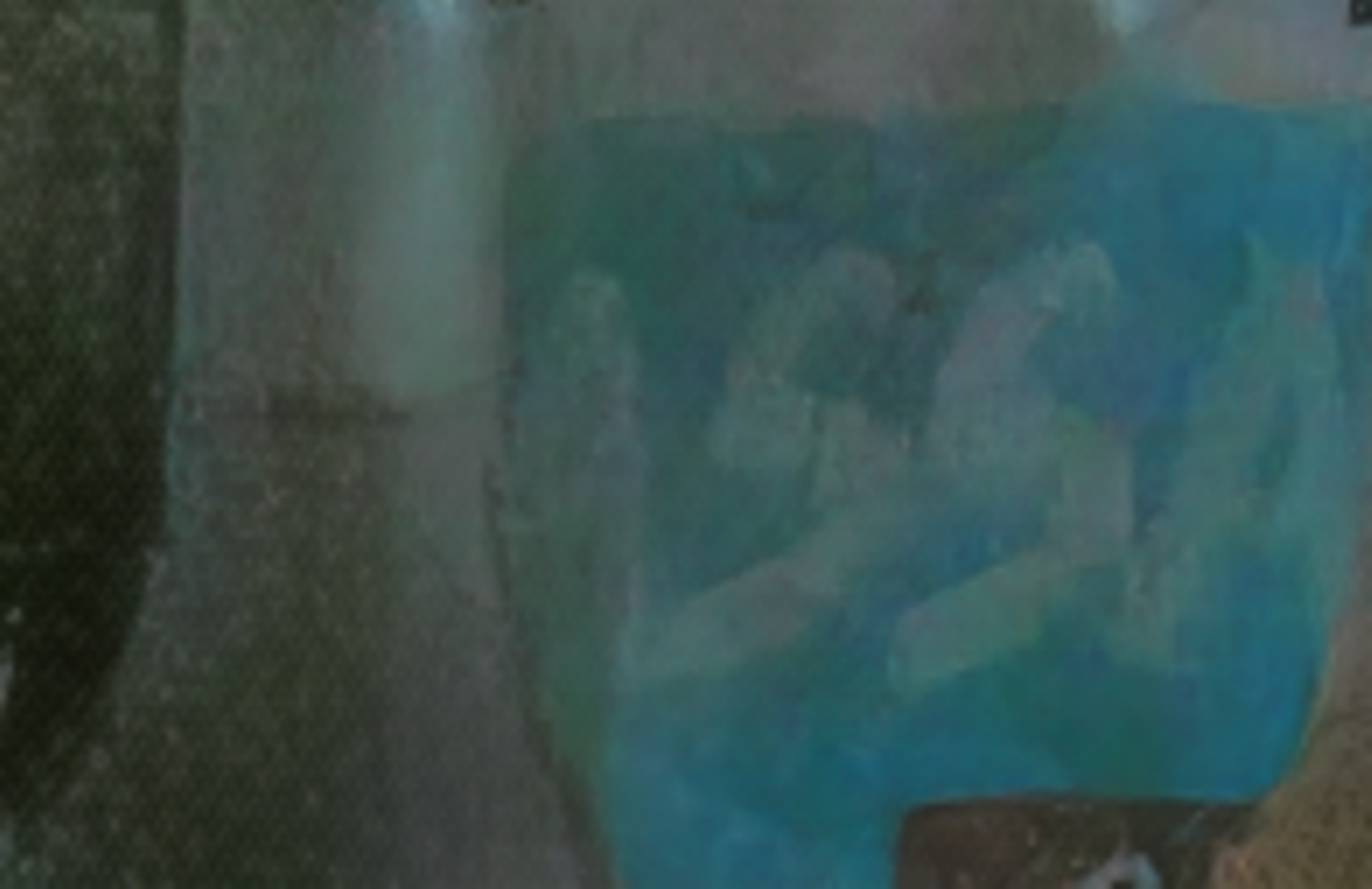}\vspace{0.1cm}
        \end{minipage}\hspace{-0.3mm}
        
                 \begin{minipage}[b]{0.16\linewidth}
            \centering
            \includegraphics[width=1\linewidth,height=1.8cm]{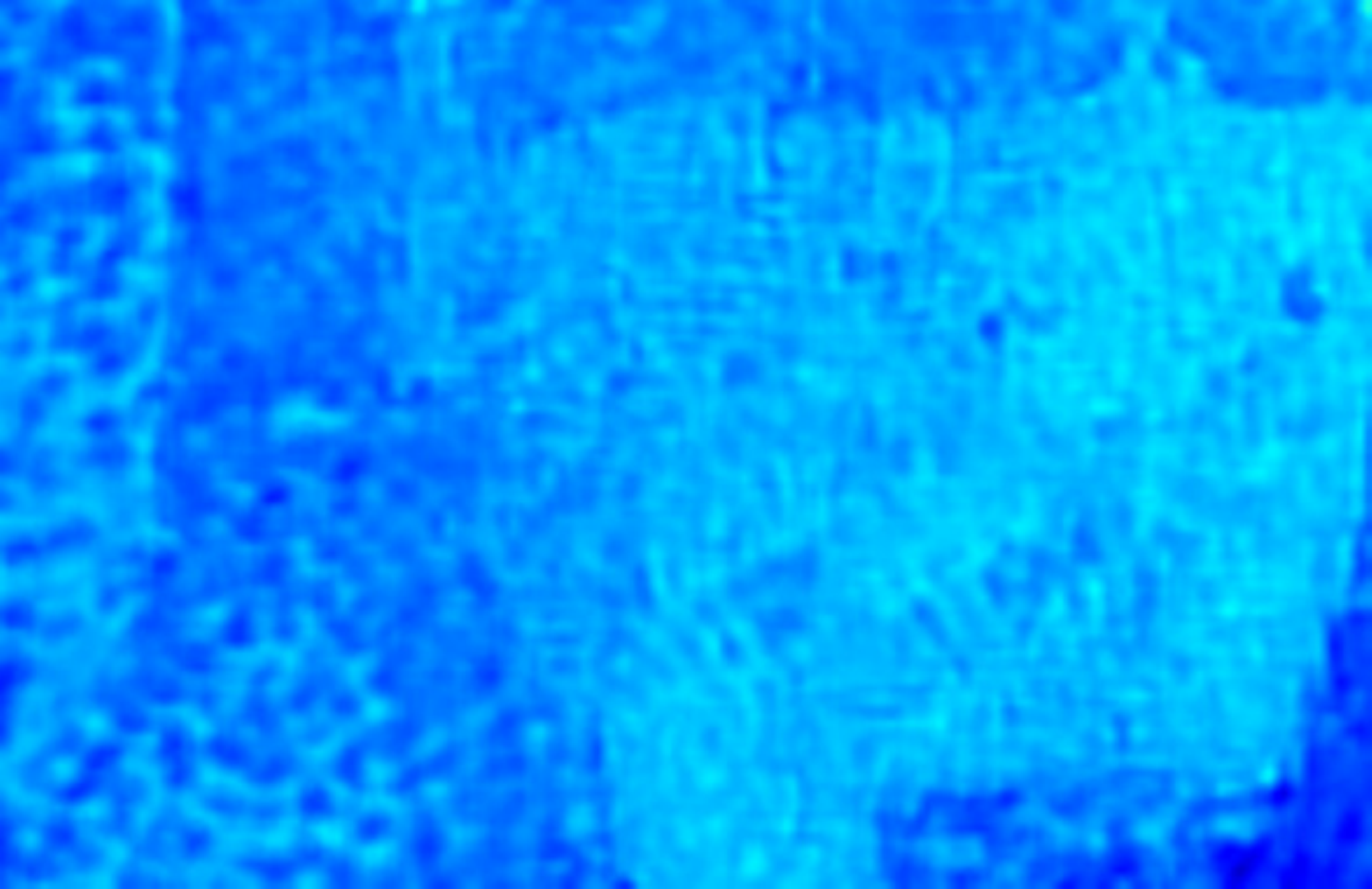}\vspace{0.1cm}
        \end{minipage}
       }
       
           \vspace{-0.25cm}

    \subfigure 
    {   \footnotesize
        \begin{minipage}[b]{0.16\textwidth}
            \centering
             \footnotesize
                 \stackunder[5pt]{\includegraphics[width=1\linewidth,height=1.8cm]{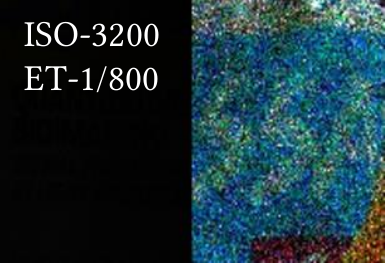}}{Input / LIME}
        \end{minipage}\hspace{-0.1mm}

        \begin{minipage}[b]{0.16\textwidth}
            \centering
\stackunder[5pt]{\includegraphics[width=1\linewidth,height=1.8cm]{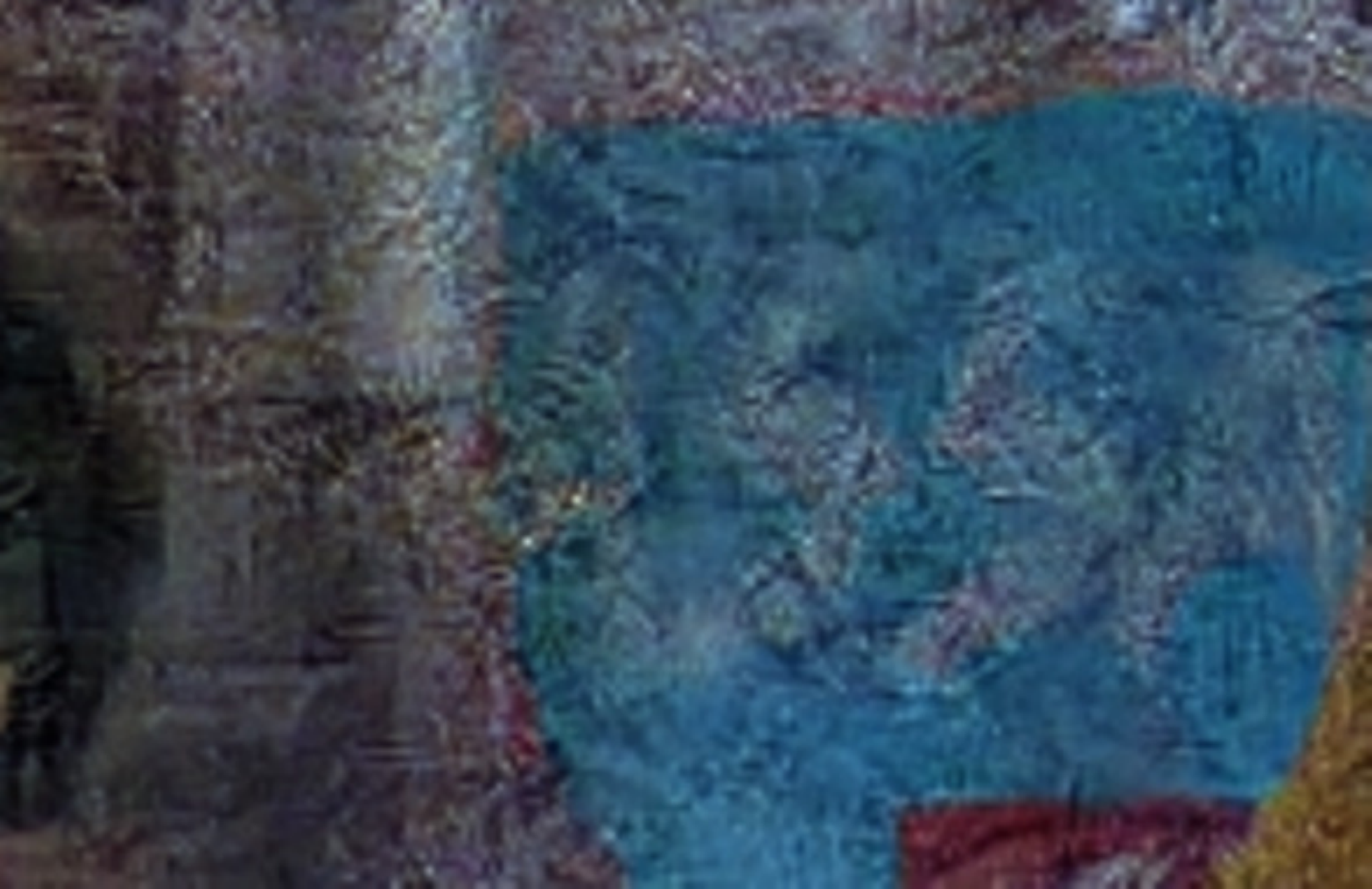}}{AGLLNet}
        \end{minipage}\hspace{-0.1mm}
   
         \begin{minipage}[b]{0.16\textwidth}
            \centering
\stackunder[5pt]{\includegraphics[width=1\linewidth,height=1.8cm]{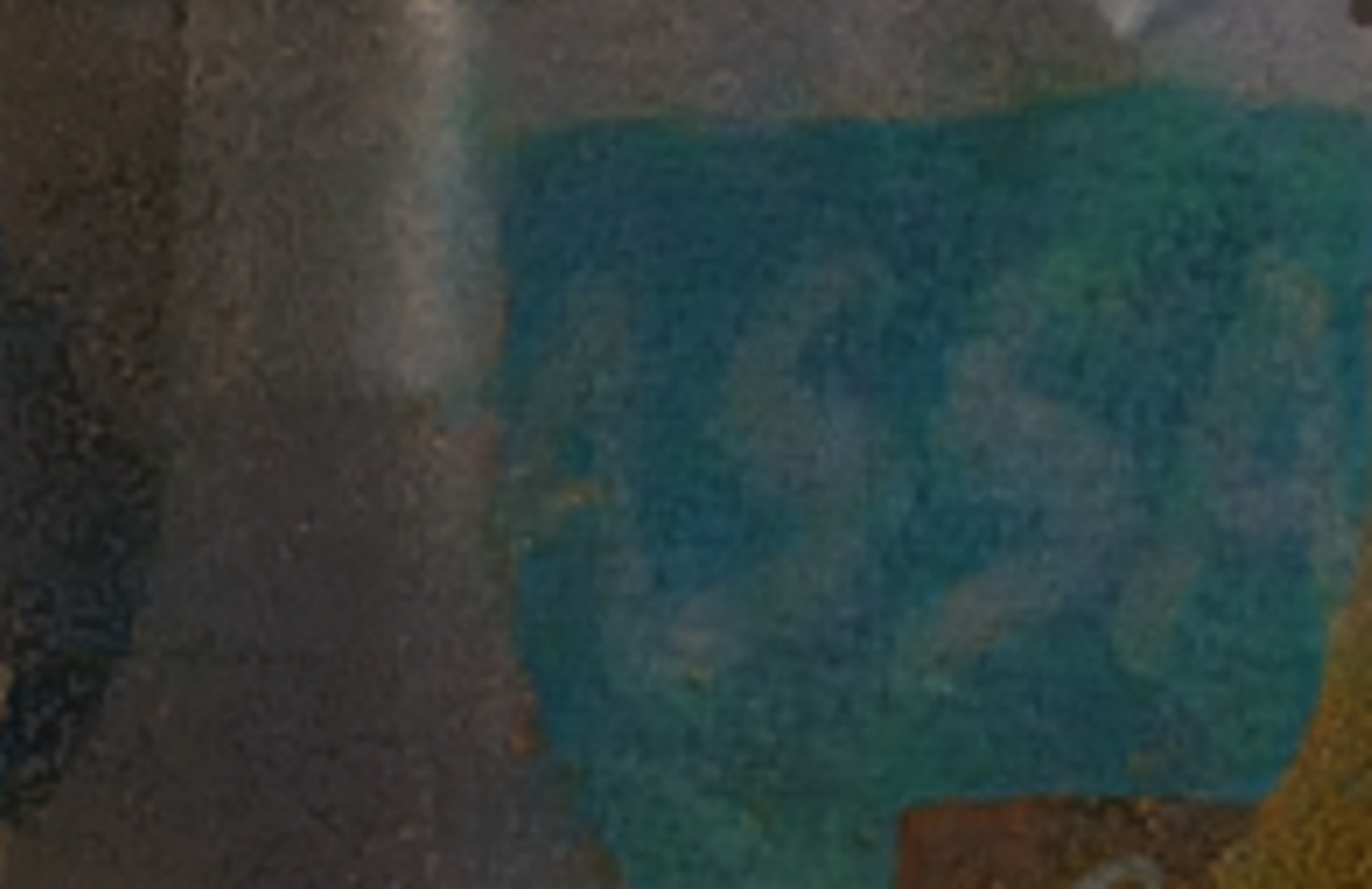}}{MIRNet}
        \end{minipage}\hspace{-0.1mm}

         \begin{minipage}[b]{0.16\textwidth}
            \centering
\stackunder[5pt]{\includegraphics[width=1\linewidth,height=1.8cm]{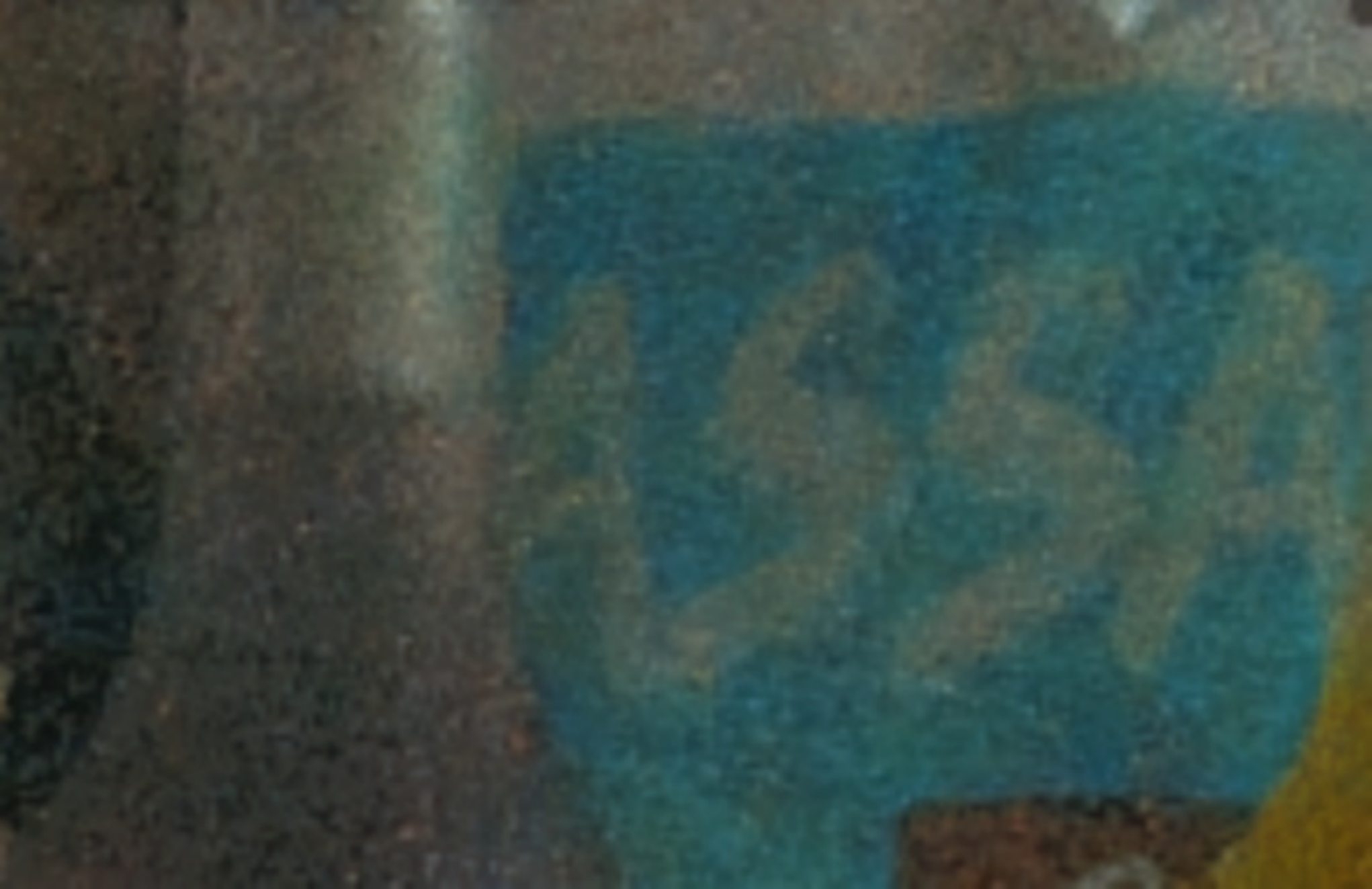}}{D\&E}
        \end{minipage}\hspace{-0.1mm}
   
         \begin{minipage}[b]{0.16\textwidth}
            \centering
\stackunder[5pt]{\includegraphics[width=1\linewidth,height=1.8cm]{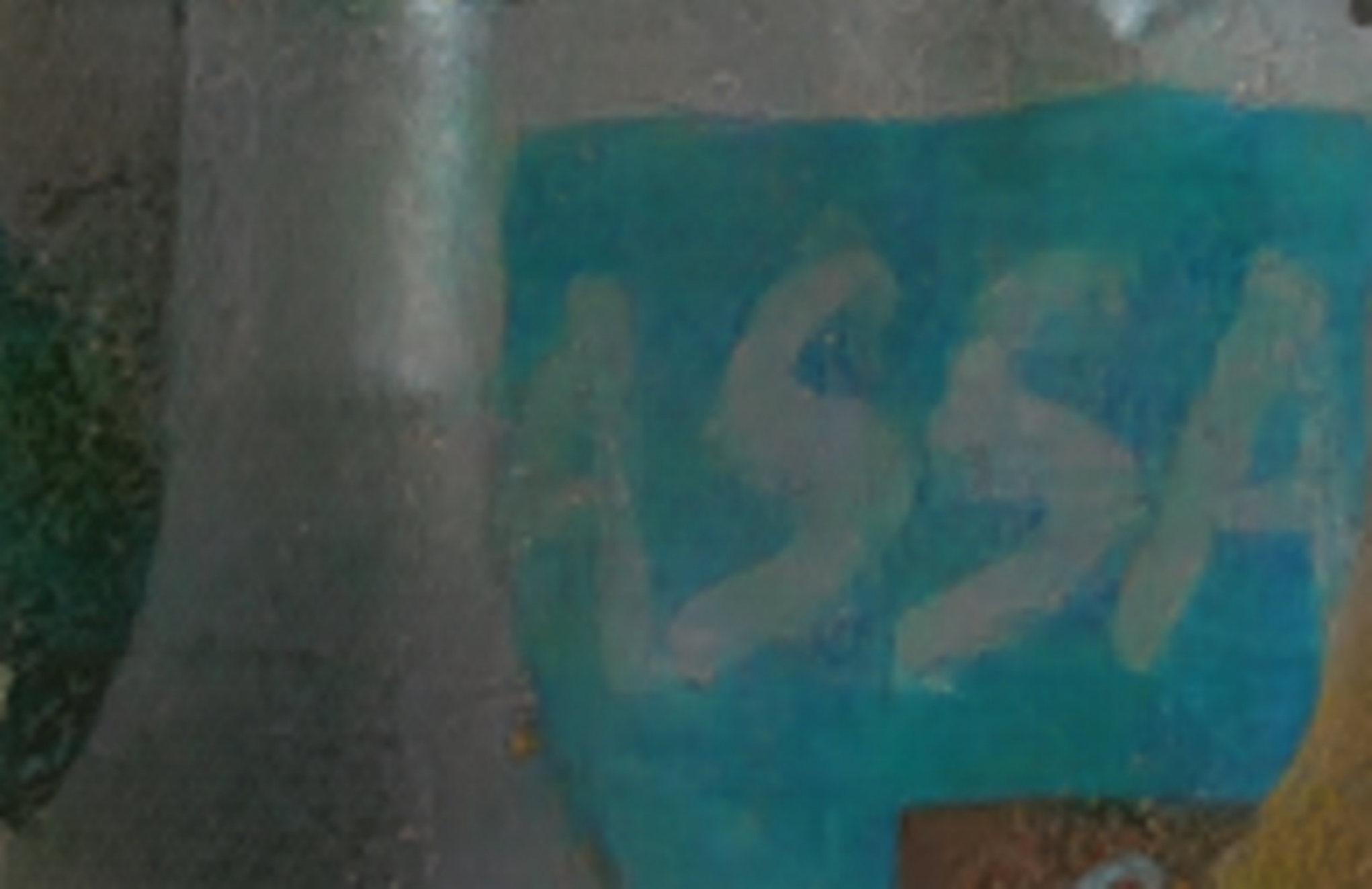}}{\textbf{UTVNet (Ours)}}
        \end{minipage}\hspace{-0.1mm}
        
        \begin{minipage}[b]{0.160\textwidth}
            \centering
\stackunder[5pt]{\includegraphics[width=1\linewidth,height=1.8cm]{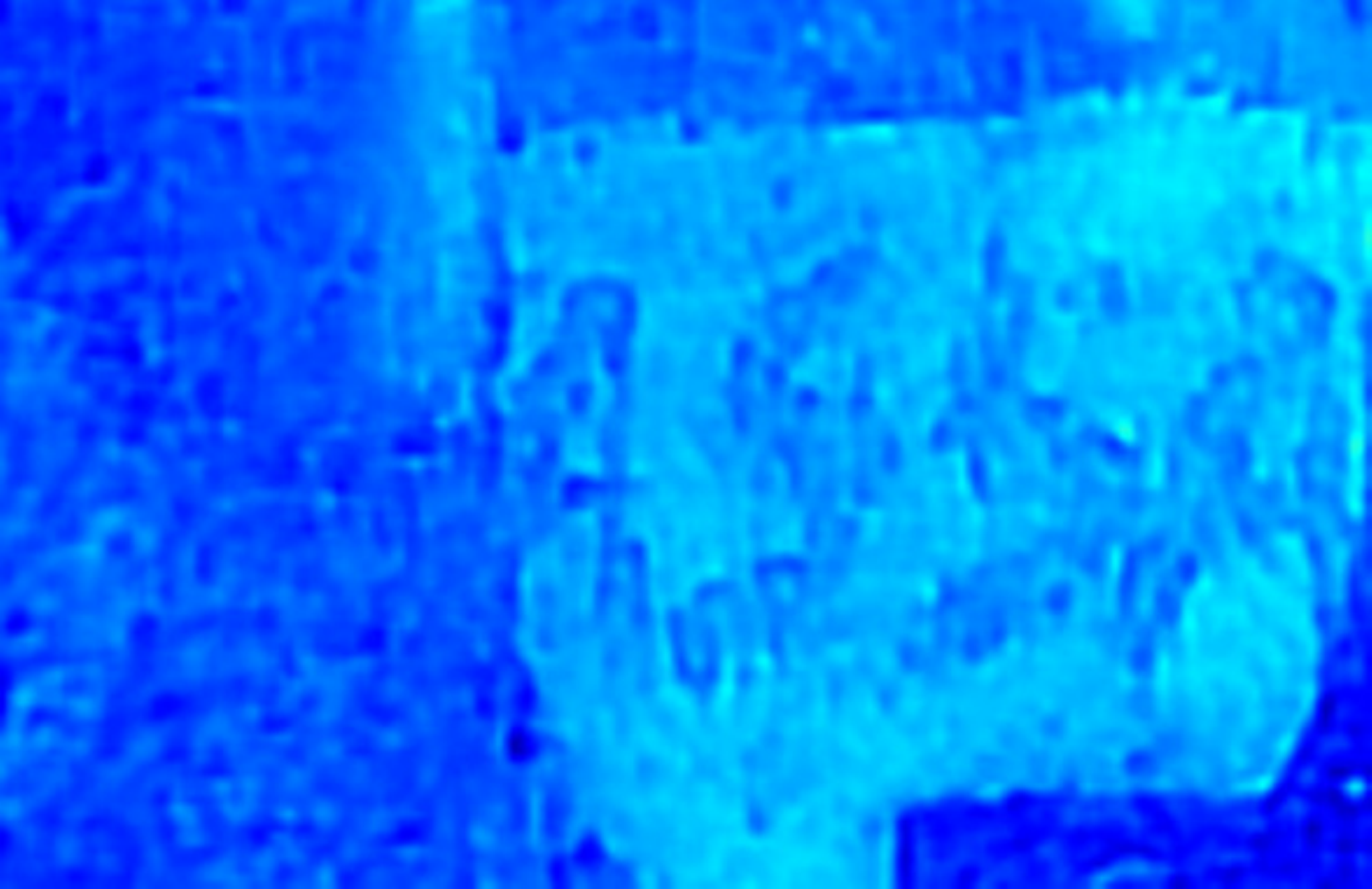}}{$\mathbf{M}^{avg}$}
        \end{minipage}
        
       }\end{minipage}\hspace{-1.1mm}\begin{minipage}{0.21\linewidth}
       \vspace{-0.36cm}
              \subfigure 
    {\footnotesize
        \begin{minipage}[b]{1.62\textwidth}
            \centering   \hspace{-46.2mm}  
                 \includegraphics[width=1\linewidth,height=0.25cm]{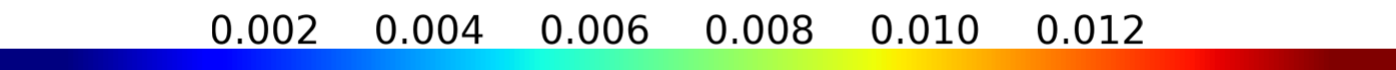}
        \end{minipage}
     }
\subfigure 
    {\footnotesize
        \begin{minipage}[b]{1\textwidth}
            \centering  \vspace{-0.16cm}
\stackunder[5pt]{\includegraphics[width=1\linewidth,height=3.66cm]{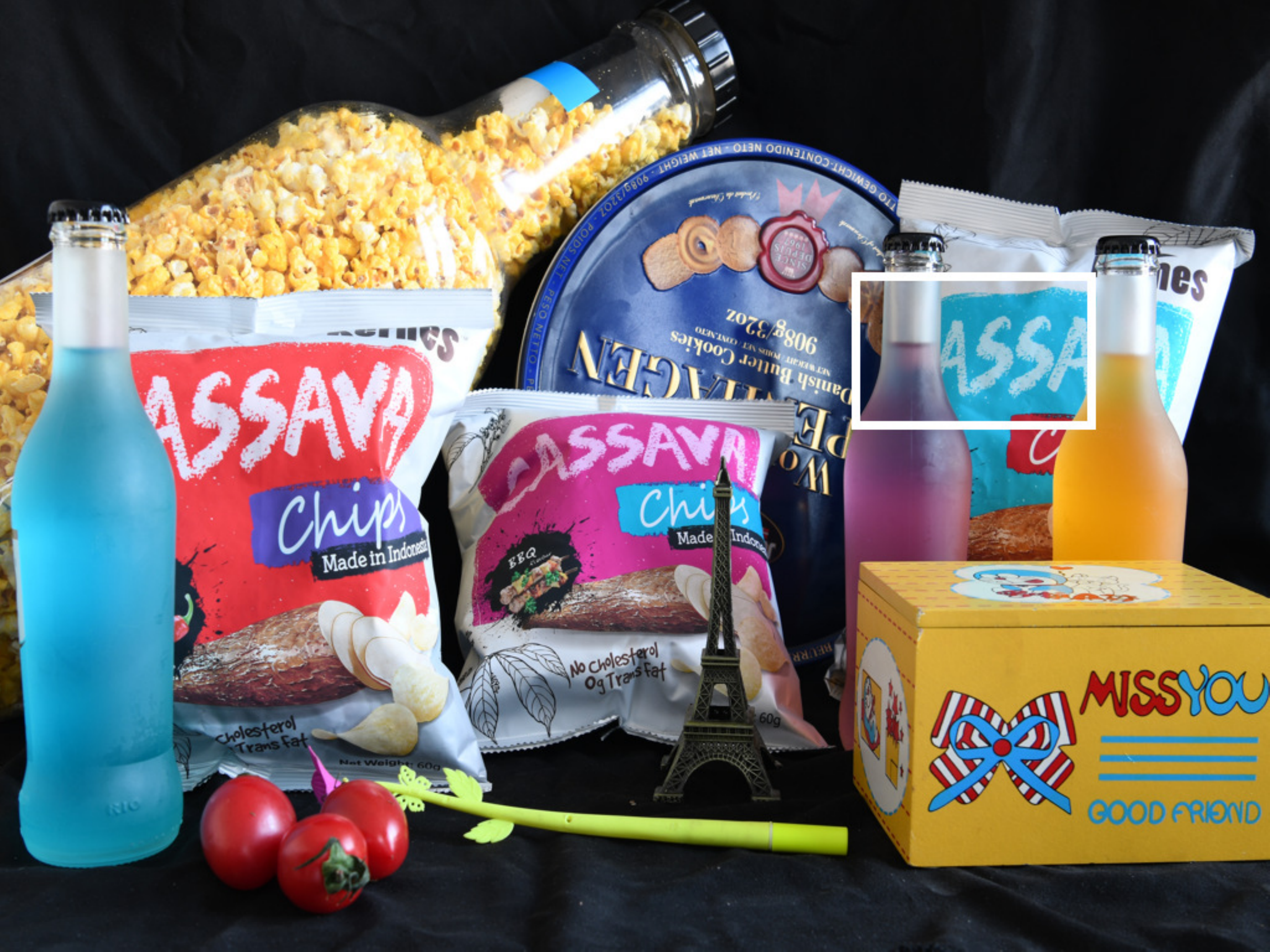}}{Ground Truth}
        \end{minipage}
     }

    \end{minipage}

\vspace{0.2cm} 
\caption{ Evaluating the performance at different ISO levels ($1600$ vs.\;$3200$) under exposure time (ET) $1/800s$ from the ELD dataset.  $\mathbf{M}^{avg}$ is the approximated average of noise level maps in all iterations by the proposed UTVNet. \textit{ (Best viewed with zoom)}}

\vspace{0.4cm} 
    \begin{minipage}{0.79\linewidth}
    \subfigure 
    {
        \begin{minipage}[b]{0.16\linewidth}
            \centering
            \includegraphics[width=1\linewidth,height=1.8cm]{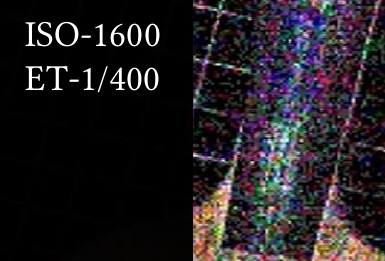}\vspace{0.1cm}
        \end{minipage}\hspace{-0.3mm}
   
        \begin{minipage}[b]{0.16\linewidth}
            \centering
            \includegraphics[width=1\linewidth,height=1.8cm]{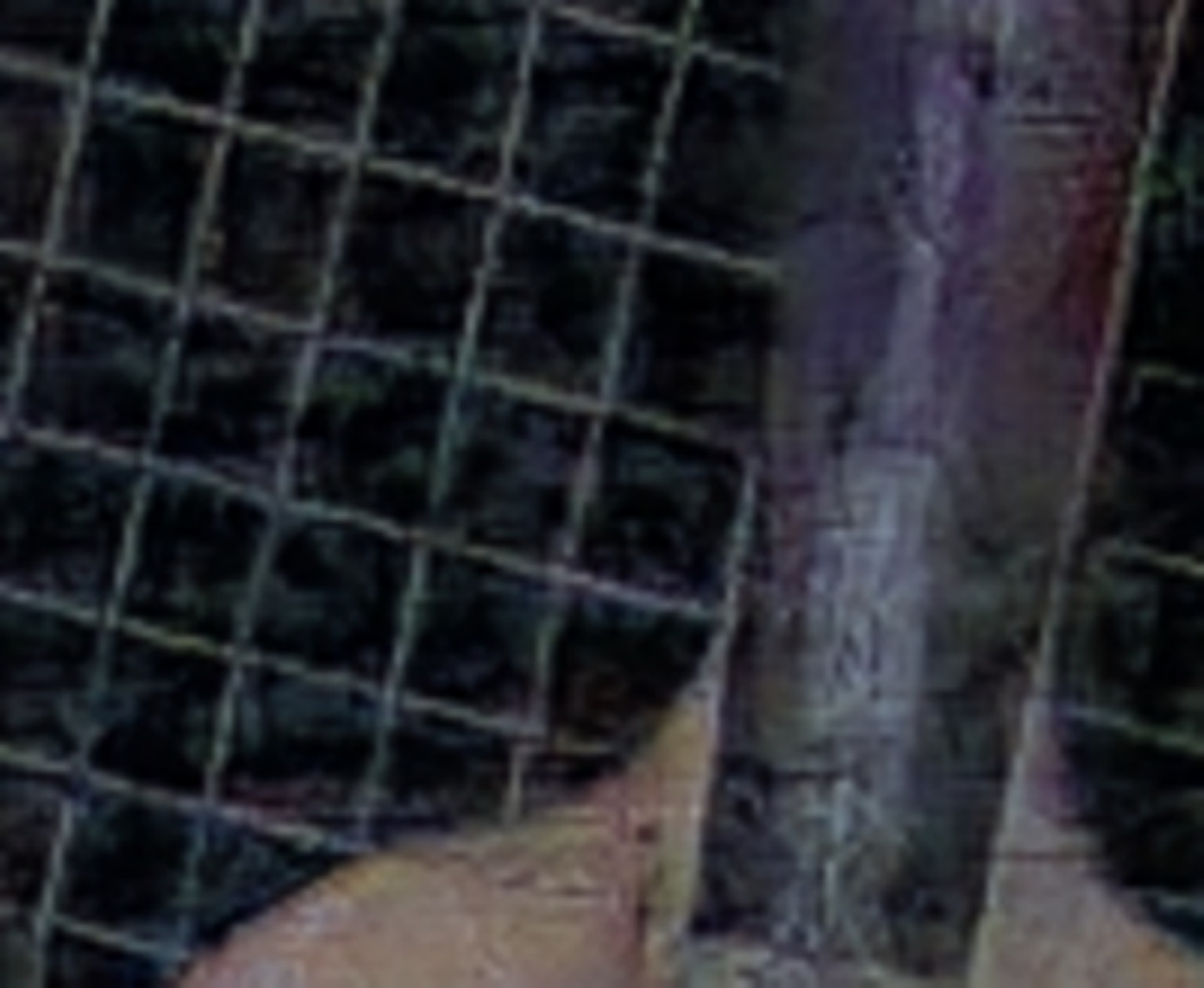}\vspace{0.1cm}
        \end{minipage}\hspace{-0.3mm}
   
         \begin{minipage}[b]{0.16\linewidth}
            \centering
            \includegraphics[width=1\linewidth,height=1.8cm]{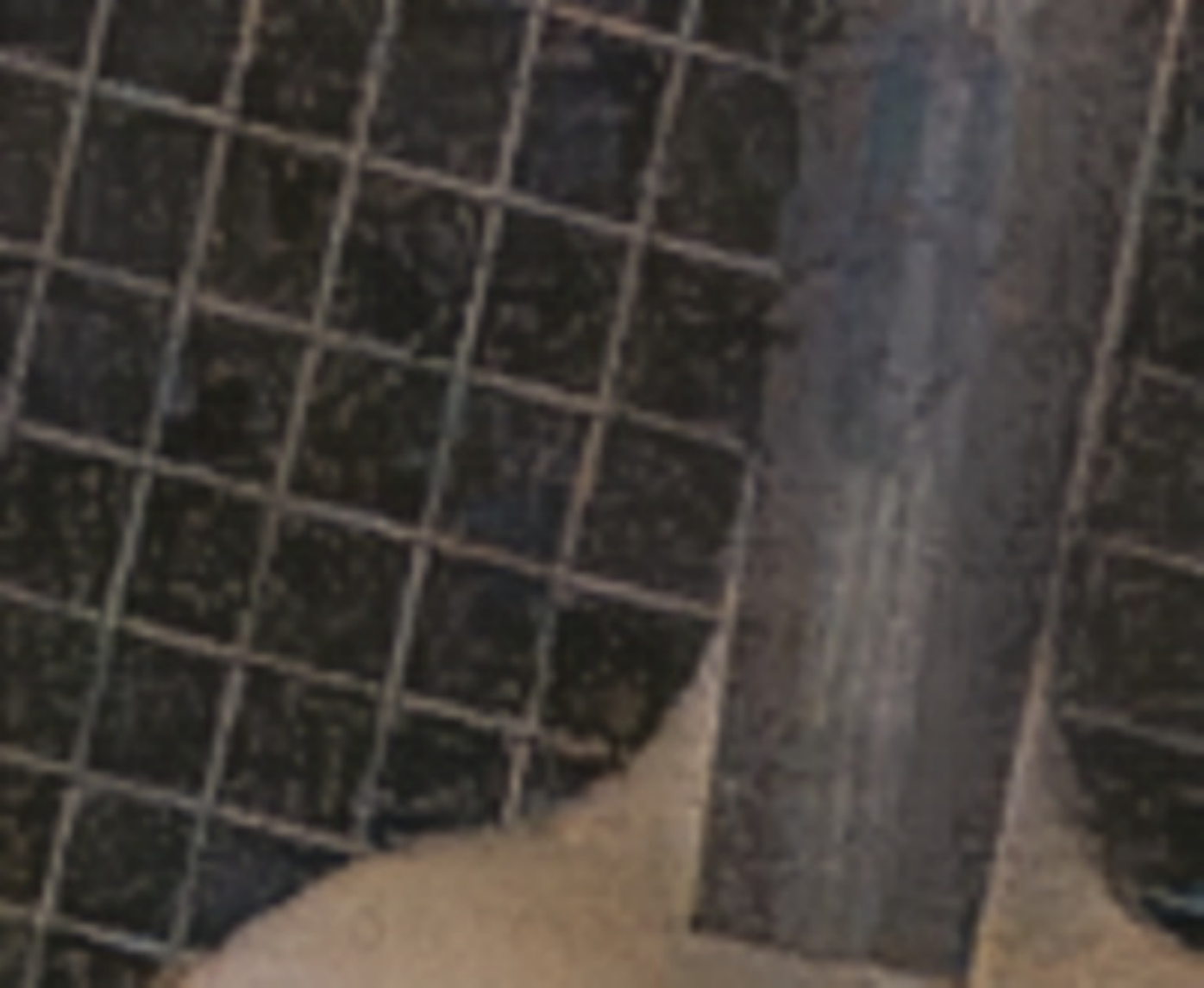}\vspace{0.1cm}
        \end{minipage}\hspace{-0.3mm}
   
         \begin{minipage}[b]{0.16\linewidth}
            \centering
            \includegraphics[width=1\linewidth,height=1.8cm]{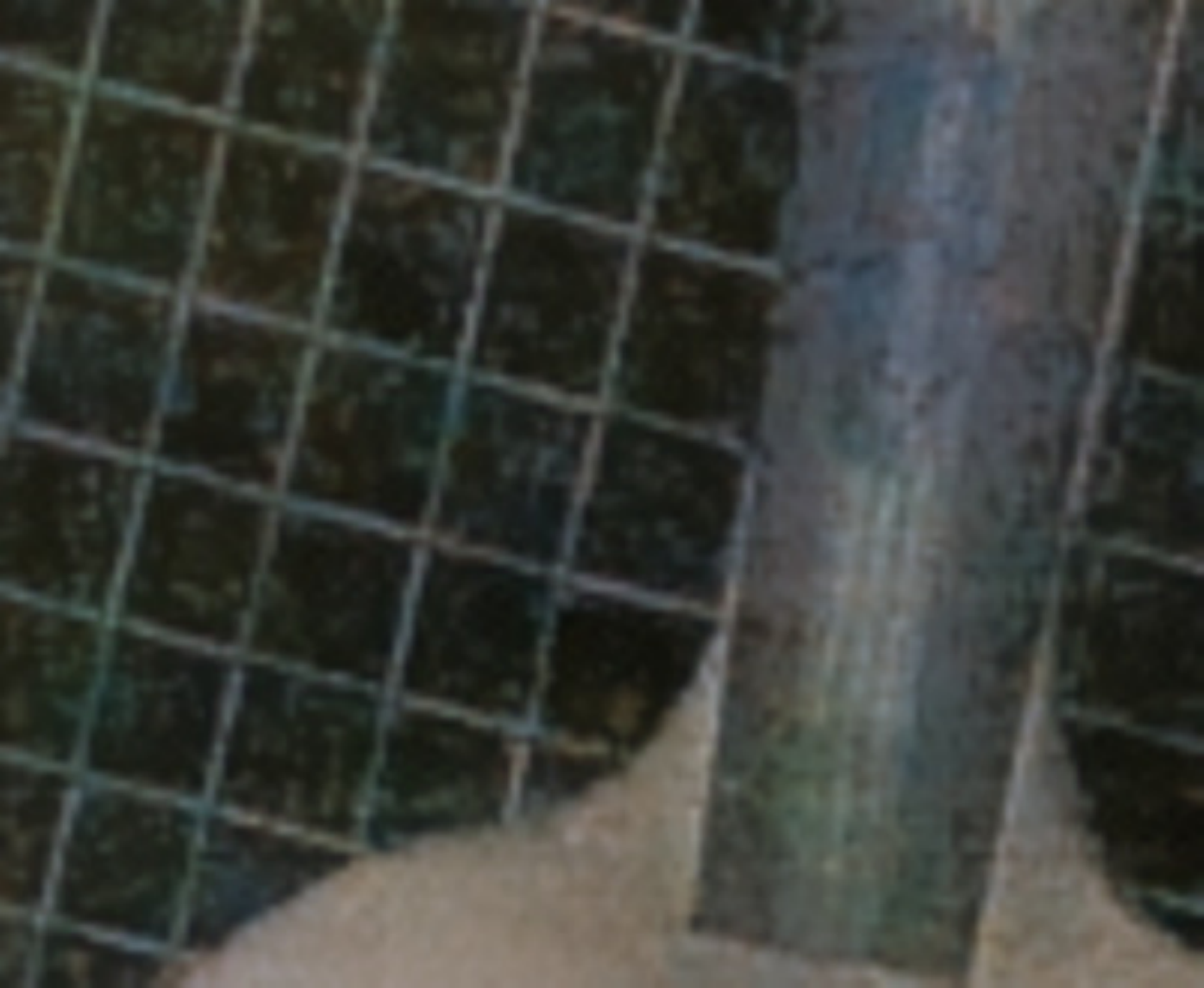}\vspace{0.1cm}
        \end{minipage}\hspace{-0.3mm}
   
         \begin{minipage}[b]{0.16\linewidth}
            \centering
            \includegraphics[width=1\linewidth,height=1.8cm]{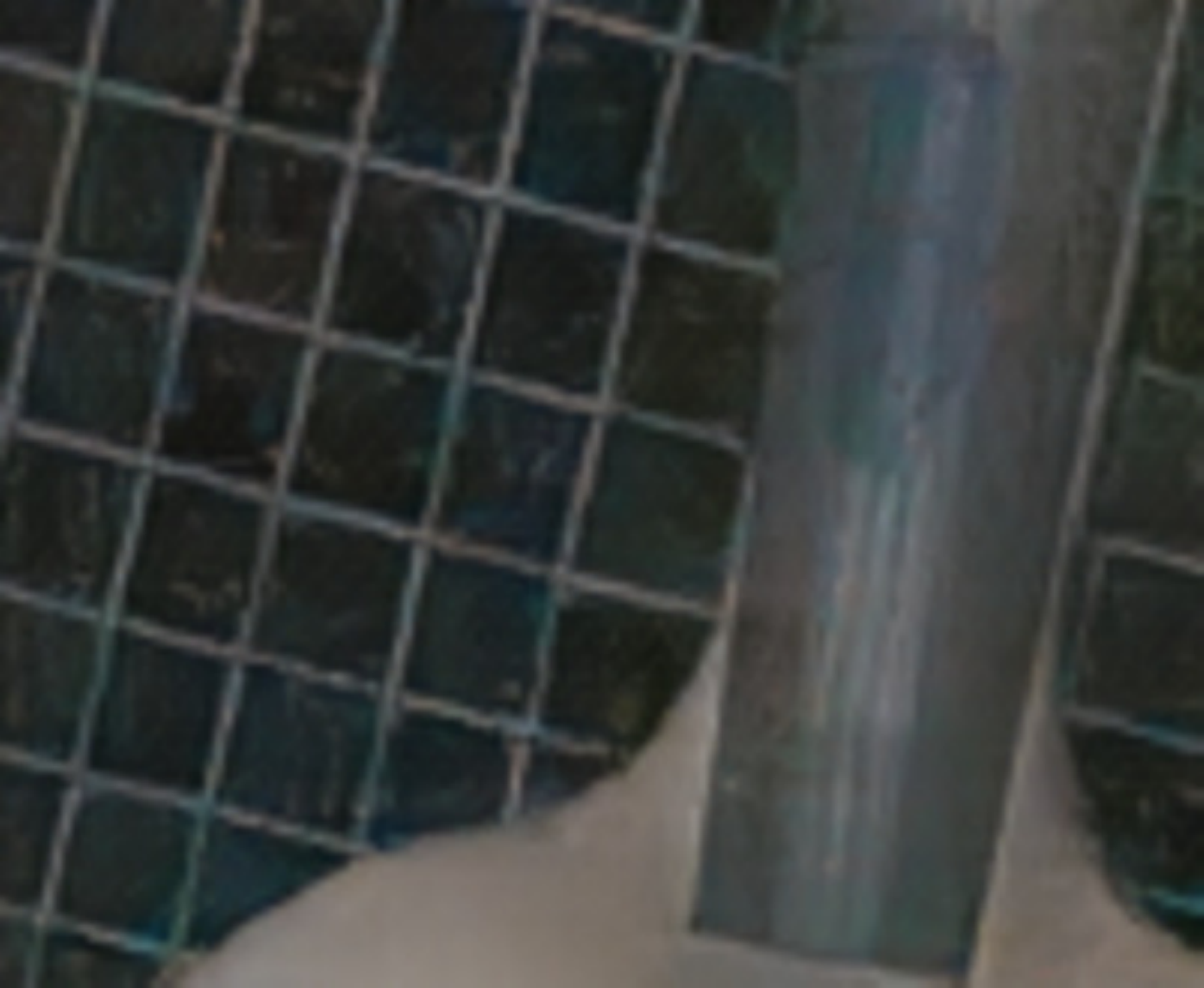}\vspace{0.1cm}
        \end{minipage}\hspace{-0.3mm}
        
                 \begin{minipage}[b]{0.16\linewidth}
            \centering
            \includegraphics[width=1\linewidth,height=1.8cm]{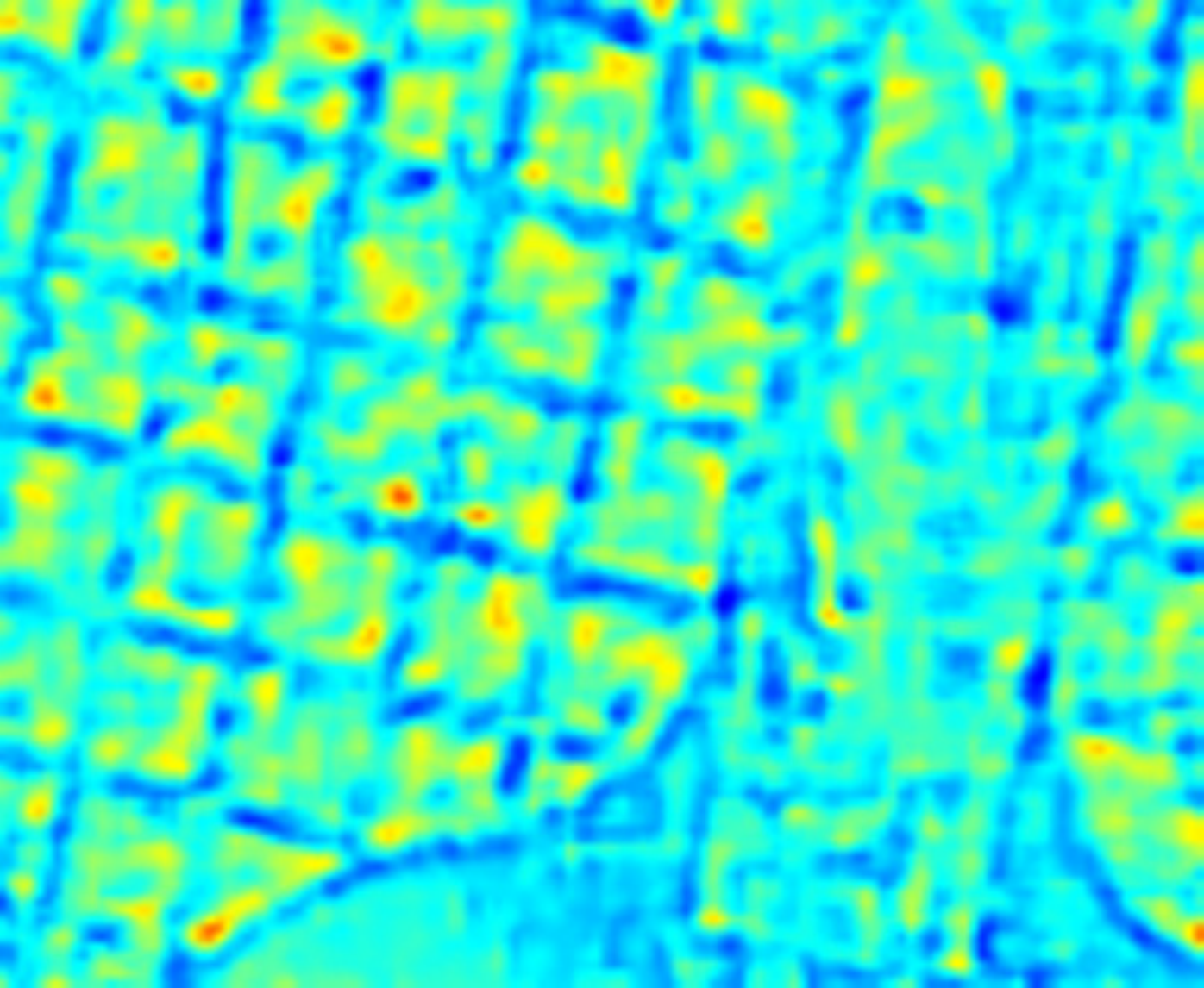}\vspace{0.1cm}
        \end{minipage}
       }
       
           \vspace{-0.25cm}

    \subfigure 
    {   \footnotesize
        \begin{minipage}[b]{0.16\textwidth}
            \centering
             \footnotesize
                 \stackunder[5pt]{\includegraphics[width=1\linewidth,height=1.8cm]{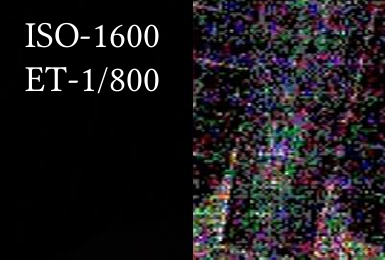}}{Input / LIME}
        \end{minipage}\hspace{-0.1mm}

        \begin{minipage}[b]{0.16\textwidth}
            \centering
\stackunder[5pt]{\includegraphics[width=1\linewidth,height=1.8cm]{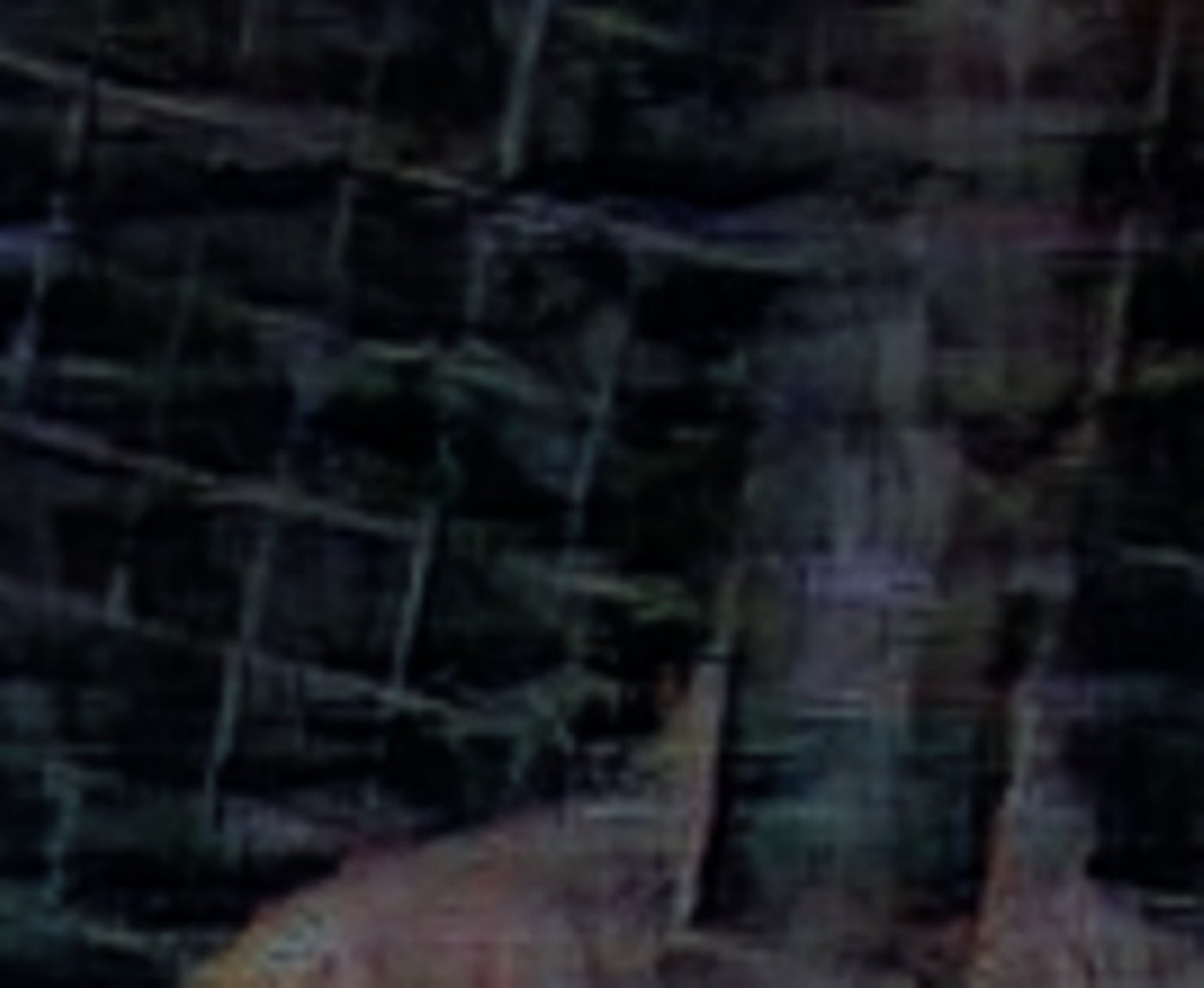}}{AGLLNet}
        \end{minipage}\hspace{-0.1mm}
   
         \begin{minipage}[b]{0.16\textwidth}
            \centering
\stackunder[5pt]{\includegraphics[width=1\linewidth,height=1.8cm]{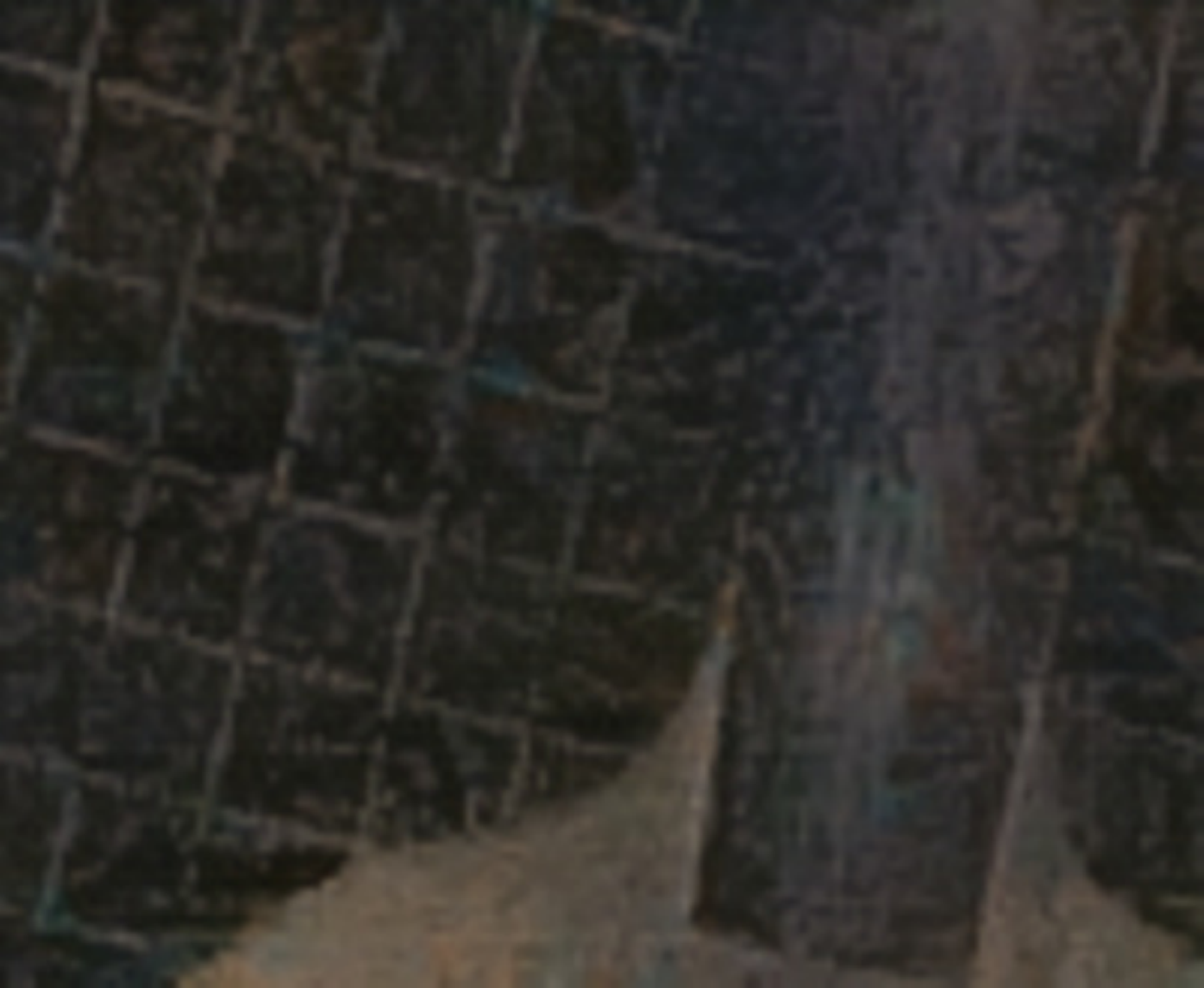}}{MIRNet}
        \end{minipage}\hspace{-0.1mm}

         \begin{minipage}[b]{0.16\textwidth}
            \centering
\stackunder[5pt]{\includegraphics[width=1\linewidth,height=1.8cm]{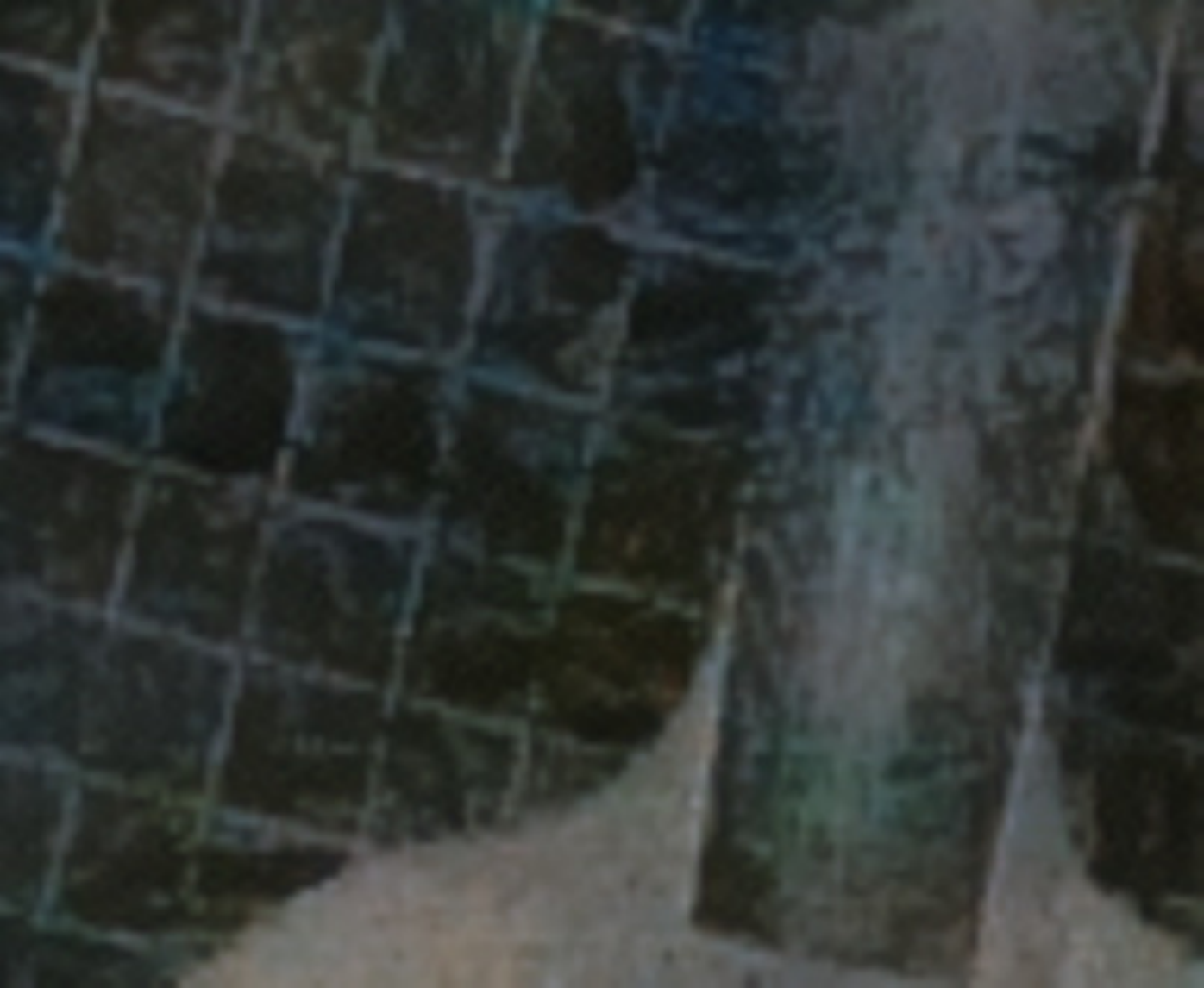}}{D\&E}
        \end{minipage}\hspace{-0.1mm}
   
         \begin{minipage}[b]{0.160\textwidth}
            \centering
\stackunder[5pt]{\includegraphics[width=1\linewidth,height=1.8cm]{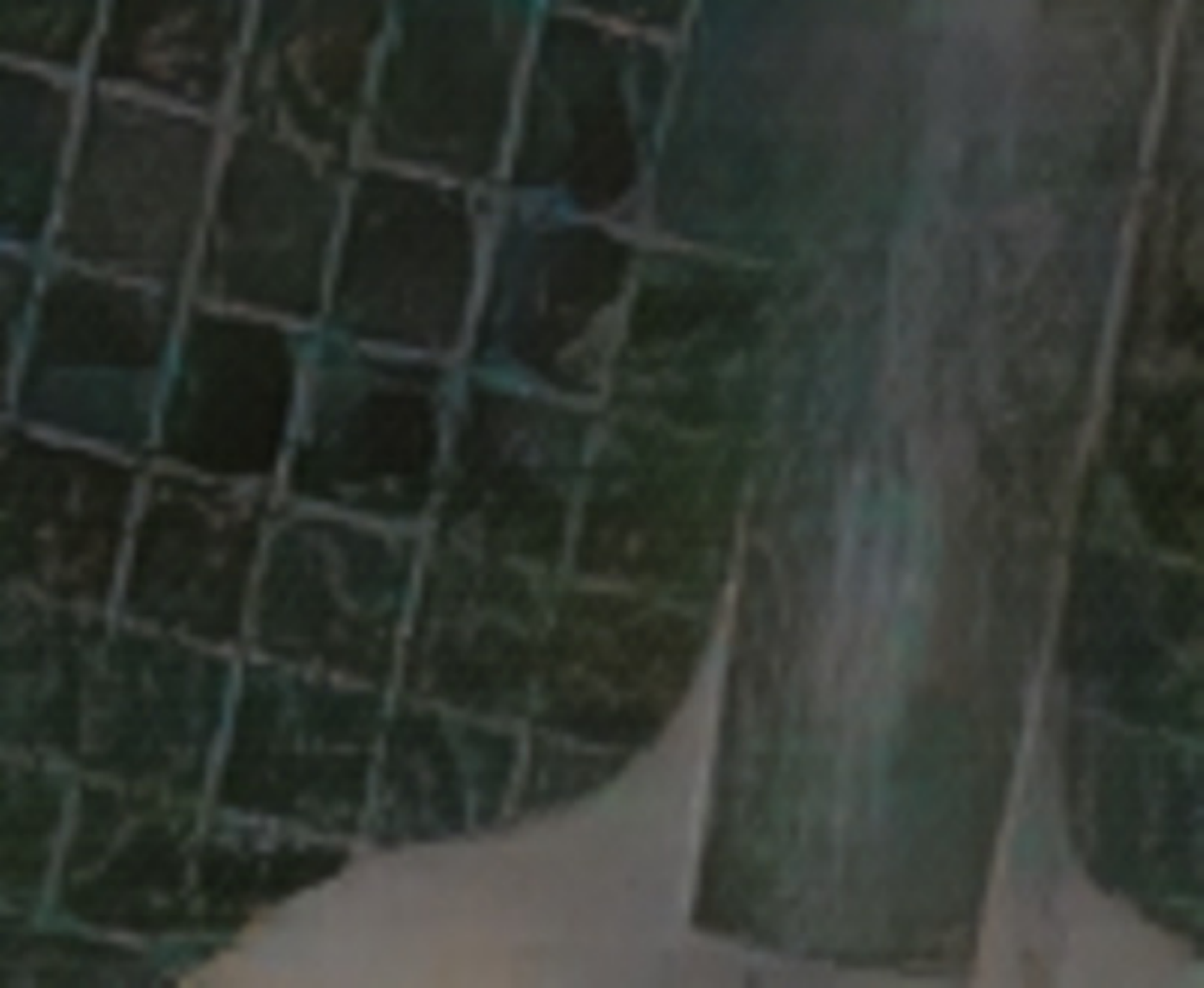}}{\textbf{UTVNet (Ours)}}
        \end{minipage}\hspace{-0.1mm}
        
        \begin{minipage}[b]{0.160\textwidth}
            \centering
\stackunder[5pt]{\includegraphics[width=1\linewidth,height=1.8cm]{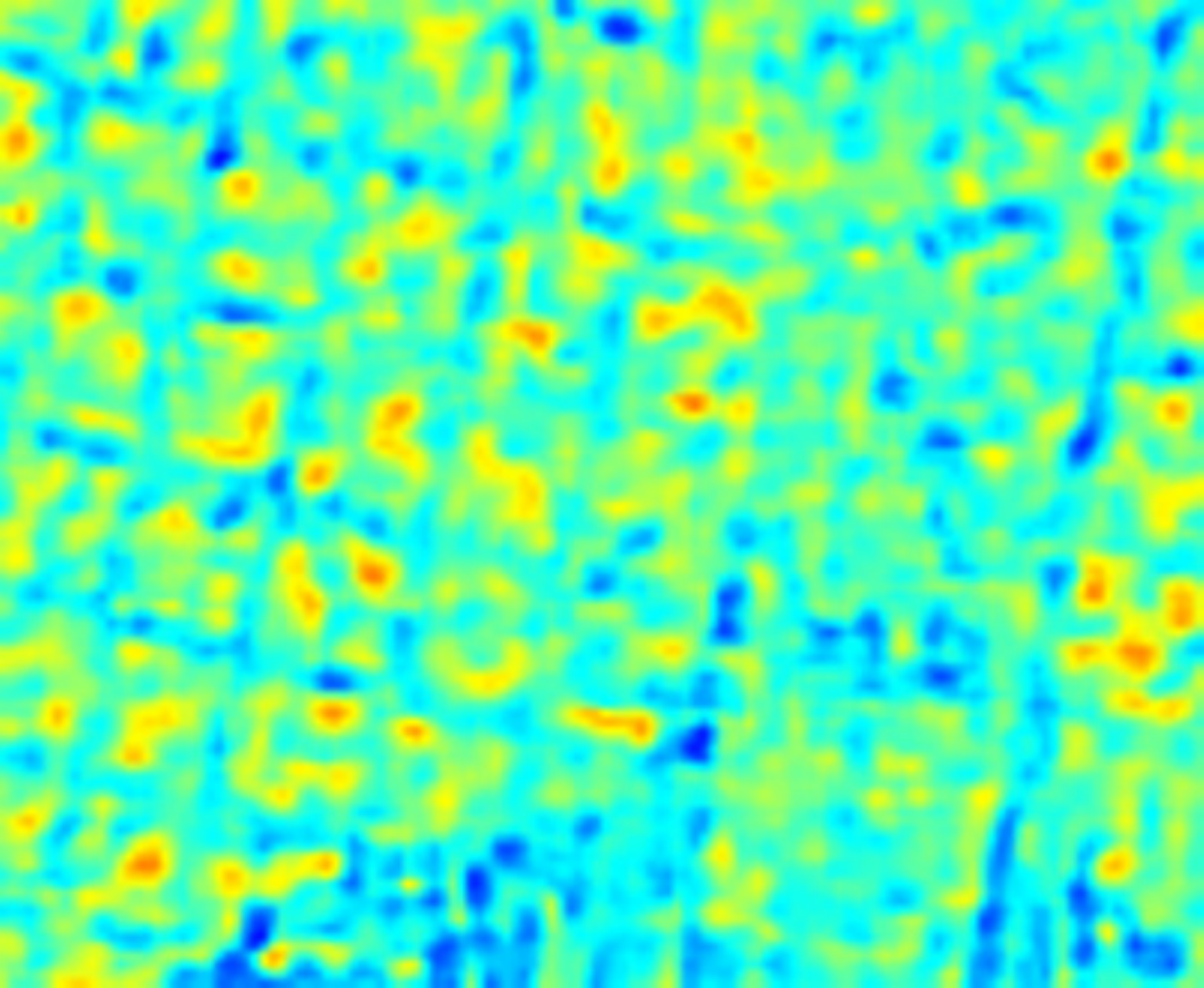}}{$\mathbf{M}^{avg}$}
        \end{minipage}
        
       }\end{minipage}\hspace{-1.1mm}\begin{minipage}{0.21\linewidth}
       \vspace{-0.36cm}
              \subfigure 
    {\footnotesize
        \begin{minipage}[b]{1.62\textwidth}
            \centering   \hspace{-46.2mm}  
                 \includegraphics[width=1\linewidth,height=0.25cm]{iso-1-level1-eps-converted-to.pdf}
        \end{minipage}
     }
\subfigure 
    {\footnotesize
        \begin{minipage}[b]{1\textwidth}
            \centering  \vspace{-0.15cm}
\stackunder[5pt]{\includegraphics[width=1\linewidth,height=3.66cm]{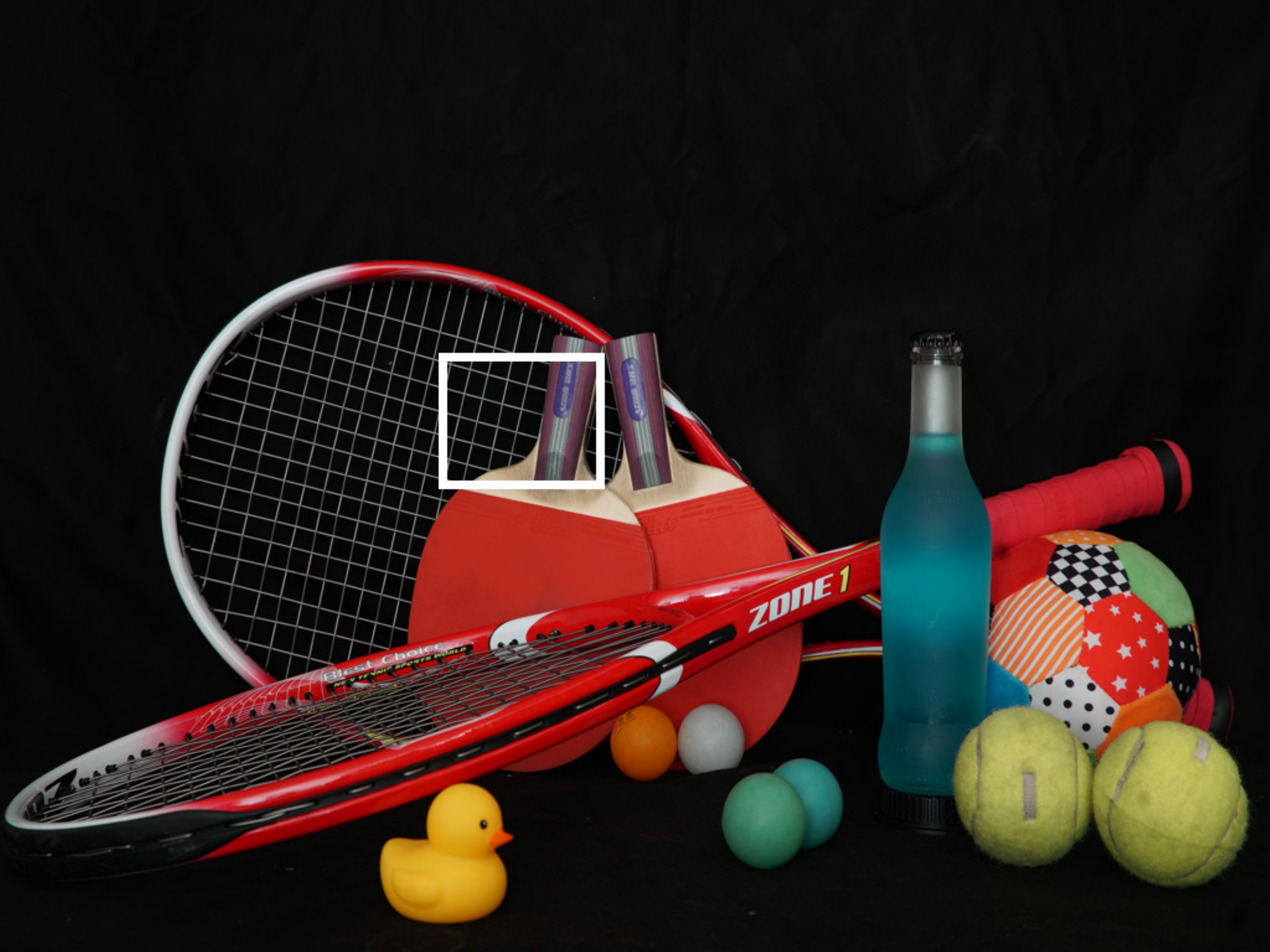}}{Ground Truth}
        \end{minipage}
     }

    \end{minipage}

\vspace{0.2cm} 
\caption{Evaluating the performance at different exposure times ($1/400s$ vs.\;$1/800s$) under ISO $1600$ from the ELD dataset. $\mathbf{M}^{avg}$ is the approximated average of noise level maps in all iterations by the proposed UTVNet. \textit{ (Best viewed with zoom) } 
   }
\end{figure*}

\begin{table*}
\caption{Quantitative results of different methods that trained on sRGB-SID or other synthetic datasets and tested on ELD dataset. }
\vspace{1mm}
\begin{center}
\footnotesize
\setlength{\extrarowheight}{1.5pt}
\setlength{\tabcolsep}{4.2mm}
\begin{tabular}{c|ccc|ccc|ccc}
\hline
\multirow{2}{*}{\textbf{Method}} & \multicolumn{3}{c|}{\textbf{SonyA7S2}} & \multicolumn{3}{c|}{\textbf{NikonD850}}    & \multicolumn{3}{c}{\textbf{CanonEOS700D}}     \\ 
    & \textbf{PSNR}   & \textbf{SSIM}     & \textbf{LPIPS}  & \textbf{PSNR}   & \textbf{SSIM}    & \textbf{LPIPS}  & \textbf{PSNR}   & \textbf{SSIM}  & \textbf{LPIPS}  \\ \hline
LIME \cite{LIME}                    & 16.467 & 0.3960  & 0.5384 & 15.155 & 0.3034  &0.5616 & 15.423& 0.3177  & 0.5854    \\ 
AGLLNet \cite{atl}                 & 17.274 & 0.4915  & 0.4653 & 16.047 & 0.4564  & 0.5072 & 16.399 & 0.3569  & 0.6145 \\ 
MIRNet \cite{MIRNet}                  & 17.301 & 0.5103  & 0.4643 & 17.692 & 0.5045  & 0.5020 & 16.201 & 0.3778  & 0.5926  \\ 
D\&E   \cite{LDE}                 & 18.571 & 0.5783  & 0.4612 & 17.926 & 0.5532  & 0.4536 & \textbf{16.639} & 0.3729 & 0.5868 \\ \hline
\textbf{UTVNet (Ours)}         & \textbf{18.675} & \textbf{0.6054}  & \textbf{0.4524} & \textbf{18.161} & \textbf{0.5910}  & \textbf{0.4446} & 16.496 & \textbf{0.4233}   & \textbf{0.5601} \\ \hline
\end{tabular}
\end{center}
\vspace{-0.3cm}
\end{table*}

\textbf{Quantitative and visual comparisons.} We compare UTVNet with state-of-the-art methods in terms of PSNR, SSIM \cite{ssim}, Mean $L2$ error in L*a*b* space (lower is better) and LPIPS \cite{lpips} (lower is better).\;The results are presented in Table 1.\;The first six rows of Table 1 show that the exposure correction methods provide unsatisfactory performance even when combined with the real-world denoising method NBNet \cite{nbn}.  Compared with all learning-based exposure correction methods and image restoration approaches, our UTVNet achieves the best performance in four metrics.  The visual results of the proposed approach and the state-of-the-art methods can be seen in Fig.\;4. One can see that only correcting the exposure by LIME may fail to recover images with noise, because it is hard to recover original details from small pixel values disrupted by noise. SID and Pixel2Pixel both introduce artifacts. Without noise level estimation, D\&E and MIRNet may provide over-smooth or still noisy results. Our UTVNet outperforms all methods in recovering finer details and colors.

\textbf{Computational cost.} We analyze the computational complexity of some recently proposed methods and our UTVNet on sRGB-SID dataset with a resolution of 1024 × 768 pixels. The parameter size, MACs, execution time, and PSNR results are shown in Table 3. Compared with D\&E and MIRNet, our UTVNet achieves the best performance with a balanced parameter size and lower MAC operations.

\begin{figure*}[h]
 \hsize=\textwidth
 \subfigcapskip=-3pt
   \subfigcapskip=-3pt
 \subfigbottomskip=-3pt
 \begin{center}
 \hspace{-1mm}
  \subfigure 
    { \footnotesize
         \begin{minipage}[b]{0.14\textwidth}
            \centering
         \stackunder[5pt]{   \includegraphics[width=1\linewidth,height=1.8cm]{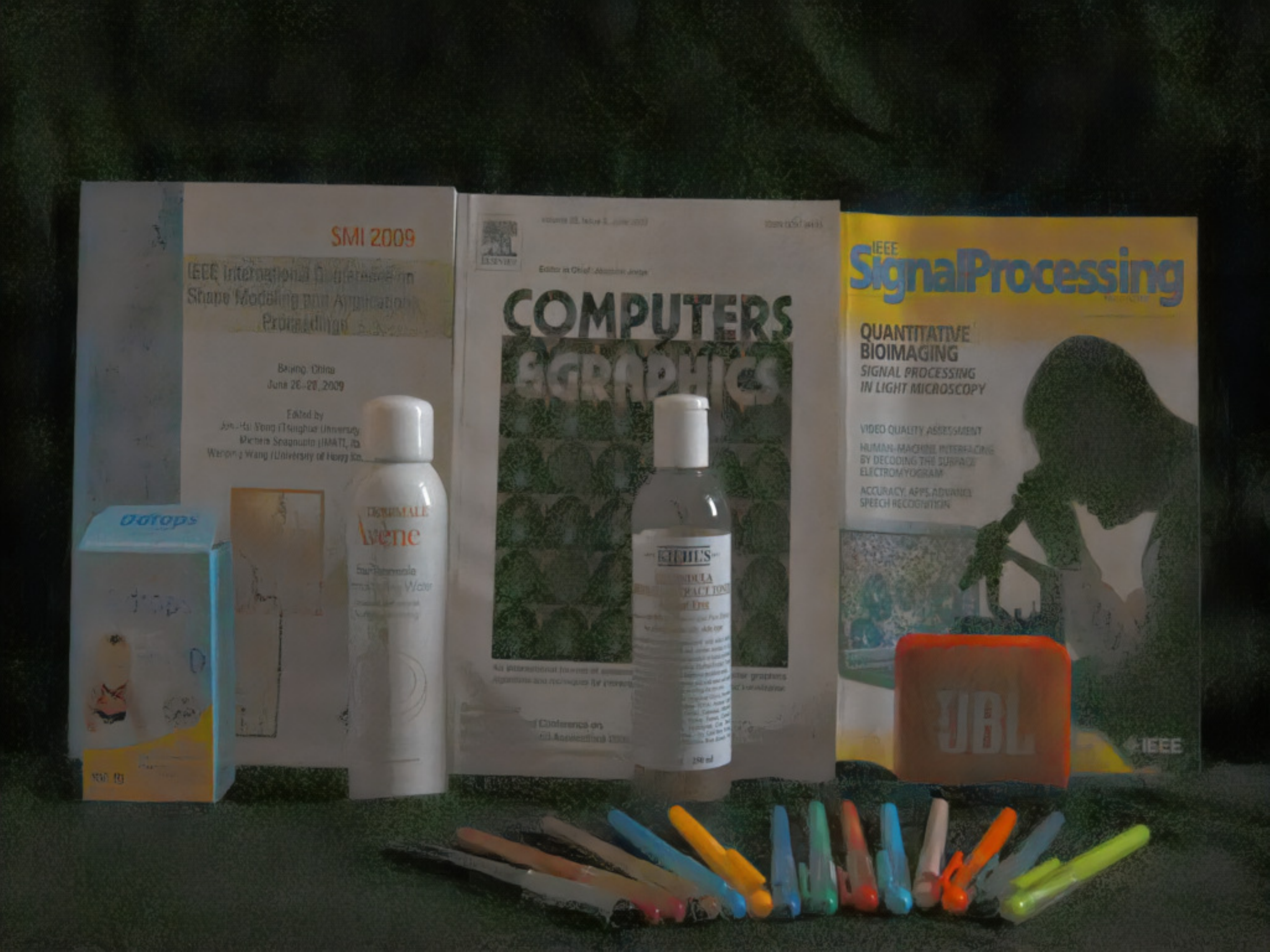}}{Output} 
        \end{minipage}
    
         \begin{minipage}[b]{0.14\textwidth}
            \centering
       \stackunder[5pt]{     \includegraphics[width=1\linewidth,height=1.8cm]{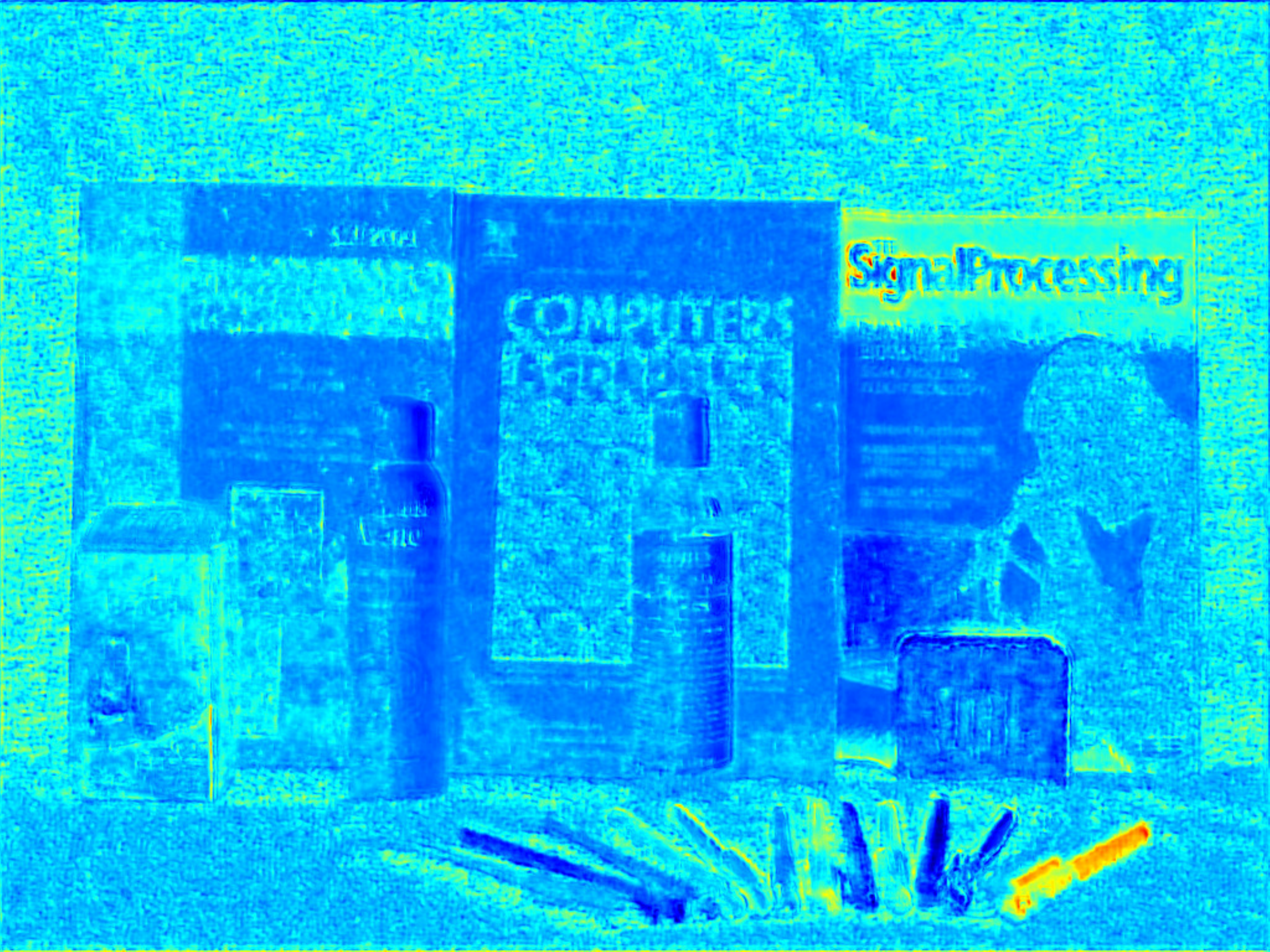}}{$\mathbf{M}_{1}$}
        \end{minipage}
   
         \centering
        \begin{minipage}[b]{0.14\textwidth}
            \centering
        \stackunder[5pt]{    \includegraphics[width=1\linewidth,height=1.8cm]{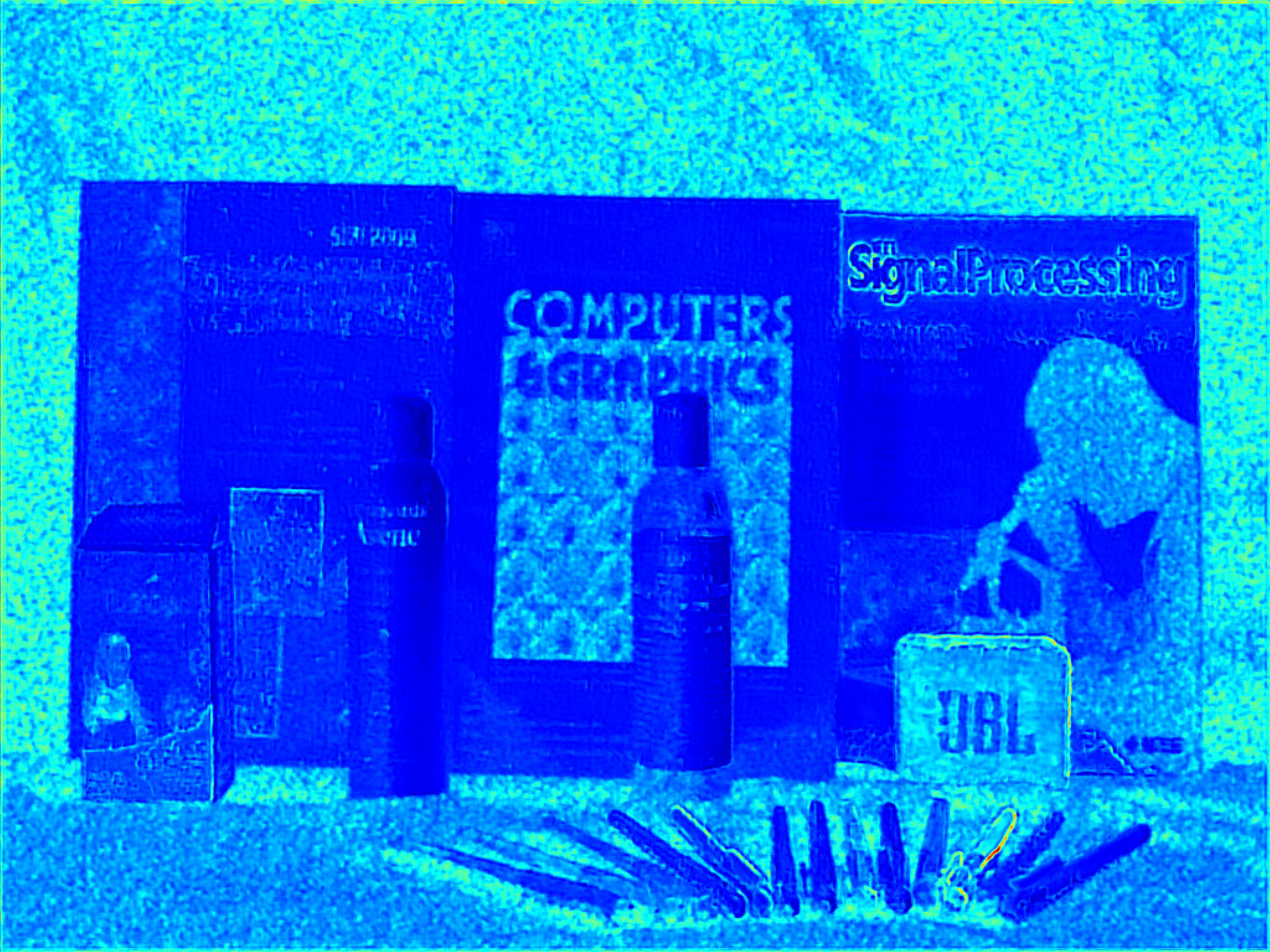}}{$\mathbf{M}_{2}$}
        \end{minipage}
    
        \begin{minipage}[b]{0.14\textwidth}
            \centering
       \stackunder[5pt]{     \includegraphics[width=1\linewidth,height=1.8cm]{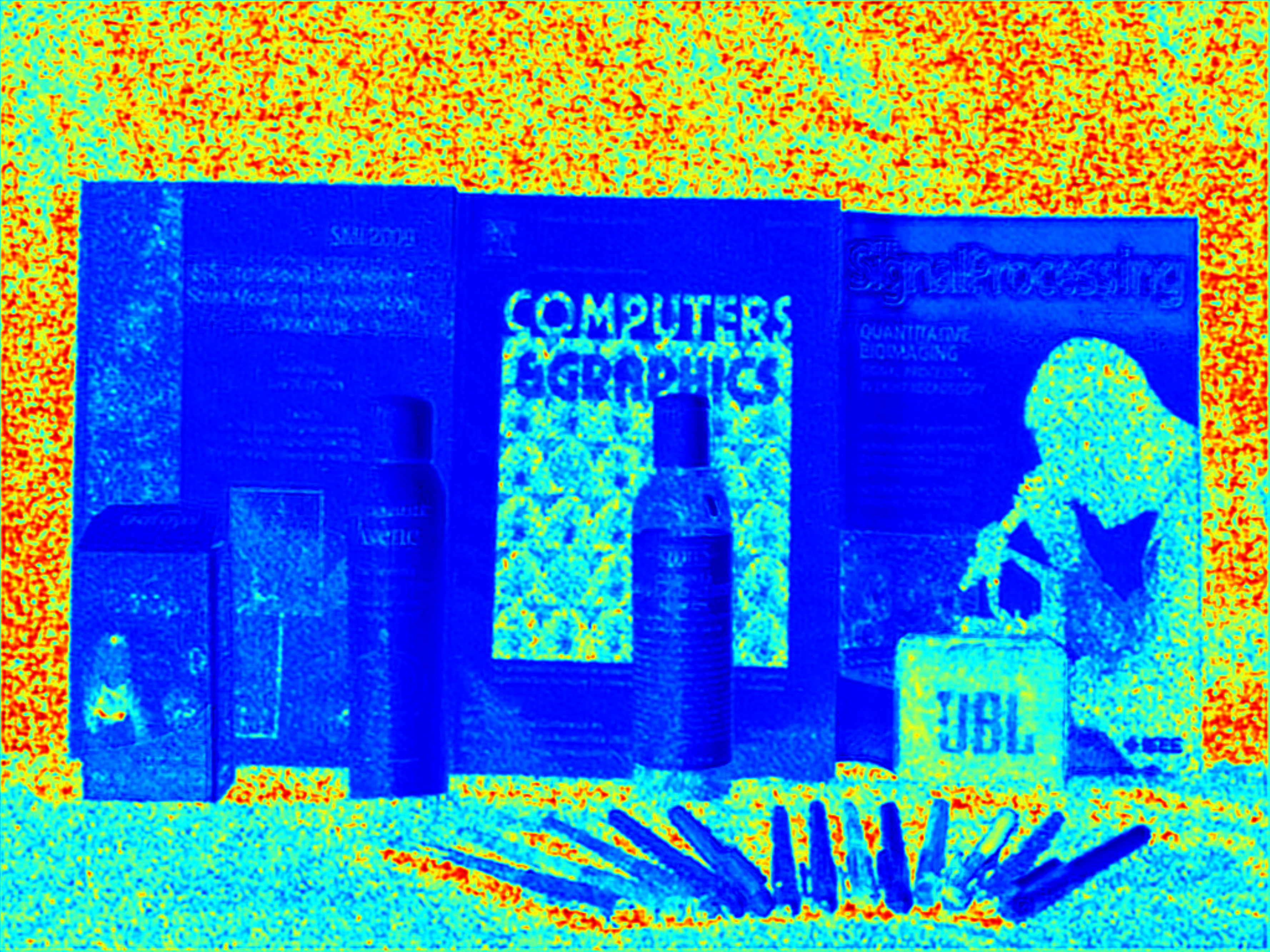}}{$\mathbf{M}_{3}$}
        \end{minipage}
   
        \begin{minipage}[b]{0.14\textwidth}
            \centering
        \stackunder[5pt]{    \includegraphics[width=1\linewidth,height=1.8cm]{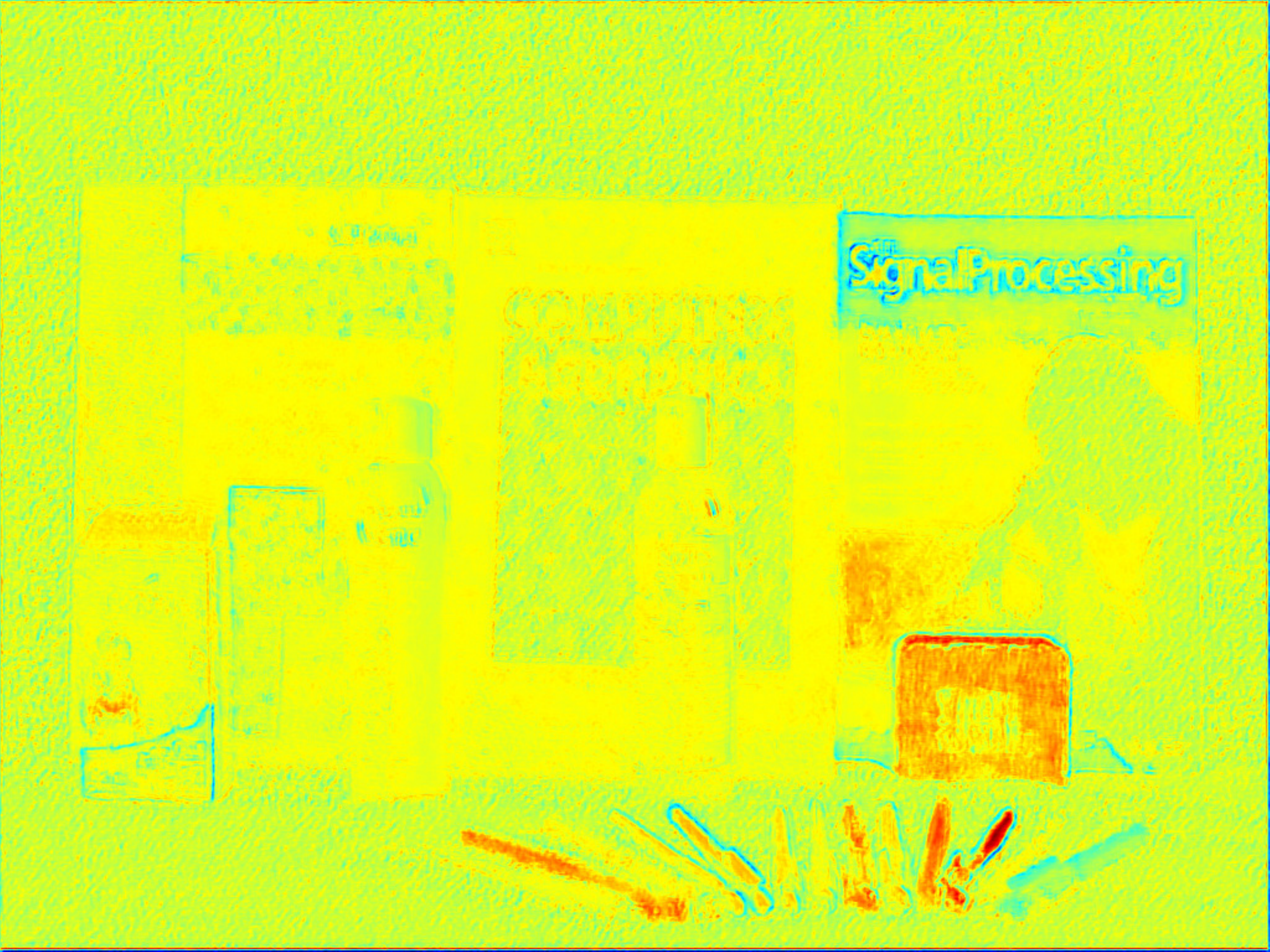}}{$\mathbf{M}_{6}$}
        \end{minipage}
    
        \begin{minipage}[b]{0.14\textwidth}
            \centering
       \stackunder[5pt]{     \includegraphics[width=1\linewidth,height=1.8cm]{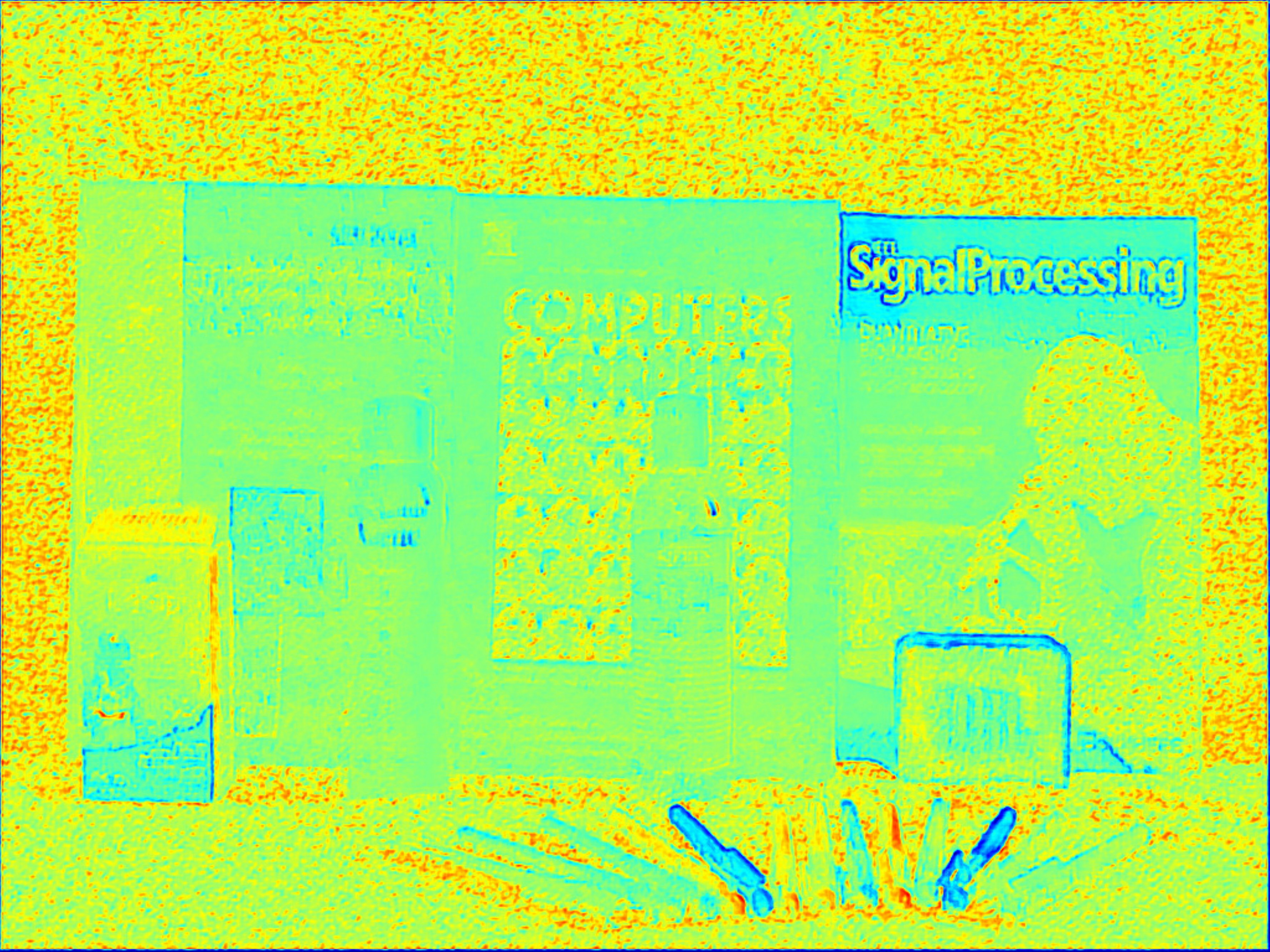}}{$\mathbf{M}_{7}$}
        \end{minipage}
    
        \begin{minipage}[b]{0.14\textwidth}
            \centering
      \stackunder[5pt]{      \includegraphics[width=1\linewidth,height=1.8cm]{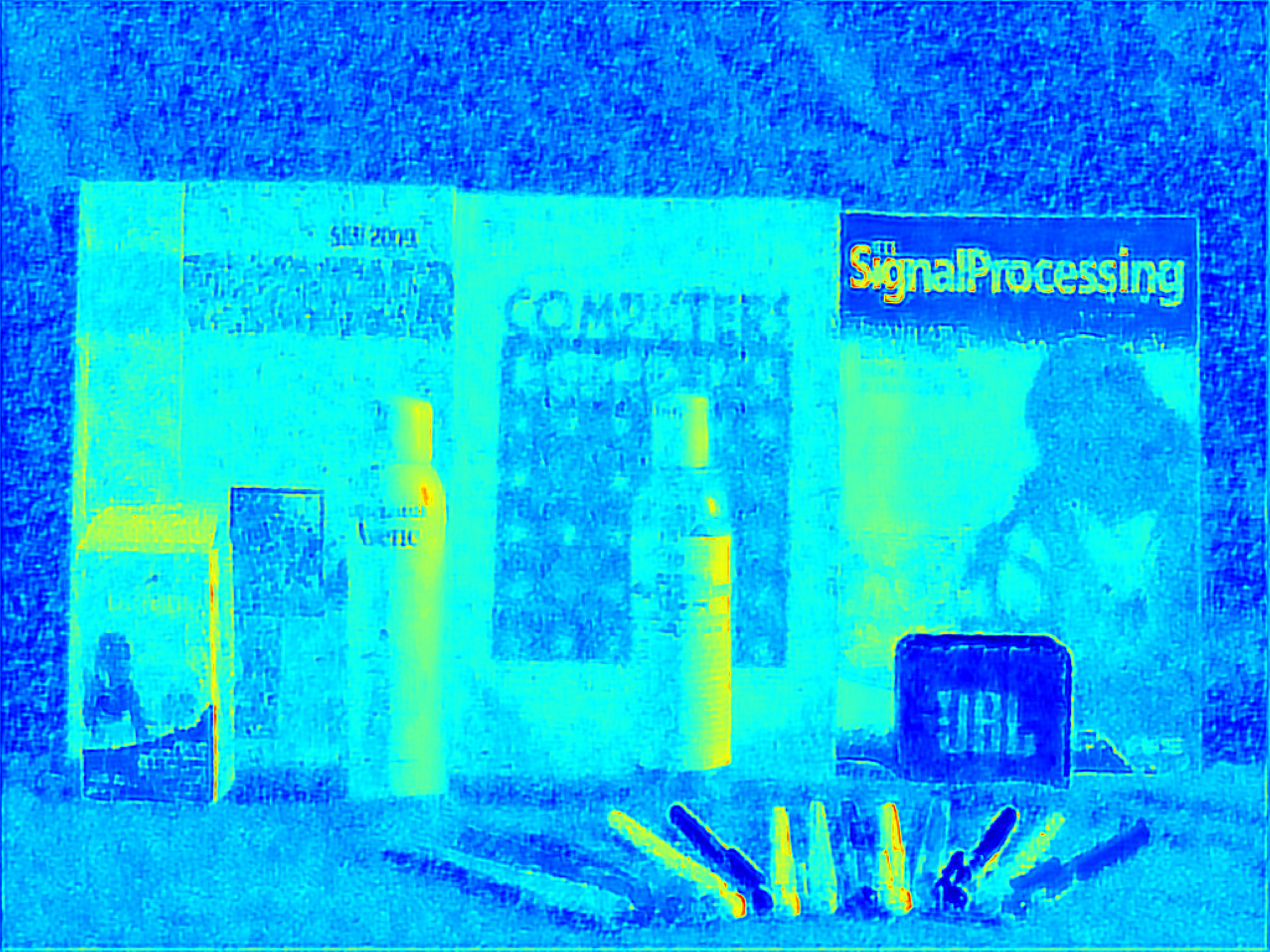}}{$\mathbf{M}_{8}$}
        \end{minipage}
    
        \begin{minipage}[b]{0.03\textwidth}
            \centering
       \stackunder[5pt]{     \includegraphics[width=1\linewidth,height=1.8cm]{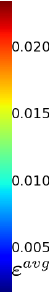}}{}
        \end{minipage}
       }

      \vspace{-0.1cm} 
       
\end{center}
\caption{ We visualize the average noise level maps of R, G, B channels in the first three and the last three iterations respectively. $\boldsymbol{\mathbf{\varepsilon}^{avg}}$ in the color coordinate axis is the average global noise variance. \textit{ (Best viewed with zoom) } 
   }

\end{figure*}
\subsection{  Performance in Real Captured Scenes }
To substantiate the robustness in real captured low-light scenes of proposed UTVNet, we evaluate the performance of different methods with ELD \cite{rewd} dataset. The extremely low-light images in ELD dataset covers 10 indoor scenes and were taken by 4 camera devices. In each scene, the noisy images were captured with different  ISO levels and exposure times to guarantee noise and low-light conditions. Only the sRGB images were used in our experiment. We compare UTVNet with LIME, AGLLNet, MIRNet, and D\&E. We adopted these module that were trained on other datasets to restore the images in the ELD dataset, where MIRNet, D\&E, and our UTVNet were trained on the sRGB-SID dataset, AGLLNet was trained on the synthetic low-light dataset with Gaussian-Poisson mixed noise model. 

\textbf{Quantitative and visual comparisons.} 
The quantitative results  are shown in Table 2. Our UTVNet obtained competitive performance on PSNR, SSIM, and LPIPS scores. 

In visual comparison,  we visualize the average noise level map $\mathbf{M}^{avg}$ of all iterations in UTVNet. First, we evaluate the results under different ISO levels in Fig.\;5. The outputs of AGLLNet suffer from artifacts.\;D\&E achieves better color quality at ISO $1600$ but is still noisy. When increasing the ISO to $3200$, guided by the estimated noise level map, our UTVNet generates clearer result than the outputs of MIRNet and D\&E. Next, we compare the visual results at different exposure times as shown in Fig.\;6. The color is not recovered correctly by MIRNet.  When reducing the exposure time from $1/400$ to $1/800$, fewer photons can be gathered by the camera. The proposed UTVNet introduces fewer artifacts than the state-of-the-art methods.

\subsection{Analysis and Discussions}

\textbf{Ablation study.}
The results of the ablation study are shown in Table 4. The first two rows in Table 4 demonstrate the importance of the noise level map $\mathbf{M}$. To estimate the effectiveness of the NLI-Block, we first remove the global noise variance $\boldsymbol{\mathbf{\varepsilon}}$, so that parameter $\mathbf{M}$ is directly learned from the convolution layers in the NLI-Block, which cause a performance drop as shown in the third row of Table 4. Because the pixel value is small in extremely low-light images, modifying the global noise variance is a better way to estimate the differences in noise levels among pixels. Similarly, replacing $\mathcal{A}(\cdot)$ with ReLU also causes a performance drop. The results verify the effectiveness of the way to approximate the noise level map and demonstrate that $\mathcal{A}(\cdot)$ provides a larger noise level value for guaranteeing the role of $\mathbf{M}$ as the balancing parameter.

\begin{table}
\caption{ Analysis of computational cost.  Our UTVNet achieves  SOTA performance with a balanced computational requirement.  }
\vspace{0.4mm}
\footnotesize
\setlength{\extrarowheight}{1.2pt}
\setlength{\tabcolsep}{2.5mm}
\begin{tabular}{ccccc}
\toprule
\textbf{Method}     & \textbf{Params (M)} & \textbf{Time (s)} & \textbf{MACs (G)} & \textbf{PSNR (dB)} \\
\midrule
DeepLPF     & 1.716      & 0.154    & 49.930   & 19.645    \\
Pixel2Pixel & 11.38      & 0.047    & 217.71   & 21.182    \\
SID         & 7.760      & 0.060    & 164.62   & 21.212    \\
D\&E        & 8.621      & 0.234    & 417.36   & 22.157    \\
MIRNet      & 3.107      & 0.756    & 929.43   & 22.345    \\ \hline
\textbf{UTVNet}      & 7.745      & 0.199    & 273.76   & 22.691  \\
\bottomrule 
\end{tabular}
\vspace{-0.2cm}
\end{table}

\vspace{0.5cm}
\begin{figure}
 \subfigcapskip=-3pt
 \subfigbottomskip=-3pt
 
    \subfigure 
    {  \footnotesize \centering
        \begin{minipage}[b]{0.115\textwidth}
            \centering
     \stackunder[5pt]{\includegraphics[width=1\linewidth,height=1.3cm]{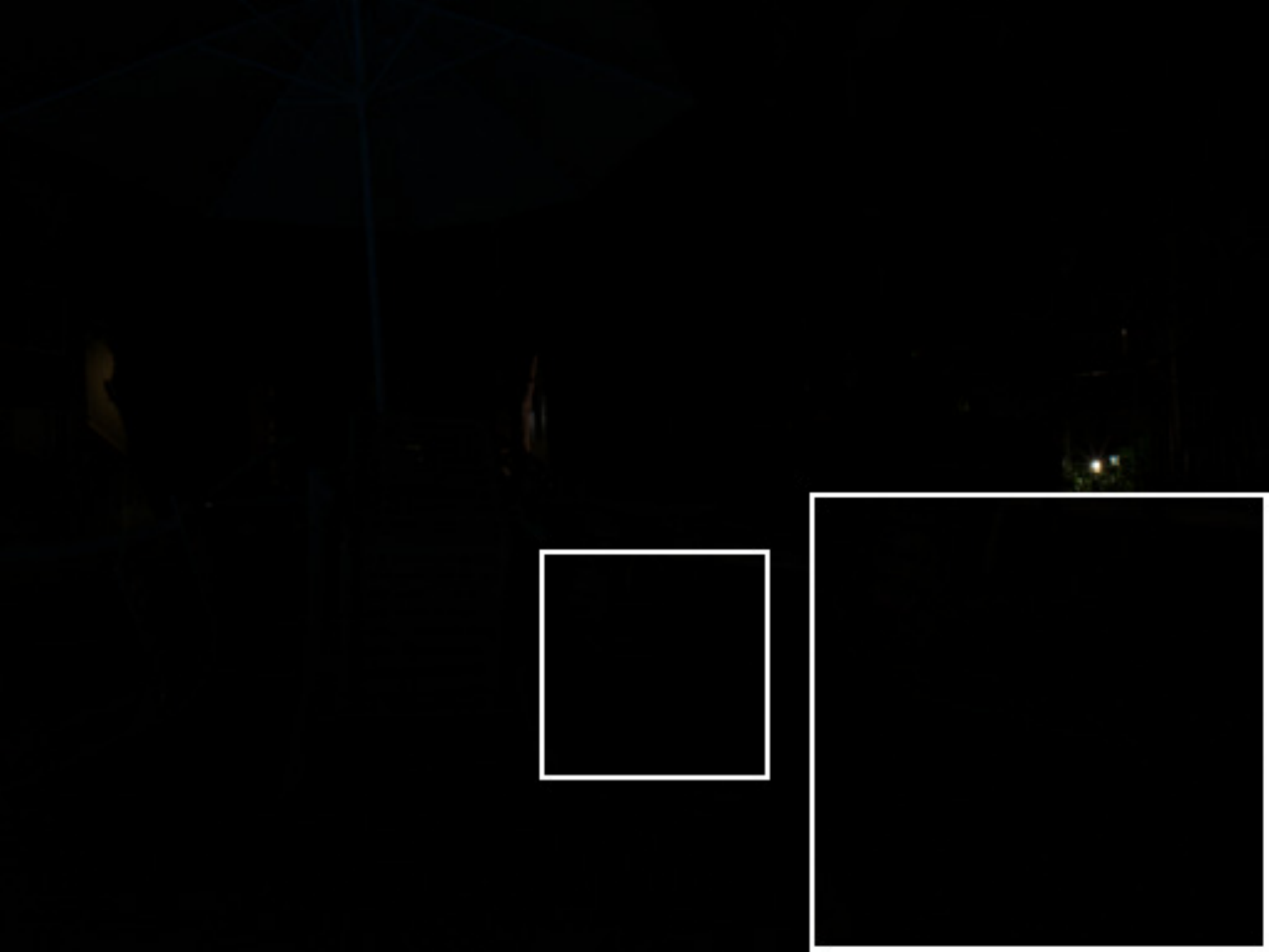}}{Input}
        \end{minipage}\hspace{-1.7mm}
   \hspace{1mm}\centering
        \begin{minipage}[b]{0.115\textwidth}
            \centering
     \stackunder[5pt]{       \includegraphics[width=1\linewidth,height=1.3cm]{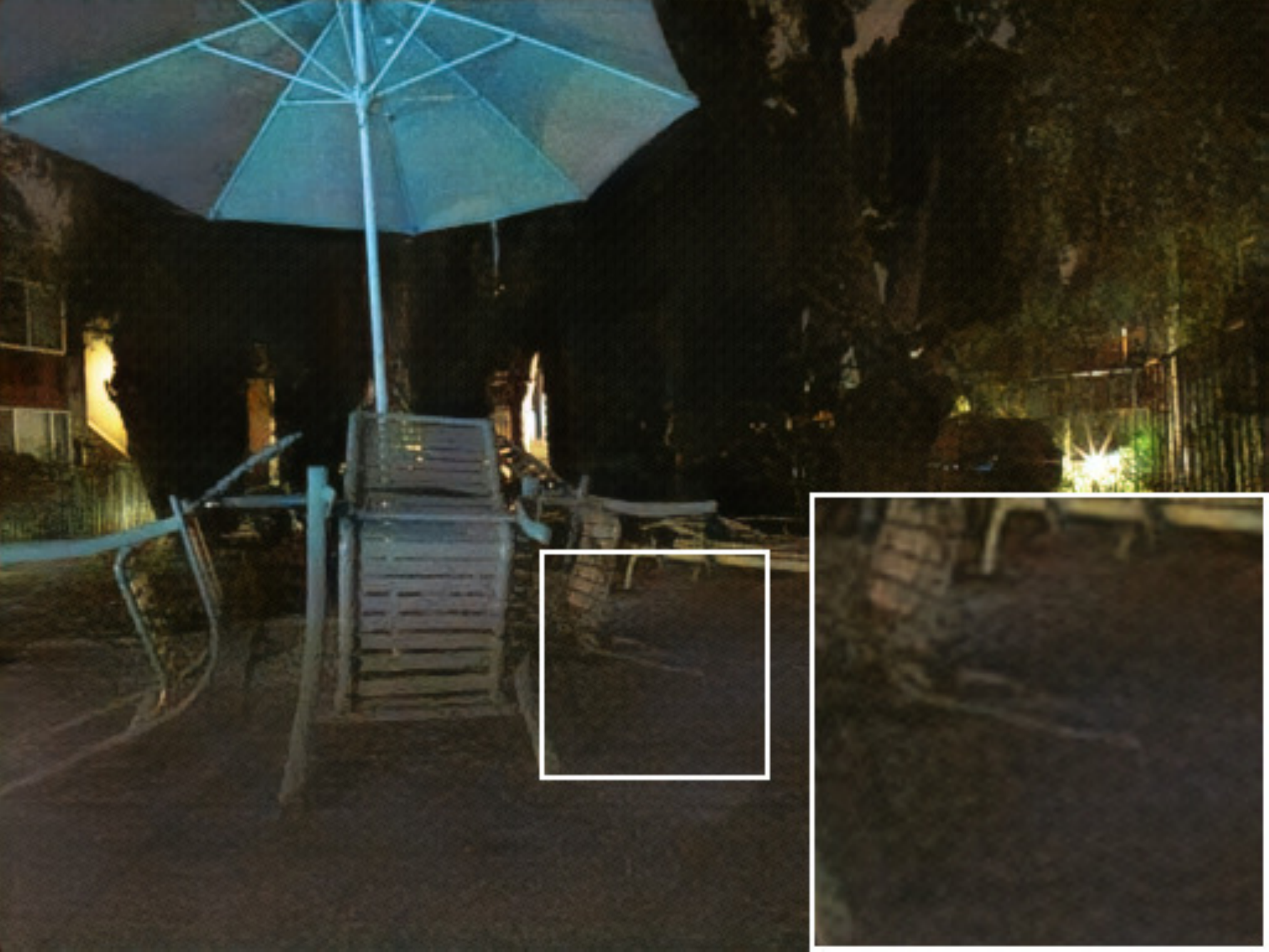}}{D\&E}
        \end{minipage}\hspace{-1mm}
   \hspace{1mm}\centering
        \begin{minipage}[b]{0.115\textwidth}
            \centering
 \stackunder[5pt] {          \includegraphics[width=1\linewidth,height=1.3cm]{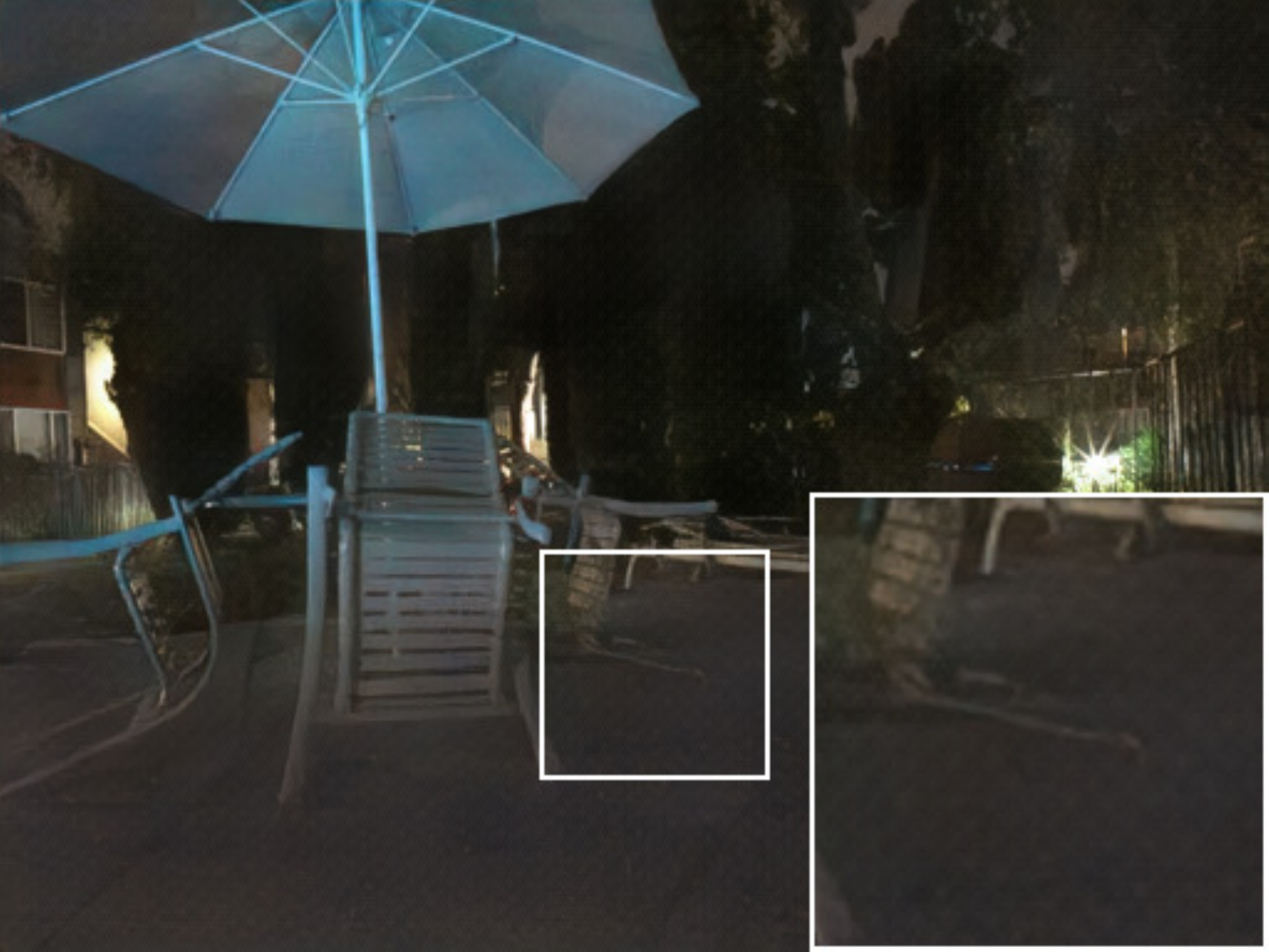}}{UTVNet}
        \end{minipage}\hspace{-1mm}
   \hspace{1mm}\centering
        \begin{minipage}[b]{0.115\textwidth}
            \centering
 \stackunder[5pt] {          \includegraphics[width=1\linewidth,height=1.3cm]{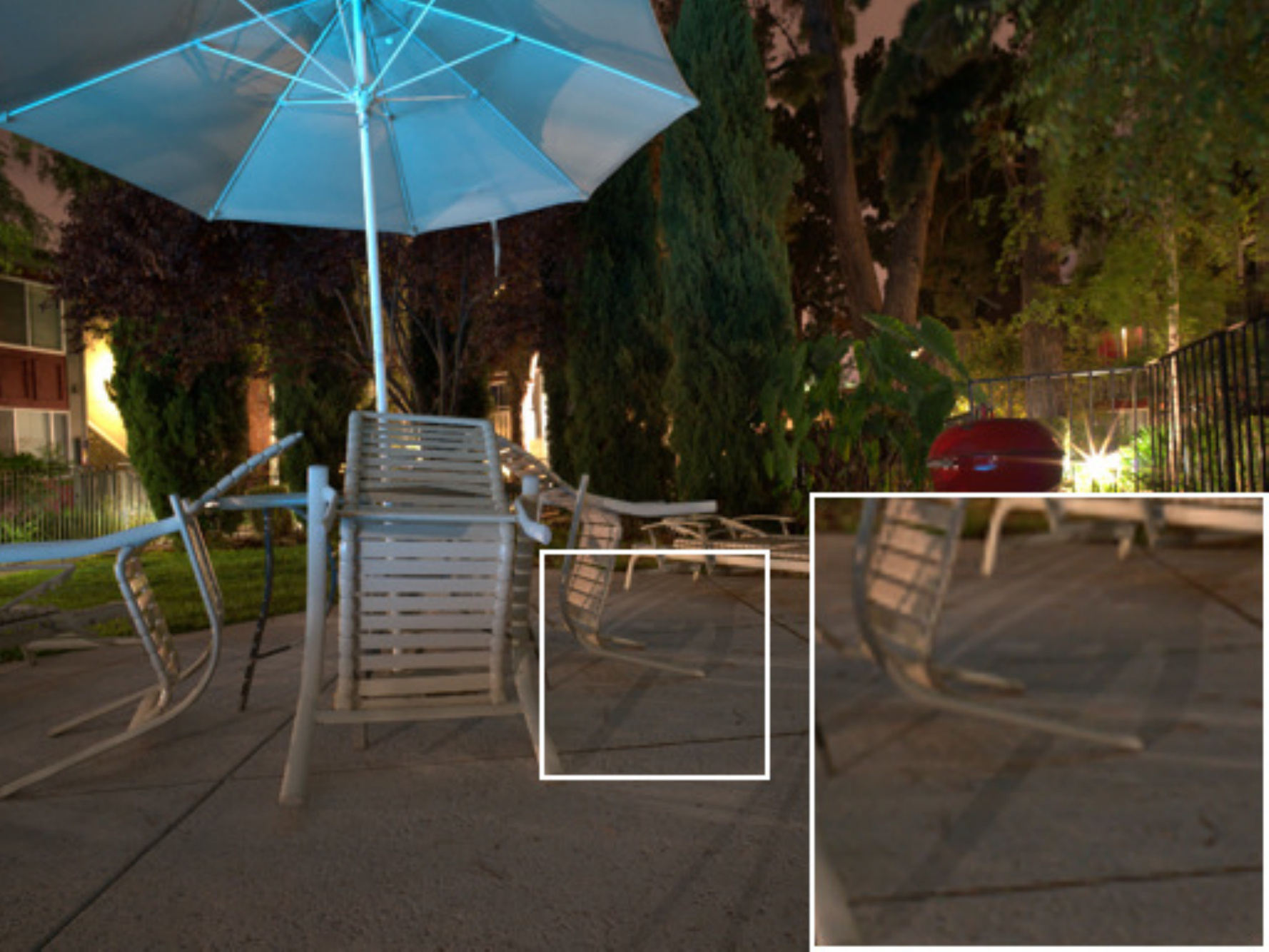}}{Ground Truth}
        \end{minipage}
     }
\vspace{-0.05cm}
\caption{The example of failure case when restoring images. 
   }  \vspace{-0.5cm}
\end{figure}

\vspace{-0.4cm}
 \textbf{Analysis of the unfolding TV module.}
 We investigate the estimation of the noise level map in each iteration. The average noise level maps of three R, G, B channels are calculated for visualization in Fig.\;7. In most of the iterations, the noise level values of sharp edges or object boundaries is lower than that of other regions, which means when smoothing the regions containing edges and boundaries, the output is mainly determined by the fidelity term. Such noise level maps are essential for noise suppression, as they guarantee the detail preservation after noise suppression, which is the main reason that our UTVNet can recover finer details compared with the state-of-the-art methods. We also analyze the iteration times, with results shown in Table 4. For a speed-accuracy trade-off, we adopt UTVNet with 8 iterations.

\vspace{1mm}

\textbf{Limitation.} As shown in Fig.\;8, our UTVNet as well as other state-of-the-art approaches may fail to restore the details with similar color as the background, and introduce artificial halos. Such a problem is more difficult to tackle in sRGB color space compared with  raw Bayer space.

 \begin{table}
\caption{Ablation study and execution time with different numbers of iteration on sRGB-SID dataset.}
\vspace{1mm}
\footnotesize
\setlength{\extrarowheight}{1.2pt}
\setlength{\tabcolsep}{2.8mm}
\begin{tabular}{lcccc}
\toprule
 \textbf{Method} &  \textbf{PSNR} &  \textbf{SSIM}   &  \textbf{LPIPS} & \textbf{Time(s)}\\
\midrule
w/o $\boldsymbol{\mathbf{M}}$        &21.858&0.6938 &0.3519&-   \\
$\boldsymbol{\mathbf{M}}$ $\rightarrow$ ${\boldsymbol{\mathbf{\varepsilon}}}$  &21.483&0.6930&0.3547&-\\
w/o ${\boldsymbol{\mathbf{\varepsilon}}}$ in NLI-Block  &21.909&0.6974&0.3511 &-\\
$\mathcal{A}(\cdot)$ $\rightarrow$ ReLU &22.136&0.7050 &0.3432&-\\\hline
$k=4$&22.476&0.6897&0.3479&\textbf{0.159}\\
$k=6$&22.539&0.6985&0.3422&0.171\\
$k=10$&22.647&\textbf{0.7203}&0.3379&0.220\\
$k=12$&22.521&0.7124&\textbf{0.3368}&0.246\\\hline
\textbf{UTVNet} ($k=8$) &\textbf{22.691}&0.7179&0.3417& 0.199 \\ 
\bottomrule
\end{tabular}
\vspace{-0.6cm}
\end{table}

\section{Conclusion}
In this paper, we propose an adaptive unfolding total variation network, referred to as UTVNet, for real-world low-light noisy image enhancement in the sRGB domain. Inspired by the relationship between the noise level and the balancing parameter in model-based denoising methods, we formulate an adaptive total variation regularization by introducing a learnable noise level map that serves as the balancing parameters for each pixel to control the trade-off between noise reduction and detail preservation. The TV minimization problem is unfolded with fidelity and smoothness constraints to learn the noise level map in an end-to-end manner, and generating a noise-free smooth layer. Furthermore, we also design a  noise level initialization block to predict noise level maps in a coarse-to-fine way before the unfolding TV architecture. Extensive experimental results have verified the capability of the proposed UTVNet. We expect future work to make further improvements in image quality, especially  in real captured low-light scenes.\vspace{4mm}

\noindent\textbf{Acknowledgments.} This work is supported by Ministry of Science and Technology China (MOST) Major Program on New Generation of Artificial Intelligence 2030 No. 2018AAA0102200. It is also supported by Natural Science Foundation China (NSFC) Major Project No. 61827814 and Shenzhen Science and Technology Innovation Commission (SZSTI) Project No. JCYJ20190808153619413.

{\small
\bibliography{egbib}

\begin{thebibliography}{10}

\bibitem{LIME}
X.~{Guo}, Y.~{Li}, and H.~{Ling}.
\newblock Lime: Low-light image enhancement via illumination map estimation.
\newblock {\em IEEE Transactions on Image Processing}, 26(2):982--993, 2017.

\bibitem{nbn}
Shen Cheng, Yuzhi Wang, Haibin Huang, Donghao Liu, Haoqiang Fan, and Shuaicheng
  Liu.
\newblock Nbnet: Noise basis learning for image denoising with subspace
  projection.
\newblock In {\em Proceedings of the IEEE/CVF Conference on Computer Vision and
  Pattern Recognition (CVPR)}, pages 4896--4906, June 2021.

\bibitem{atl}
Feifan Lv, Yu~Li, and Feng Lu.
\newblock Attention guided low-light image enhancement with a large scale
  low-light simulation dataset.
\newblock {\em International Journal of Computer Vision}, 129(7):2175--2193,
  2021.

\bibitem{LDE}
K.~{Xu}, X.~{Yang}, B.~{Yin}, and R.~W.~H. {Lau}.
\newblock Learning to restore low-light images via
  decomposition-and-enhancement.
\newblock In {\em 2020 IEEE/CVF Conference on Computer Vision and Pattern
  Recognition (CVPR)}, pages 2278--2287, 2020.

\bibitem{MIRNet}
Syed~Waqas Zamir, Aditya Arora, Salman Khan, Munawar Hayat, Fahad~Shahbaz Khan,
  Ming-Hsuan Yang, and Ling Shao.
\newblock Learning enriched features for real image restoration and
  enhancement.
\newblock In {\em ECCV}, 2020.

\bibitem{rewd}
Kaixuan Wei, Ying Fu, Jiaolong Yang, and Hua Huang.
\newblock A physics-based noise formation model for extreme low-light raw
  denoising.
\newblock In {\em Proceedings of the IEEE/CVF Conference on Computer Vision and
  Pattern Recognition (CVPR)}, June 2020.

\bibitem{googlo}
Samuel~W Hasinoff, Dillon Sharlet, Ryan Geiss, Andrew Adams, Jonathan~T Barron,
  Florian Kainz, Jiawen Chen, and Marc Levoy.
\newblock Burst photography for high dynamic range and low-light imaging on
  mobile cameras.
\newblock {\em ACM Transactions on Graphics (TOG)}, 35(6):1--12, 2016.

\bibitem{whitebox}
Y.~{Hu}, H.~{He}, C.~{Xu}, and and S.~{Lin} B.~{Wang}.
\newblock Exposure: A white-box photo post-processing framework.
\newblock {\em ACM Trans. Graph.}, 37(2), May 2018.

\bibitem{upe}
R.~{Wang}, Q.~{Zhang}, C.~{Fu}, X.~{Shen}, W.~{Zheng}, and J.~{Jia}.
\newblock Underexposed photo enhancement using deep illumination estimation.
\newblock pages 6842--6850, 2019.

\bibitem{FFDnet}
K.~{Zhang}, W.~{Zuo}, and L.~{Zhang}.
\newblock Ffdnet: Toward a fast and flexible solution for cnn-based image
  denoising.
\newblock {\em IEEE Transactions on Image Processing}, 27(9):4608--4622, 2018.

\bibitem{cbdnet}
S.~{Guo}, Z.~{Yan}, K.~{Zhang}, W.~{Zuo}, and L.~{Zhang}.
\newblock Toward convolutional blind denoising of real photographs.
\newblock In {\em 2019 IEEE/CVF Conference on Computer Vision and Pattern
  Recognition (CVPR)}, pages 1712--1722, 2019.

\bibitem{adam}
D.~P. Kingma and J.~Ba.
\newblock Adam: A method for stochastic optimization.
\newblock {\em arXiv preprint arXiv:1412.6980}, 2014.

\bibitem{csrnet}
Jingwen He, Yihao Liu, Yu~Qiao, and Chao Dong.
\newblock Conditional sequential modulation for efficient global image
  retouching.
\newblock In {\em European Conference on Computer Vision}, pages 679--695.
  Springer, 2020.

\bibitem{hdrnet}
M.~Gharbi, andJ. T.~{Barron} J.~Chen, S.~W. Hasinoff, and F.~{Durand}.
\newblock Deep bilateral learning for real-time image enhancement.
\newblock {\em ACM Transactions on Graphics (TOG)}, 36(4):1--12, 2017.

\bibitem{crm}
Zhenqiang Ying, Ge~Li, Yurui Ren, Ronggang Wang, and Wenmin Wang.
\newblock A new low-light image enhancement algorithm using camera response
  model.
\newblock In {\em Proceedings of the IEEE International Conference on Computer
  Vision (ICCV) Workshops}, Oct 2017.

\bibitem{learningtosee}
C.~{Chen}, Q.~{Chen}, J.~{Xu}, and V.~{Koltun}.
\newblock Learning to see in the dark.
\newblock In {\em 2018 IEEE/CVF Conference on Computer Vision and Pattern
  Recognition (CVPR)}, pages 3291--3300, 2018.

\bibitem{guided}
K.~{He}, J.~{Sun}, and X.~{Tang}.
\newblock Guided image filtering.
\newblock {\em IEEE Transactions on Pattern Analysis and Machine Intelligence},
  35(6):1397--1409, 2013.

\bibitem{nlic}
J.~{Xu}, L.~{Zhang}, D.~{Zhang}, and X.~{Feng}.
\newblock Multi-channel weighted nuclear norm minimization for real color image
  denoising.
\newblock pages 1105--1113, 2017.

\bibitem{nan}
S.~{Nam}, Y.~{Hwang}, Y.~{Matsushita}, and S.~J. {Kim}.
\newblock A holistic approach to cross-channel image noise modeling and its
  application to image denoising.
\newblock pages 1683--1691, 2016.

\bibitem{nonelocal}
A.~{Buades}, B.~{Coll}, and J.~. {Morel}.
\newblock A non-local algorithm for image denoising.
\newblock In {\em 2005 IEEE Computer Society Conference on Computer Vision and
  Pattern Recognition (CVPR'05)}, volume~2, pages 60--65 vol. 2, 2005.

\bibitem{sparsity}
M.~{Elad} and M.~{Aharon}.
\newblock Image denoising via sparse and redundant representations over learned
  dictionaries.
\newblock {\em IEEE Transactions on Image Processing}, 15(12):3736--3745, 2006.

\bibitem{rank}
S.~{Gu}, L.~{Zhang}, W.~{Zuo}, and X.~{Feng}.
\newblock Weighted nuclear norm minimization with application to image
  denoising.
\newblock In {\em 2014 IEEE Conference on Computer Vision and Pattern
  Recognition}, pages 2862--2869, 2014.

\bibitem{admm}
Mingyi Hong and Zhi-Quan Luo.
\newblock On the linear convergence of the alternating direction method of
  multipliers.
\newblock {\em Mathematical Programming}, 162(1-2):165--199, 2017.

\bibitem{rui}
R.~{Liu}, Z.~{Jiang}, X.~{Fan}, and Z.~{Luo}.
\newblock Knowledge-driven deep unrolling for robust image layer separation.
\newblock {\em IEEE Transactions on Neural Networks and Learning Systems},
  31(5):1653--1666, 2020.

\bibitem{adcov}
R.~{Liu}, S.~{Cheng}, L.~{Ma}, X.~{Fan}, and Z.~{Luo}.
\newblock Deep proximal unrolling: Algorithmic framework, convergence analysis
  and applications.
\newblock {\em IEEE Transactions on Image Processing}, 28(10):5013--5026, 2019.

\bibitem{hqsp}
M.~V. {Afonso}, J.~M. {Bioucas-Dias}, and M.~A.~T. {Figueiredo}.
\newblock Fast image recovery using variable splitting and constrained
  optimization.
\newblock {\em IEEE Transactions on Image Processing}, 19(9):2345--2356, 2010.

\bibitem{dnopi}
W.~{Dong}, P.~{Wang}, W.~{Yin}, G.~{Shi}, F.~{Wu}, and X.~{Lu}.
\newblock Denoising prior driven deep neural network for image restoration.
\newblock {\em IEEE Transactions on Pattern Analysis and Machine Intelligence},
  41(10):2305--2318, 2019.

\bibitem{bygu}
K.~{Zhang}, W.~{Zuo}, Y.~{Chen}, D.~{Meng}, and L.~{Zhang}.
\newblock Beyond a gaussian denoiser: Residual learning of deep cnn for image
  denoising.
\newblock {\em IEEE Transactions on Image Processing}, 26(7):3142--3155, 2017.

\bibitem{USRnet}
K.~{Zhang}, L.~{Van Gool}, and R.~{Timofte}.
\newblock Deep unfolding network for image super-resolution.
\newblock In {\em 2020 IEEE/CVF Conference on Computer Vision and Pattern
  Recognition (CVPR)}, pages 3214--3223, 2020.

\bibitem{tved}
Robert Acar and Curtis~R Vogel.
\newblock Analysis of bounded variation penalty methods for ill-posed problems.
\newblock {\em Inverse problems}, 10(6):1217, 1994.

\bibitem{newyv}
Yilun Wang, Junfeng Yang, Wotao Yin, and Yin Zhang.
\newblock A new alternating minimization algorithm for total variation image
  reconstruction.
\newblock {\em SIAM Journal on Imaging Sciences}, 1(3):248--272, 2008.

\bibitem{tipbio}
Kai-Fu Yang, Xian-Shi Zhang, and Yong-Jie Li.
\newblock A biological vision inspired framework for image enhancement in poor
  visibility conditions.
\newblock {\em IEEE Transactions on Image Processing}, 29:1493--1506, 2020.

\bibitem{ladmmtv}
S.~H. {Chan}, R.~{Khoshabeh}, K.~B. {Gibson}, P.~E. {Gill}, and T.~Q. {Nguyen}.
\newblock An augmented lagrangian method for total variation video restoration.
\newblock {\em IEEE Transactions on Image Processing}, 20(11):3097--3111, 2011.

\bibitem{nes}
John Immerkær.
\newblock Fast noise variance estimation.
\newblock {\em Computer Vision and Image Understanding}, 64(2):300--302, 1996.

\bibitem{bd}
B.~{Mildenhall}, J.~T. {Barron}, J.~{Chen}, D.~{Sharlet}, R.~{Ng}, and
  R.~{Carroll}.
\newblock Burst denoising with kernel prediction networks.
\newblock In {\em 2018 IEEE/CVF Conference on Computer Vision and Pattern
  Recognition}, pages 2502--2510, 2018.

\bibitem{unet}
O.~{Ronneberger}, P.~{Fischer}, and T.~{Brox}.
\newblock U-net: Convolutional networks for biomedical image segmentation.
\newblock In {\em International Conference on Medical image computing and
  computer-assisted intervention}, pages 234--241. Springer, 2015.

\bibitem{dped}
A.~{Ignatov}, N.~{Kobyshev}, R.~{Timofte}, and K.~{Vanhoey}.
\newblock Dslr-quality photos on mobile devices with deep convolutional
  networks.
\newblock In {\em 2017 IEEE International Conference on Computer Vision
  (ICCV)}, pages 3297--3305, 2017.

\bibitem{lpf}
S.~{Moran}, P.~{Marza}, S.~{McDonagh}, S.~{Parisot}, and G.~{Slabaugh}.
\newblock Deeplpf: Deep local parametric filters for image enhancement.
\newblock In {\em 2020 IEEE/CVF Conference on Computer Vision and Pattern
  Recognition (CVPR)}, pages 12823--12832, 2020.

\bibitem{p2p}
Phillip Isola, Jun-Yan Zhu, Tinghui Zhou, and Alexei~A Efros.
\newblock Image-to-image translation with conditional adversarial networks.
\newblock In {\em Proceedings of the IEEE conference on computer vision and
  pattern recognition}, pages 1125--1134, 2017.

\bibitem{ssim}
W.~{Zhou }, A.~C. {Bovik}, H.~R. {Sheikh}, and E.~P. {Simoncelli}.
\newblock Image quality assessment: from error visibility to structural
  similarity.
\newblock {\em IEEE Transactions on Image Processing}, 13(4):600--612, 2004.

\bibitem{lpips}
Richard R.~{Zhang}, P.~{Isola}, A.~A. {Efros}, E.~{Shechtman}, and O.~{Wang}.
\newblock The unreasonable effectiveness of deep features as a perceptual
  metric.
\newblock In {\em Proceedings of the IEEE Conference on Computer Vision and
  Pattern Recognition (CVPR)}, June 2018.

\end{thebibliography}
}

\end{document}